\documentclass[twocolumn,superscriptaddress,amssymb,amsmath,footinbib,aps,prd,showpacs]{revtex4-1}
\pdfoutput=1

\usepackage{graphicx}
\usepackage{dcolumn}
\usepackage{hyperref}

\usepackage{amsfonts}
\usepackage{resizegather}
\usepackage{caption}
\usepackage{color}
\usepackage{subfigure}
\usepackage{ulem}
\usepackage{blindtext}
\usepackage{url}

\usepackage{lipsum}
\usepackage{graphicx}
\usepackage[font=footnotesize,labelfont=bf,
   justification=justified,
   format=plain]{caption} 
   
\usepackage{array,mathtools,amssymb,booktabs}
\newcolumntype{C}{>{$}c<{$}}

\newcommand{\luis}[1]{\textcolor{black}{#1}}
\newcommand{\rev}[1]{\textcolor{black}{#1}}

\newcommand{\wde}[1]{$w_{\tt DE}(z)$#1}
\newcommand{\rde}[1]{$\rho_{\tt DE}(z)/\rho_{\tt c,0}$#1}

\begin{document}

\title{Model selection applied to 
reconstructions of the Dark Energy \\
}

 


\author{Luis A. Escamilla}
\email{luis.escamilla@icf.unam.mx}
\affiliation{Instituto de Ciencias F\'isicas, Universidad Nacional Aut\'onoma de M\'exico, Cuernavaca, Morelos, 62210, M\'exico}

\author{J. Alberto Vazquez}
\email{javazquez@icf.unam.mx}
\affiliation{Instituto de Ciencias F\'isicas, Universidad Nacional Aut\'onoma de M\'exico, Cuernavaca, Morelos, 62210, M\'exico}

\begin{abstract}


The main aim of this paper is to perform a model comparison for 
some reconstructions of the key properties that describe the dark energy of the Universe i.e. energy density and the equation of state (EoS). 
We carry out this process by using a binning and a linear interpolation methodologies, and on top of that, we incorporate a correlation function mechanism. An extension of the two of them was also introduced, where internal amplitudes are allowed to vary in height as well as in position. 
The reconstructions were made with data from the Hubble parameter, Supernovae Type Ia and Baryon Acoustic Oscillations (H+SN+BAO), all of which span a range from $z=0.01$ to $z=2.34$. First we perform the parameter estimation for each of the reconstructions to then provide a model selection through the Bayesian Evidence. Throughout our process we found a better fit to the data, up to $4\sigma$ compared to $\Lambda$CDM, and the presence of some interesting features, i.e. an oscillatory behaviour at late times, a decrease in the dark energy density component at early times and a transition to the phantom divide-line in the EoS. 
To discern these features from noisy contributions, we include a principal component analysis  and found that some of these characteristics should be taken into account to satisfy current observations. 

\end{abstract}

\maketitle

\section{INTRODUCTION}\label{section:intro}

Since the discovery of the current accelerated expansion of the Universe, cosmologists have been challenged to explain the cause of this mysterious phenomenon.
In order to provide a possible explanation, the idea of a new exotic source was introduced, called dark energy (DE). 
In addition to the cosmological constant $\Lambda$, being the simplest assumption to be the dark energy, there is also a still-unknown key component for structure formation in the Universe, coined as Cold Dark Matter (hereafter CDM). These two dark components \luis{comprise} the basis of the standard cosmological model or $\Lambda$CDM. 
Even though the $\Lambda$CDM model has had remarkable achievements, i.e. satisfies most of the current cosmological observations on different scales, it still presents several shortcomings. 
On the theoretical side, for instance, a disagreement encountered by cosmology and quantum field theory about the vacuum predictions   \cite{sahni2002cosmological}; on the observational side several problems began to arise, for example the discrepancy on the measurements 
of the Hubble constant $H_0$ among datasets \cite{bull2016beyond}. This so-called Hubble tension seems to be enhancing as observations become more precise \cite{zhao2017dynamical} \rev{and even a variable $H_0$ (also named ``running'') has also been found \cite{Wong:2019kwg, Krishnan:2020vaf, Hu:2022kes, Krishnan:2022fzz, Colgain:2022rxy}, further indicating the standard model's inability to reconcile late-time data with high-redshift data.}
The presence of these issues opens up the possibility to consider alternatives beyond the $\Lambda$CDM model. A natural extension to the cosmological constant is to allow a redshift dependency either through the equation of state (EoS) \wde{}  or the energy density $\rho_{\tt DE}(z)$  for the dark energy. 
Different proposals with a dynamical behaviour have been presented. Some of them include scalar fields in the form of Quintessence, Phantom \cite{Vazquez:2020ani, caldwell2002phantom, caldwell2005limits} or the two of them combined, called Quintom models \cite{guo2005cosmological}; others intent to modify the theory of general gravity known as $f(R)$ models \cite{nojiri2011unified}, anisotropic massive Brans-Dicke gravity \cite{Akarsu:2019pvi} or theories with extra dimensions such as brane world models \cite{ida2000brane, brax2004brane}. 
\\

On the other hand, in the absence of a fundamental and well defined theory, several parameterizations to cosmological functions
have been suggested to get insights of the general DE behaviour and hence to look for possible deviations from the cosmological constant \cite{sahni2006reconstructing}. Some of them include quantities such as the deceleration parameter $q(z)$ \cite{gong2007reconstruction, al2017parametric, yu2010reconstructing}, the jerk parameter $j(z)$ \cite{hernandez2020generalized, mukherjee2016parametric} or, more predominantly, the dark energy EoS \wde. 
Focusing on the EoS we can find a plethora of parameterizations to express \wde{} (or $w_{\tt DE}(t)$), for instance in terms of power-law, logarithmic, exponential and trigonometric components, or combinations of them \cite{sahni2006reconstructing, Akarsu:2015yea, Arciniega:2021ffa}. These approaches can even be classified depending on the number of parameters that describe \wde, i.e. with a single, two or more parameters \cite{yang2019observational, chevallier2001accelerating}.
Similarly, there have been, although not as many as for the equation of state (EoS), parameterizations for the dark energy density, i.e. the Generalised Emergent Dark Energy (GEDE) \cite{li2020evidence,yang2021generalized, hernandez2020generalized}, the Graduated Dark Energy (gDE) \cite{akarsu2020graduated, Acquaviva:2021jov, Akarsu:2021fol, Akarsu:2022typ}, the Early Dark Energy (EDE) \cite{poulin2018cosmological} and the Energy-Momentum Log-gravity (EMLG) \cite{Akarsu:2019ygx}. 
Generally, in these models it can be obtained an associated EoS but the main assumptions come from the density, that is, the GEDE considers directly the evolution of the density parameter, and in the gDE model an inertial mass density is introduced which allows a possible transition to negative effective energy densities.
Even though these parametric forms usually provide a better fit to the data,  they have the limitation of assuming an \textit{a priori} functional form  which may lead to some bias or misleading model-dependent results, regardless of the DE nature.  
To avoid these possible issues, non-parametric and model-independent techniques are used. They allow us to extract information directly from the data to detect features within cosmological functions. That is, the goal of non-parametric and model-independent approaches is to reconstruct (infer) an unknown quantity without a predefined shape. Non-parametric reconstructions may include Artificial Neural Networks or Gaussian processes, as they have proven to be useful in cosmology as more data become available
\cite{Keeley:2020aym, seikel2012reconstruction, holsclaw2010nonparametric, velasquez2020growth, aljaf2020constraints, yang2020evidence, gerardi2019reconstruction, yang2015reconstructing, holsclaw2013gaussian, wang2017improved, wang2020reconstructing, Gomez-Vargas:2021zyl}. Examples of model-independent ones are higher order terms on a Taylor series \cite{sahni2006reconstructing}, Fourier series expansion \cite{tamayo2019fourier} and the Pad\'e approximation \cite{basilakos2018dark}, just to mention a few.
Another example of a model-independent reconstruction (relevant to this work) was introduced by \cite{crittenden2012fables} that considers adding a correlation function term to the data. An interesting result, by using this technique, found that a dynamical form of \wde{} is preferred over the constant value with a 3.5$\sigma$ significance level \cite{zhao2017dynamical}. \rev{Also, in \cite{wang2018evolution} this same method was used directly with the DE energy-density, finding a dynamic behavior and even evidence in its favor when compared with the standard model, although this only happened when reducing some of the parameters' priors.}
Moreover, a similar path consists of combining the Principal Component Analysis (PCA) with the Goodness of Fit \cite{yu2013nonparametric}, where PCA helps to remove noisy oscillations induced by the reconstruction method to find deviations from the cosmological constant \cite{huterer2003parametrization}.  
In a similar fashion, the study in \cite{vazquez2012model} introduced the nodal reconstruction in which a piecewise linear interpolation (or cubic splines) represents the main functional form to be
reconstructed. This approach has been utilized to recover the EoS directly from the data \cite{vazquez2012model} and its final form presented a well constrained redshift dependency with a small bump at $z\approx 1.3$, but beyond $z\approx 1.5$ the data was too inaccurate to provide a good prescription. Later on, a similar form of this method was also applied in \cite{Vazquez:2013dva, handley2019bayesian}  and in \cite{aslanyan2014knotted} (although here it is referred as knot-spline reconstruction) and showed that this type of model-independent approach can be used to find specific features by allowing some central knots to vary both in height and position. More recently an improvement of these methods has been applied to the dark energy EoS \cite{hee2017constraining}.

The main aim of this work is to perform model-independent reconstructions of the dark energy features and to provide a model comparison through the Bayesian evidence and goodness-of-fit. Our analysis is carried out among the nodal reconstruction \cite{hee2017constraining}, the correlation function method \cite{crittenden2012fables}, an extension of the two of them where internal amplitudes are allowed to vary in height as well as in position, and finally, for comparison, to include some of the parametric proposals. Even though these techniques are applicable to any function describing the dark energy, we focus on the EoS and the energy density. After the reconstruction is carried out, some of the cosmological functions can therefore be derived, i.e. Hubble parameter $H(z)$, deceleration parameter $q(z)$ and the $Om(z)$ diagnostic. Finally we perform the PCA to discern possible important features from noise contributions. The novelty in this work is the joint study of the EoS and the energy density by using the model-independent approaches (nodal and step functions) to then perform model comparison through the Bayesian evidence, derived some functions in terms of the results and carried out the PCA to find preferred behaviours of DE to then keep, discard or propose other models/parameterizations. \\

The paper is organized as follows: in section \ref{section:recons_metods}
we describe the reconstruction methodology, the PCA method is also introduced and explained. Then in section \ref{section:theory} we provide a brief review of the underlying theory, datasets and some specifications about the parameter estimation and model selection. In section \ref{section:results} we present the main results, and finally in section \ref{section:conclusions} we give our conclusions. 
%

\section{Reconstruction methodology}\label{section:recons_metods}

One of the primary reconstruction methods used in this paper is built on a set of step functions. In this approach steps or bins are utilized to describe any function $f$, where the steps are connected  together via hyperbolic tangents to preserve continuity. The target function looks like this:
\begin{equation}
    f(z)= f_1+\sum^{N-1}_{i=1}\frac{f_{i+1} - f_i}{2}\bigg(1+\tanh{\Big(\frac{z-z_i}{\xi}\Big)} \bigg),
\label{bin_equation}
\end{equation}
where $N$ is the number of bins, $f_i$ the amplitude of the bin value, $z_i$ the position where the bin begins in the $z$ axis and $\xi$ a smoothness parameter. 
\\

Another approach used in this paper is the nodal reconstruction \cite{hee2017constraining}. This type of reconstruction consists of performing interpolations, either by using linear, cubic or higher order splines, to fill in the gaps amongst  certain number of nodes. The simplest example is the linear interpolation, that is, given two coordinates $(z_i, f_i)$ and $(z_{i+1}, f_{i+1})$, the interpolated function is as follows 
\begin{equation}
    f(z) = f_i + \frac{f_{i+1}- f_i}{z_{i+1}-z_i}(z-z_i), \quad z \in [z_i, z_{i+1}].
\label{linear_interp}
\end{equation}
The interpolation could also be made with higher order polynomials (or splines) to preserve smoothness, nonetheless it may present heavy correlations between nodes or introduce unwanted noise to the reconstruction, it may also present numerical problems when the amplitude values are changing too abruptly or when the nodes are positioned too close, as shown in the appendix of \cite{vazquez2012model}. 
\\

\begin{figure}[t!]
    \begin{center}
     \includegraphics[ width=9.cm, height=5.cm]{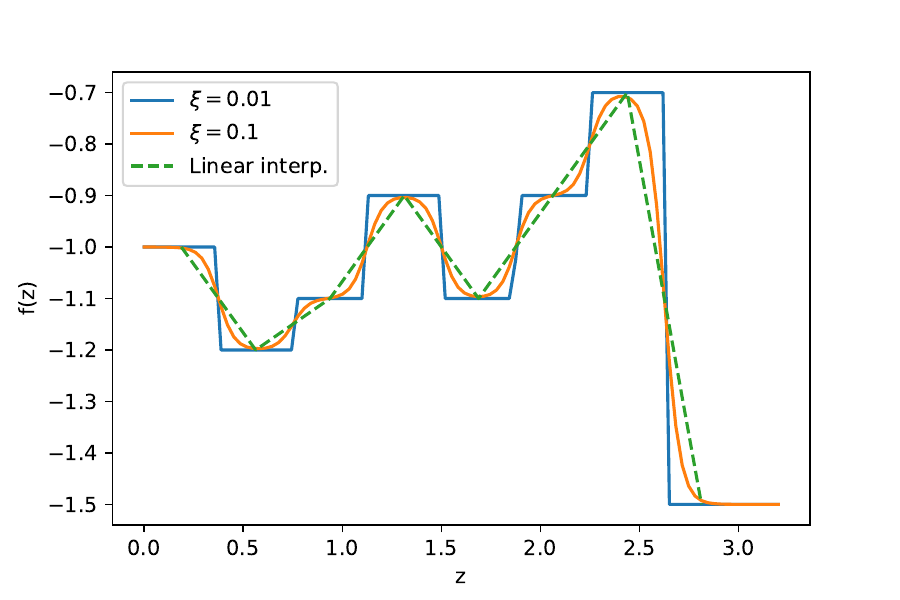}
    \end{center}
    \caption{Example of the reconstruction plot for the $f(z)$ function with 8 nodes and bins. We observe the difference in the reconstruction when using two values for the smoothness parameter $\xi$. 
    }\label{fig:tanh}
\end{figure}

In both types of reconstructions the nodes are located in space at certain positions $z_i$ and with variable amplitudes $f_i$. 
For instance, Figure \ref{fig:tanh} displays an arbitrary function $f(z)$ with amplitudes randomly  selected ($N=8$) at fixed equally distant positions and for two values of the smoothness parameter, along with the interpolation method. 

A modified version of the binning and nodal reconstruction techniques is to consider the internal positions $z_i$  be free parameters, which will allow us to capture more specific features at certain places \cite{handley2019bayesian, aslanyan2014knotted}.
Notice that in this version, the internal variable positions have to be sorted in a way to avoid possible overlapping in the reconstructions.
This approach gives more degrees of freedom (one for each variable $z_i$) to the reconstruction.
When using the linear interpolation the expected behavior is straightforward, as it only varies the lines between nodes, however in the binning process it will affect the width of the bins, so it would be easier to find specific features on the positions rather than on the amplitudes. 

Finally, there exists the possibility of either overfitting by using a very complex model with a large amount of bins (nodes) or underfitting by not capturing enough features due to the use of just few bins (nodes). 
Both of these possible issues can be managed by performing a model comparison among reconstructions through the Bayesian Evidence, that is, we modulate the impact of additional parameters and their priors to find out the most suitable to the data. This method follows the principle of simplicity and economy of explanation known as Occam's Razor \cite{trotta2008bayes} which states: \textit{``the simplest theory compatible with the available evidence ought to be preferred''}.

\subsection*{Correlation Function method}

This method is applied on top of the binning approach in order to obtain a  function that evolves smoothly \cite{crittenden2009investigating}.
The idea behind it is to treat the function in place as a random field evolving along with a correlation function $\xi$, for instance
\begin{equation} \label{eq:CPZ}
    \xi(\delta z) = \frac{\xi(0)}{1+ \left(\frac{\delta z}{z_c}\right)^2},
\end{equation}
with  $\xi(0)$ being the normalization factor and $z_c$ represents a smoothing distance. The correlation function (\ref{eq:CPZ}), named CPZ \cite{crittenden2012fables}, has a characteristic correlation length after which its contribution decreases, hence providing stronger correlations between neighboring bins when they are located at distances smaller than $z_c$. 
There exist several alternatives that can reproduce this behaviour up to a certain degree, like the exponential fall-off $\xi (\delta z) = \xi (0)e^{-\delta z/ z_c}$ or the power law $\xi (\delta z) = (\delta z/ z_c)^{-n}$, but the CPZ has a more transparent dependency in the parameters $f_i$, a relatively simpler behaviour and it constrains the high frequency modes better \cite{crittenden2012fables}.
Throughout this work, we use the following values $z_c = 0.3$ and $\xi(0) = 0.1$ since they normalize the shorter wavelength modes of the data \cite{crittenden2009investigating}.

Assuming every amplitude $f_i$ are equally distributed with the same width-location  $\Delta = z_{i+1} - z_i$, then the average of $f(z)$ over each  $f_i$ is
\begin{equation}
    f_i = \int^{z_i+\Delta}_{z_i}f(z) dz.
\end{equation}
The variation from the fiducial model averaged over the bin is $\delta f_i = f_i - f_{\rm fid}$,
where the fiducial model is the underlying scheme upon which our reconstruction will be dependant on. In this way the covariance matrix can be obtained by 
\begin{equation}
    C_{ij} \equiv \langle \delta f_i \delta f_j \rangle  = \int^{z_i +\Delta}_{z_i} dz \int^{z_j+\Delta}_{z_j} dz' \xi(|z-z'|),
\end{equation}
and therefore the associated prior
\begin{equation}
    P_{\rm prior} \propto e^{-\frac{1}{2}(f-f_{\rm fid})^{T} C^{-1} (f-f_{\rm fid})}. 
\end{equation}
Finally, once the prior is given (or equivalently $\chi_{\rm prior}^2 = -2 \ln P_{\rm prior}$),
our total $\chi^2$ to minimize becomes 
\begin{equation}
    \chi^2 = \chi_{\rm data}^2 + \chi_{\rm prior}^2.
\end{equation}
The fiducial model could be one previously known, for instance in the case of the dark energy EoS, it could be the cosmological constant with \wde $=-1$ or $\rho_{\tt DE}(z)$=constant. However, demanding a behaviour like the cosmological constant may create a bias when performing the reconstruction.
The same would be true for any other fixed fiducial model, so we opted for the floating prior proposed by \cite{crittenden2012fables}, where
\begin{equation}
    f_{i}^{\rm fid}=\sum_{|z_j-z_i|\leq z_c} \frac{f_{j}^{\rm recons}}{N_j},
\end{equation}
here $z_j$ is the position of the amplitude $f_j$ and $N_j$ is the number of amplitudes that fulfill the condition $|z_j-z_i|\leq z_c$. This floating prior makes sure that the parameter values stay continuous by evaluating the mean value of each parameter with its immediate neighbors.

\rev{It is important to mention that the correlation function method applied here is used as an ``agnostic'' way of reducing overfitting, agnostic in the sense that our imposed correlations are mainly obtained via a data-driven approach, not a theory-driven one, i.e. effective field theories (EFTs) such as Quintessence or Horndeski \cite{Colgain:2021pmf, Raveri:2017qvt, Espejo:2018hxa, Pogosian:2021mcs, Raveri:2021dbu}. This could provide a small bias against EFTs, particularly Quintessence, as explained in \cite{Colgain:2021pmf}. Also, since this method is being used as a way to diminish overfitting it could even prevent some interesting features to appear or even wither existing ones, such as the disappearance of wiggles in \cite{Pogosian:2021mcs, Raveri:2021dbu} when using a theory prior. To see if any possible bias may arise we will have an instance where we do not use the Correlation Function method (as seen in Fig. \ref{fig:rho_eos_corr_4y2x}).}

\subsection*{Bayesian statistics}

In order to perform the parameter estimation we follow
the definition of the Bayes Theorem
\begin{equation}
    P(u|D,M) = \frac{\mathcal{L}(D|u,M)P(u|M)}{E(D|M)},
\end{equation}
being $u$ the vector of parameters of our hypothesis $M$  (or model) to assess, $D$ is the experimental (observational) data, $P(u|D,M)$ the posterior probability distribution, $\mathcal{L}(D|u,M)$ the likelihood, $P(u|M)$ the prior distribution and $E(D|M)$ the Bayesian evidence. 
Once the Bayesian evidence is computed for two models $M_1$ and $M_2$, the Bayes factor
is defined as 
\begin{equation}
    B_{12} \equiv \frac{E(D|M_1)}{E(D|M_2)}.
\end{equation}
This quotient, together with the Jeffrey's scale shown in Table \ref{jeffreys} \cite{trotta2008bayes,efron2001scales}, is a great empirical tool for performing model selection, that is, it gives an insight of how good a model $M_1$ is when compared to model $M_2$. An example of this method can be seen in \cite{cedeno2019bayesian} where the authors compared some DE models to $\Lambda$CDM. In this work, $M_1$ will correspond to $\Lambda$CDM and $M_2$ will be any of our reconstructions in order to make a direct comparison with the standard model. \luis{Even so, it is important to mention that Jeffreys' scale is empirical in nature and sometimes rule out the true model \cite{Nesseris:2012cq}; added to this we have the dependence of the Bayesian evidence on priors and on model constrains \cite{Keeley:2021dmx}. So, even though it is a great tool for comparison, it should not be taken as completely decisive when performing model selection.} 

\begin{table}
\captionsetup{justification=raggedright,singlelinecheck=false,font=footnotesize}
\footnotesize
\scalebox{1.2}{%
\begin{tabular}{cccc} 
\cline{1-4}\noalign{\smallskip}
 \vspace{0.15cm}

 $\ln{B_{12}}$ & Odds  &   Probability  &  Strength of evidence \\
\hline
 
\hline
\vspace{0.15cm}
$<$ 1.0 & $<$3:1 & $<$0.75 & Inconclusive  \\
\vspace{0.15cm}
1.0 &  $\sim$3:1 & 0.750 & Weak evidence  \\
\vspace{0.15cm}
2.5 & $\sim$12:1  &  0.923  & Moderate evidence  \\
\vspace{0.15cm}
5.0 & $\sim$150:1 &  0.993  & Strong evidence  \\

\hline
\hline
\end{tabular}}
\caption{Jeffreys' scale for model selection with the logarithm of the Bayes' factor. Using the convention from \cite{trotta2008bayes}.}
\label{jeffreys}
\end{table}

\subsection*{Principal Component Analysis}\label{section:pca}

After doing the reconstruction we can perform a process known as Principal Component Analysis (PCA) on the parameter space to draw conclusions about the data and how well they are constraining the parameter space. Once the parameter inference is complete, we compute the Fisher matrix of the parameters $w_i$, which is $F=C^{-1}$, where $C$ is the covariance matrix. Then, we diagonalize $F$ to find a basis where the parameters are uncorrelated, so that 
 \begin{equation}
     F=W^T D W,
 \end{equation}
where the rows of $W$ are the eigenvectors $e_i(z)$ of the basis in which our parameters are uncorrelated and $D$ is a diagonal matrix. If $\Vec{p}$ is the vector of the best-fit values of our $f_i$ then the new uncorrelated parameters are $\Vec{q}=W\Vec{p}$. Being $d_i$ the diagonal elements of $D$ sorted out such that $d_1>d_2>...>d_N$, and also their corresponding $e_i(z)$ and $q_i$. These $d_i$ are related to the errors as $\sigma_i =\frac{1}{\sqrt{d_i}}$. Then, we can reconstruct the function in place 
\begin{equation}
    f(z) = \sum^{N}_{i=1} q_i e_i(z),
\end{equation}
with standard deviation 
\begin{equation}
    \sigma(f(z_n)) = \bigg( \sum^{N}_{i=1} \sigma^2(q_i) e_{i}^{2}(z_n) \bigg)^{1/2},
\end{equation}
where $z_n$ is the location for each parameter $f_i$. 
Now we can choose any number of principal components (PCs) to reconstruct the function (from one PC up to the original number of parameters in the reconstruction). If we have the same number of PCs as there are bins then the reconstruction will look the same as no variance has been removed, but by removing the PCs with smaller $d_i$ contribution, then we remove the noisiest aspects of the reconstruction (with the biggest errors $\sigma_i$). A common practice is to remove enough PCs to maintain $95\%$ of the information or variance.

\section{Theory and datasets}\label{section:theory}

For a homogeneous and isotropic flat universe given by the Friedmann-Robertson-Walker metric, the Friedmann equation describing the dynamical evolution, in terms of redshift $z$, can be written as 
\begin{equation}
\frac{H^2(z)}{H^{2}_{0}}= \Omega_{\rm r,0}( 1+z )^4 + \Omega_{\rm m,0}(1+z)^3 + \Omega_{\Lambda,0},
\label{friedmann}
\end{equation}
where the Hubble parameter $H(z)$ represents the expansion rate of the Universe, with $H_0$ being its value today. Here, we have the components of the Universe written in terms of the dimensionless density parameters $\Omega_i(z)\equiv\rho_i(z)/\rho_{\tt c}(z)$, where their contribution at the present time (represented by subscript 0) are $\Omega_{\rm r,0}$ for the relativistic particles, $\Omega_{\rm m,0}$ describes the  matter content (baryons and dark matter) and  $\Omega_{\Lambda, 0}$ is associated to the cosmological constant;
$\rho_{\tt c}$ is the critical density of a spatially flat Universe.

By letting aside the cosmological constant and allowing a dynamical dark energy component, the last term in the equation (\ref{friedmann}) is replaced by 
$\Omega_{\Lambda,0} \to$ \rde.
Furthermore, if the dark energy is assumed to be 
a perfect fluid with a barotropic EoS, then once we compute the EoS we are able to derive the energy density through 
\begin{equation}
\frac{\rho_{\tt DE}(z)}{\rho_{\tt c,0}}=\Omega_{\tt DE,0} e^{3\int_{0}^{z} \frac{dz'}{1+z'}(1+w_{\tt DE}(z'))}.\label{param_dens}
\end{equation}
%
%
On the other hand, the dark energy density could, in general, come from an effective model contribution and not necessarily from a physical component. Hence $\rho_{\tt DE}(z)$ may have any shape (including negative values).
Therefore, if the shape of \rde{} is obtained from any process, then we are able to derive its associated equation of state:
\begin{equation}
    w_{\tt DE}(z) = -1 + \frac{1}{3}\frac{d\ln \rho_{\tt DE}(z)}{d\ln(1+z)}.
\end{equation}

Finally, if the shape of the  energy density is obtained, either directly or through \wde, we are able to compute some derived functions, for instance the $Om$ diagnostic, which provides a null test for the cosmological constant  \cite{sahni2008two}
\begin{equation}
 Om(z)=\frac{h^2(x)-1}{(x)^3-1}, \quad x=z+1, \quad h(x)=\frac{H(x)}{H_0},    \label{om_diagnostic}
\end{equation}
%
and the deceleration parameter 
\begin{equation}
 q(z)=-1 + \frac{d\ln H}{d\ln(1+z)}.    \label{deceleration_param}
\end{equation}
%

%

\subsection*{Data sets}

%
The Hubble parameter tells us the expansion rate of the Universe, and it can be expressed as a function of the redshift through the Friedmann equation. This parameter can also be assembled by gathering measurements of old stars known as cosmic chronometers as they work as ``standard clocks'' in cosmology. That is, $H(z)$ can be obtained by calculating the derivative of the cosmic time with respect to the redshift as
\begin{equation}
    H(z)\approx\frac{-1}{1+z}\frac{\Delta z}{\Delta t},
\end{equation}
where the rate $\Delta z/\Delta t$ is measured with the difference in age of the cosmic chronometers.
In this work we will use the collection of these cosmic chronometers \cite{jimenez2003constraints, simon2005constraints, stern2010cosmic, moresco2012new, zhang2014four, moresco2015raising, moresco20166} (written as H in the datasets), which can be found within the repository \cite{hz}.
\\

Similarly to standard clocks, the type Ia supernovae (SNIa) are coined as the ``standard candles''.  
The distance modulus of the SNIa is derived from the empirical relation when observing light curves
\begin{equation}
    \mu_{SN}= m_B^*+\alpha X - \beta C - M_B,
\end{equation}
where $X$ is the stretch parameter, $C$ is the color parameter, $M_B$ is the absolute magnitude, $m_B^*$ is the $B$-band apparent magnitude, $\alpha$ and $\beta$ are nuisance parameters. The Pantheon Sample measured the apparent magnitude $m_B=m_B^*+\alpha X - \beta C$, with fix absolute magnitude $M_B$.
Also, for a given cosmological model the distance modulus is  
\begin{equation}
    \mu(z)= 5\log \frac{D_L(z)}{Mpc}+25=m_B - M_B,
\end{equation}
from which the luminosity distance, $D_L(z)=H_0 d_L(z)$,
can be calculated in terms of redshift as $d_L(z)=(1+z)r(z)$, with $r(z)$ being the comoving distance 
\begin{equation}
    r(z)=\frac{1}{H_0}\int^{z}_{0}\frac{H_0}{H(z')}dz'.
\end{equation}
In this work we use the full catalogue of 1048 supernovae from the Pantheon SN Ia sample, 
 covering a redshift range of $0<z<2$ \cite{scolnic2018complete} (written as SN in the datasets). The full covariance matrix associated is comprised of a statistical and a systematic part, and along with the data, they are provided in the repository \cite{pantheon_data}.
\\

On the other hand, the baryon acoustic oscillations (BAOs) are used as ``standard rulers'' in cosmology,  as they measure the angular distance $D_A=r(z)/(1+z)$. 
The BAO scale is set by the radius of the sound horizon
\begin{equation}
    r_d=\int^{\infty}_{z_d}\frac{c_s(z')}{H(z')}dz',
\end{equation}
with $c_s(z)$ being the sound speed of the baryon-photon fluid and $z_d$ the drag epoch \cite{dodelson2003modern}. Since the size of $r_d$ depends on the cosmological model, the BAO actually constrains $D_A(z)/r_d$ (with $D_A(z)=H_0 d_A(z)$), $H(z)/r_d$ and the volume average distance $D_V(z)/r_d=[(1+z)^2 D_A^2(z)cz/H(z)]^{1/3}$. The sound horizon is calibrated by using Big Bang Nucleosynthesis \cite{aubourg2015cosmological}. 
The BAO datasets used here contain the SDSS DR12 Galaxy Consensus, BOSS DR14 quasars (and eBOSS), Lyman-$\alpha$ DR14 auto-correlation, SDSS MGS and 6dFGS located at different redshifts up to 2.36. For a more complete explanation see \cite{alam2017clustering, de2019baryon, ata2018clustering, blomqvist2019baryon, beutler20116df, anderson2014clustering} and references therein.\\

To compute the $\chi^2$ for each data sample, we have
\begin{equation}
    \chi^2_{\rm data}= (d_{i,m}-d_{i,{\rm obs}})C_{ij,{\rm data}}^{-1}(d_{j,m}-d_{j,{\rm obs}}),
\end{equation}
where $d_{m}$ and $d_{\rm obs}$ are our model predictions and the observables respectively; $C_{\rm data}$ is the covariance matrix associated to each of the datasets. Since observations of each dataset are independent from each other, the joint $\chi^2$ can be calculated as
\begin{equation}
    \chi^2_{\rm total}=\chi^2_{\rm H}+\chi^2_{\rm SN}+\chi^2_{\rm BAO}.
    \label{eq:chi_sq_components}
\end{equation}

\subsection*{Models and priors}

\begin{table}[t!]
\captionsetup{justification=justified,singlelinecheck=false,font=footnotesize}
\footnotesize
\scalebox{0.8}{%
\begin{tabular}{ccc} 
\cline{1-3}\noalign{\smallskip}
 \vspace{0.15cm}

   Model &   \wde &  \rde   \\
\hline
&&\\
Parameterizations  &  $w$CDM &  Sigmoid \\
  & $w_c: [-2.0,0.0]$  &  $z_{\rm cut}:[0.0, 3.0]$ \vspace{.2cm}\\
    & CPL  &   \\
    & $w_0: [-2.0,0.0]$, $w_a: [-2.0,2.0]$   &   \\
\hline
&&\\
Binning/Nodal & $w(z_i): [-2.5, 1.0]$ &   $\rho(z_i): [0.0, 1.5]$ if $z_i<1.5$    \\
 ($i=1\cdots 6,20$)& &   \ \qquad \quad$: [-1.5, 1.5]$ if $z_i \ge 1.5$    \\

\hline
&&\\

Binning-internal $z_i$ & $w(z_i)|_{i=4}: [-2.5, 1.0]$ &   $\rho(z_i)|_{i=3}: [0.0, 1.5]$ if $z_i<1.5$    \\
& &   \qquad \qquad \quad$: [-1.5, 1.5]$ if $z_i \ge 1.5$     \vspace{0.15cm}\\
             & $z_1: [0.2, 1.4]$   &$z_1: [0.2, 1.4]$ \\
             & $z_2: [1.6, 2.8]$   &$z_2: [1.6, 2.8]$ \\ 
\hline
\hline

\end{tabular}}
\caption{ Additional parameters and their flat prior range.}
 \label{tabla_priors}
\end{table}

Since the Bayesian evidence is very susceptible to the number of parameters and their prior distribution, it is worth to be careful when selecting them. A summary of the additional parameters along with their prior ranges is displayed in Table \ref{tabla_priors}.
\\

First, to provide a comparison to the reconstructions, we constrain some parameterization models. For instance, the $w$CDM model which corresponds to a constant EoS \wde{}$ = w_c$ and the Chevallier-Polarski-Linder (CPL) EoS \cite{chevallier2001accelerating} in which \wde{}$= w_0 + w_a\frac{z}{1+z}$, being $w_0$, $w_a$ and $w_c$ free parameters to be estimated with data. The flat priors for $w_0$ and $w_c$ are the same $[-2.0,0.0]$ and the flat prior used for $w_a$ is $[-2.0, 2.0]$. 
Then, inspired by the idea of a density capable of performing a possible transition to $\rho_{\tt DE} \le 0$ at high redshifts, similar to the one introduced by \cite{akarsu2020graduated}, we propose a simple parameterization with the shape of a sigmoid function:
\begin{equation}
\frac{\rho_{\tt DE}(z)}{\rho_{\tt c,0}} = \Omega_{\tt DE,0}\bigg(k_0 -\frac{1}{1+e^{-10(z-z_{\rm cut})}} \bigg),    
\end{equation}
with  $z_{\rm cut}$ the redshift value where the transition may occur and $k_0=1+\frac{1}{1+e^{10z_{\rm cut}}}$ is a constant which compensates the necessary amount so that $\rho_{\tt DE}(z=0)/\rho_{\tt c,0}=1-\Omega_{\rm m,0}$, to account for the Friedmann constraint. The parameter $z_{\rm cut}$ has a flat prior within the range $[0.0, 3.0]$. The sharp part of this sigmoid function comes from the argument in the exponential, if this number is larger (smaller) its transition to zero would be sharper (smoother).
\\

Throughout all the reconstructions  we let the data to decide the level of complexity of the two main functions within the range 
$0\leq z \leq 3$, that is,  we place the nodes and bins (free parameters) over this range, and for $z>3$ we adopt a constant value corresponding to the last amplitude. 

In the first set of reconstructions, the position of the nodes and bins are kept fixed and uniformly distributed within $z=[0,3]$.
In both types of reconstructions it is relatively free to choose any number of amplitudes, thus we used from 1 to 6 (and then, without loss of generality, jump to 20) to see the improvement of the fit to the data and how well the Bayesian evidence responded. Notice that the $w$CDM model is equivalent to the EoS reconstruction with one bin. 
In particular, we allow the possibility of having negative energy density, hence the amplitudes for the nodes and bins for \rde{} were set to move freely on the ordinate with flat priors $[0.0,1.5]$ if  $z<1.5$, otherwise the prior  is set to $[-1.5,1.5]$, since it is beyond $z=1.5$ where a switch to negative energy density is generally presented. 
For the amplitudes in the EoS we have flat priors of $[-2.5, 1.0]$. 
An important point is that when incorporating the correlation function method with a floating prior ($\chi^2_{\rm prior}$) or CPZ, it is recommended to choose a large number of bins. In our case, we used 20 bins and obtained consistent results.

In the first set of reconstructions, the positions $z_i$ for each parameter remained fixed, however one may argue that the location of the amplitudes could bias the results. 
To check this point, in the second set of reconstructions every amplitude varies as well as the internal positions are allowed to move freely, spanning over the $z$-direction (but without overlapping each other).
For the reconstruction of \wde{} we consider 6 parameters: 4 varying amplitudes and 2 internal positions; similarly for \rde{} we use 5 parameters: 3 amplitudes and 2 internal positions.
We will refer to them as 4y2x (3y2x) or simply  \textit{internal $z_i$}. 
Whereas the priors for the amplitudes remain the same as in the first set, the 
priors for the internal $z_i$ positions are $[0.2,1.4]$ and $[1.6,2.8]$ respectively.

Regarding the cosmological parameters, we have: the reduced dimensionless Hubble parameter $h_0=H_0/100{\rm km s}^{-1}{\rm Mpc}^{-1}$ with a flat prior of $[0.6,0.8]$, the baryon density $\Omega_{b}h_0^2$ with prior $[0.021,0.024]$ and the matter density parameter $\Omega_{\rm m,0}$ with a prior of $[0.2,0.4]$. 
\\

To perform the reconstruction analysis we used a modified version of the  
 \texttt{SimpleMC} code \cite{simplemc} along with \texttt{dynesty} \cite{speagle2020dynesty},
 a nested sampling algorithm which allows to compute efficiently the Bayesian evidence. \luis{For the number of live points we followed the general rule \cite{dynesty} of using $50\times ndim$ live points, where the dimensionality $ndim$ corresponds to the number of parameters, so the total number of live points depends on the reconstruction and the amount of bins/nodes. As for the stopping criterion we have an accuracy of 0.01, which indicates the maximum difference between samples.}
 Within the  \texttt{SimpleMC} code we have also implemented the Principal Component Analysis and the correlation function method with the floating prior. For the functional posterior confidence contour plots we used a python package named \texttt{fgivenx} \cite{handley2019fgivenx}. See Ref.~\cite{Padilla:2019mgi}, and references therein, for an extended review of the cosmological parameter inference and model selection procedure.

\begin{table}[t!]
\captionsetup{justification=justified,singlelinecheck=false,font=footnotesize}
\footnotesize
\scalebox{0.8}{%
\begin{tabular}{cccccc} 
\cline{1-6}\noalign{\smallskip}
 \vspace{0.15cm}

  Model &  Parameters  & $h_0$ &  $\Omega_{\rm m,0}$  &   $\ln B_{\Lambda \text{CDM},i}$  &  $\Delta\chi^2$ \\
\hline
 \vspace{0.15cm}
$\Lambda$CDM & - &  0.683 (0.008) &  0.306 (0.013) &  0  &  0  \\
\vspace{0.15cm}
$w$CDM & 1 &  0.680 (0.017) &  0.305 (0.015) &  2.56 (0.14)  &  -0.14  \\
\vspace{0.15cm}
Sigmoid & 1 & 0.688 (0.009) & 0.312 (0.013) &  -0.12 (0.13)  & -2.56  \\
\vspace{0.15cm}
CPL & 2 & 0.674 (0.021) & 0.299 (0.019) &  3.54 (0.15)  & -0.68  \\

\hline
\hline

\vspace{0.15cm}
\wde & Binning \\
\vspace{0.15cm}
& 2 &  0.692 (0.017) & 0.316 (0.015) &  3.02 (0.16)  & -0.83 \\
\vspace{0.15cm}
& 3 &  0.681 (0.018) & 0.307 (0.016) &  3.60 (0.15)  & -2.88 \\
\vspace{0.15cm}
& 4 &  0.681 (0.017) & 0.303 (0.015) &  2.93 (0.16)  & -5.15 \\
\vspace{0.15cm}
& 5 &  0.681 (0.017) & 0.305 (0.016) &  4.72 (0.16)  & -3.57 \\
\vspace{0.15cm}
& 6  &   0.676 (0.016) & 0.299 (0.015) & 3.11 (0.17)  & -8.86 \\
 \vspace{0.15cm}
4y2x & 6   & 0.684 (0.017) & 0.309 (0.015) &  2.86 (0.16) & -8.03  \\
 \vspace{0.15cm}
& 20 & 0.691 (0.015) & 0.298 (0.014) &  2.33 (0.18) & -15.74 \\
 \vspace{0.15cm}
+$\chi^2_{\rm prior}$& 20 & 0.688 (0.015) & 0.298 (0.015) &  5.52 (0.18) & -9.97 \\

 \hline
\vspace{0.15cm}
 & Linear-Nodal \\
 \vspace{0.15cm}
& 2 &  0.681 (0.021) & 0.307 (0.019) &  3.96 (0.15)  & -0.12 \\
\vspace{0.15cm}
& 3 &  0.677 (0.018) & 0.301 (0.018) &  3.61 (0.16)  & -2.98 \\
\vspace{0.15cm}
& 4 &  0.682 (0.018) & 0.302 (0.017) &  3.46 (0.16)  & -3.74  \\
\vspace{0.15cm}
& 5 &  0.683 (0.017) & 0.305 (0.016) &  5.03 (0.16)  & -3.19 \\
\vspace{0.15cm}
& 6  & 0.681 (0.017) & 0.302 (0.016) &  4.15 (0.17)  & -5.06 \\
\vspace{0.15cm}
4y2x & 6   & 0.681 (0.017) & 0.302 (0.016) &  3.02 (0.16)  & -6.22 \\

\hline
\hline
\vspace{0.15cm}
\rde & Binning  \\
\vspace{0.15cm}
& 1 &  0.694 (0.013) & 0.318 (0.015) &  -0.01 (0.13)  & -0.77 \\
\vspace{0.15cm}
& 2 &  0.683 (0.013) & 0.309 (0.015) &  -0.42 (0.13)  & -3.04 \\
\vspace{0.15cm}
& 3 &  0.681 (0.013) & 0.307 (0.015) &  -0.59 (0.14)  & -4.24 \\
\vspace{0.15cm}
& 4 &  0.677 (0.014) & 0.304 (0.016) &  0.53 (0.14)  & -3.61 \\
\vspace{0.15cm}
& 5 &  0.682 (0.015) & 0.309 (0.016) &  0.54 (0.14)  & -5.72 \\
\vspace{0.15cm}
& 6 &  0.683 (0.015) & 0.311 (0.016) &  1.23 (0.15)  & -8.04 \\
\vspace{0.15cm}
3y2x & 5   & 0.686 (0.012) & 0.313 (0.014) &  -0.49 (0.14)  & -4.47 \\
\vspace{0.15cm}
& 20  & 0.685 (0.014) & 0.318 (0.014) &  2.36 (0.16)  & -11.60 \\
\vspace{0.15cm}
+$\chi^2_{\rm prior}$& 20  & 0.685 (0.015) & 0.317 (0.014) &  5.20 (0.16)  & -9.96 \\

\hline
\vspace{0.15cm}
 & Linear-Nodal \\
\vspace{0.15cm}
& 2 &  0.683 (0.017) & 0.311 (0.015) &  0.57 (0.14)  & -2.72 \\
\vspace{0.15cm}
& 3 &  0.685 (0.019) & 0.311 (0.017) &  1.58 (0.14)  & -2.47 \\
\vspace{0.15cm}
& 4 &  0.686 (0.018) & 0.313 (0.016) &  1.50 (0.14) & -2.73 \\
\vspace{0.15cm}
& 5 &  0.691 (0.017) & 0.314 (0.016) &  1.65 (0.15)  & -3.71 \\
\vspace{0.15cm}
& 6 &  0.685 (0.017) & 0.308 (0.015) &  1.85 (0.15)  & -4.13 \\
\vspace{0.15cm}
3y2x& 5  &  0.691 (0.016) & 0.315 (0.015) &  1.11 (0.14)  & -3.38 \\

\hline
\hline
\end{tabular}}
\caption{ Mean values, and standard deviations,  for the cosmological parameters. For each model, the last two columns present the Bayes Factor, and the $\Delta\chi^2=\chi^2_{\Lambda \text{CDM}}- \chi^2_i$ for fitness comparison. The datasets used are BAO+H+SN. Here $\ln E_{\Lambda \text{CDM}}=-530.79 (0.12)$.
}

 \label{tabla_evidencias}
\end{table}

\section{Results}\label{section:results}

We present the results in 4 subsections: regarding the EoS, the energy density, the derived functions in terms of our results and the PCA analysis to distinguish noise from signal. The best-fit parameter values, the logarithm of the Bayes' factor ($\ln B_{\Lambda \text{CDM},i}$) and  the goodness of fit ($\Delta\chi^2$) are presented in Table \ref{tabla_evidencias}; complementary to this table, both quantities are displayed in Fig. \ref{fig:fitness_full_panth}. The regions of confidence for the parameterizations are shown in Fig. \ref{fig:parameterizations}, whilst the reconstructions are shown in Figs. \ref{fig:nodes_full_panth_eos} and \ref{fig:rho_eos_corr_4y2x}. 
%

\subsection*{Dark Energy Equation of State}

\begin{figure*}[t!]
\captionsetup{justification=raggedright,singlelinecheck=false,font=footnotesize}
    \centering
    \makebox[11cm][c]{
    \includegraphics[trim = 5mm  0mm 25mm 0mm, clip, width=5.cm, height=4.5cm]{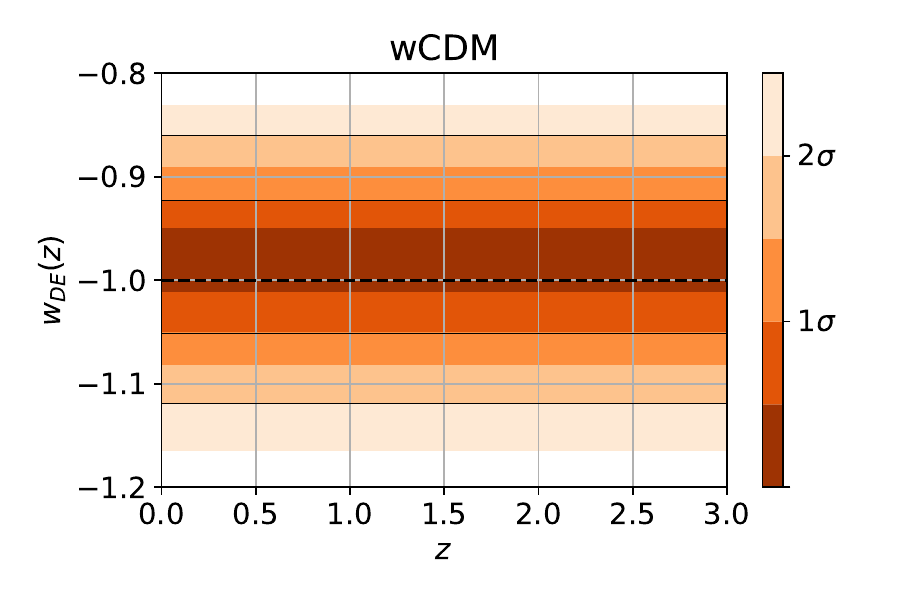}
    \includegraphics[trim = 5mm  0mm 25mm 0mm, clip, width=5.cm, height=4.5cm]{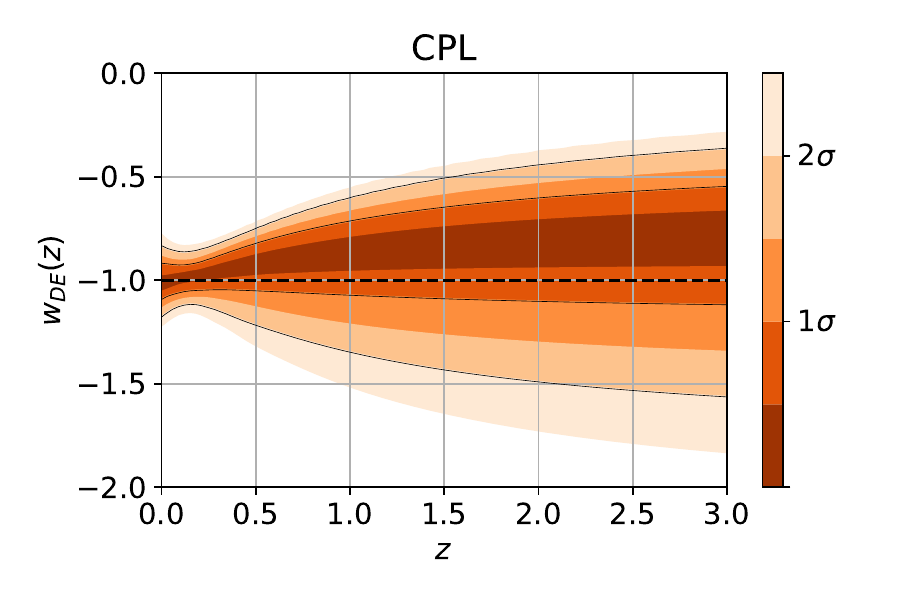}
    \includegraphics[trim = 5mm  0mm 25mm 0mm, clip, width=5.cm, height=4.5cm]{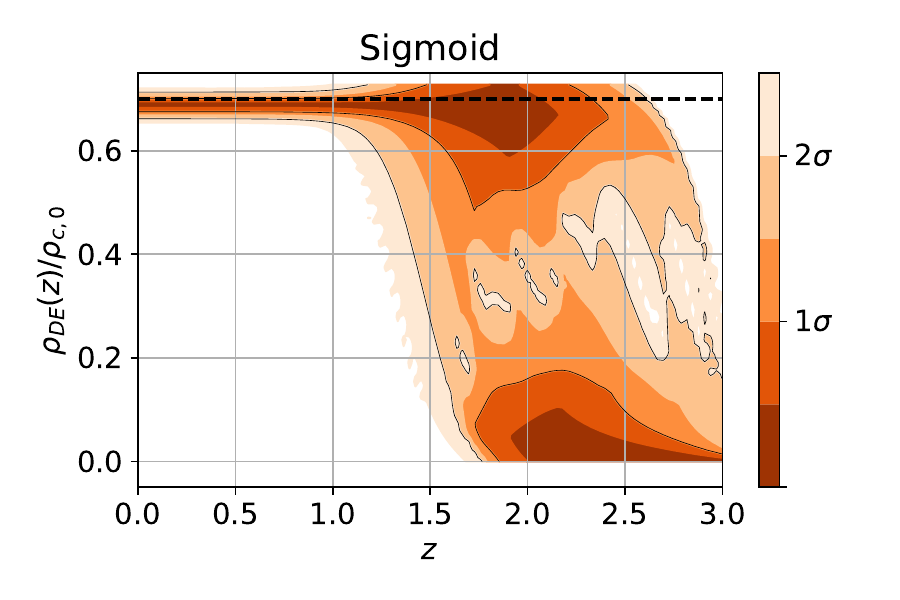}
    \includegraphics[trim = 5mm  0mm 5mm 0mm, clip, width=5.5cm, height=4.5cm]{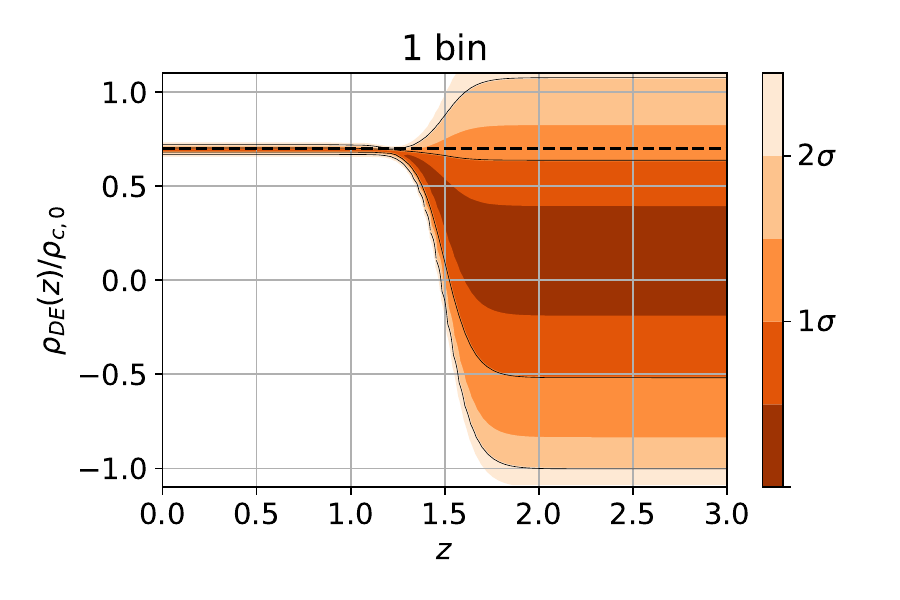}
    }
    \caption{These plots show the functional posterior probability: the probability 
of \wde{} or \rde{} as normalised in each slice of constant $z$, with colour scale in confidence interval values. 
The 68\% ($1\sigma$) and 95\% ($2\sigma$) confidence intervals are plotted as black lines. From left to right: parameterized equations of state for $w$CDM and CPL, phenomenological sigmoid and 1-bin energy density reconstructions. The dashed black line corresponds to the standard $\Lambda$CDM values. }\label{fig:parameterizations}
\end{figure*}

\begin{figure*}[t!]
\captionsetup{justification=raggedright,singlelinecheck=false,font=footnotesize}
    \makebox[11cm][c]{
     \includegraphics[trim = 5mm  0mm 25mm 0mm, clip, width=4.5cm, height=4.cm]{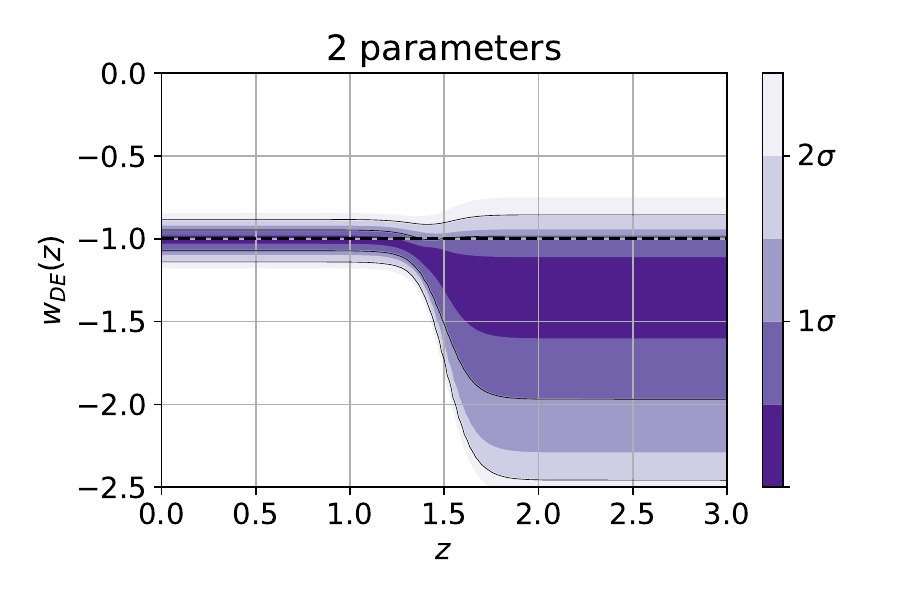}
     \includegraphics[trim = 26mm  0mm 25mm 0mm, clip, width=4.cm, height=4.cm]{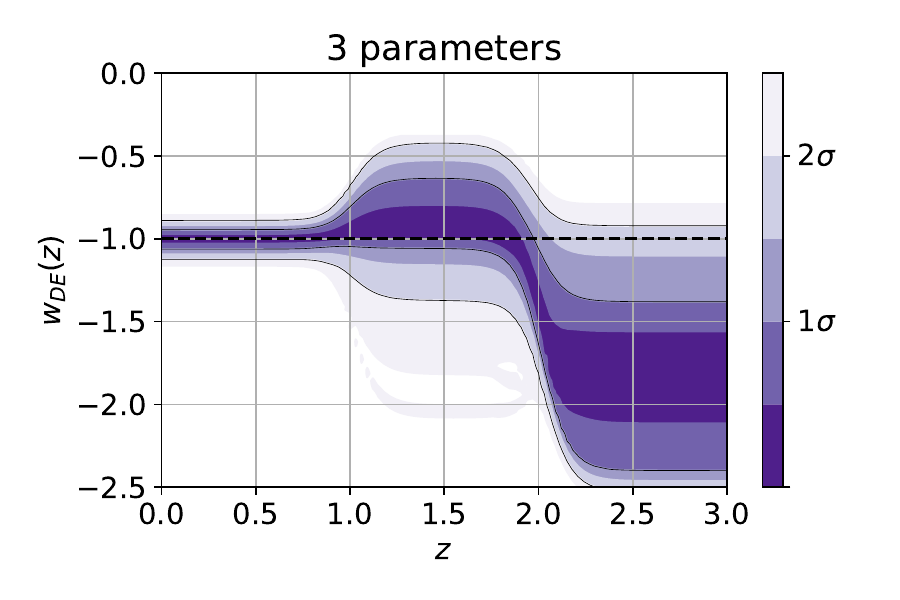}
     \includegraphics[trim = 26mm  0mm 25mm 0mm, clip, width=4.cm, height=4.cm]{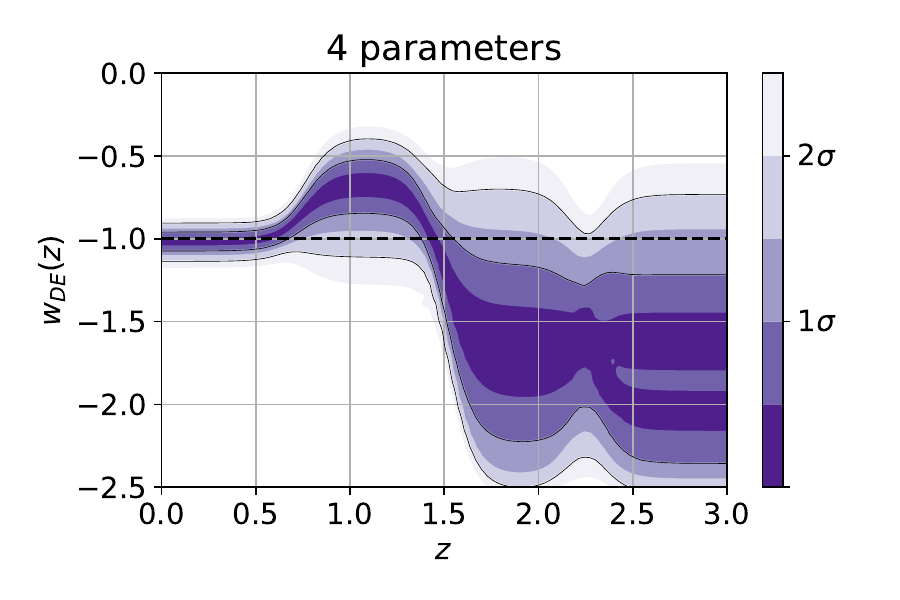}
     \includegraphics[trim = 26mm  0mm 25mm 0mm, clip, width=4.cm, height=4.cm]{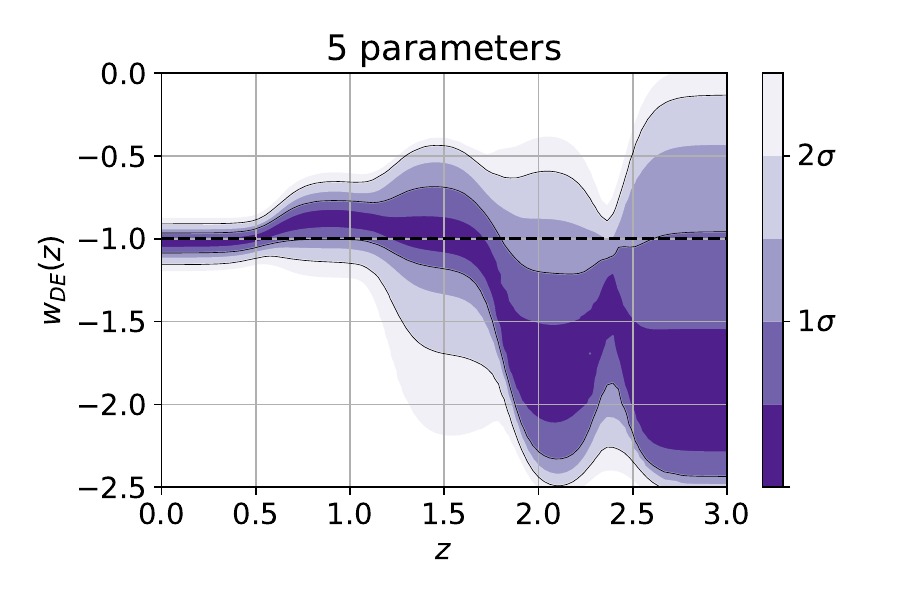}
     \includegraphics[trim = 26mm  0mm 5mm 0mm, clip, width=4.5cm, height=4.cm]{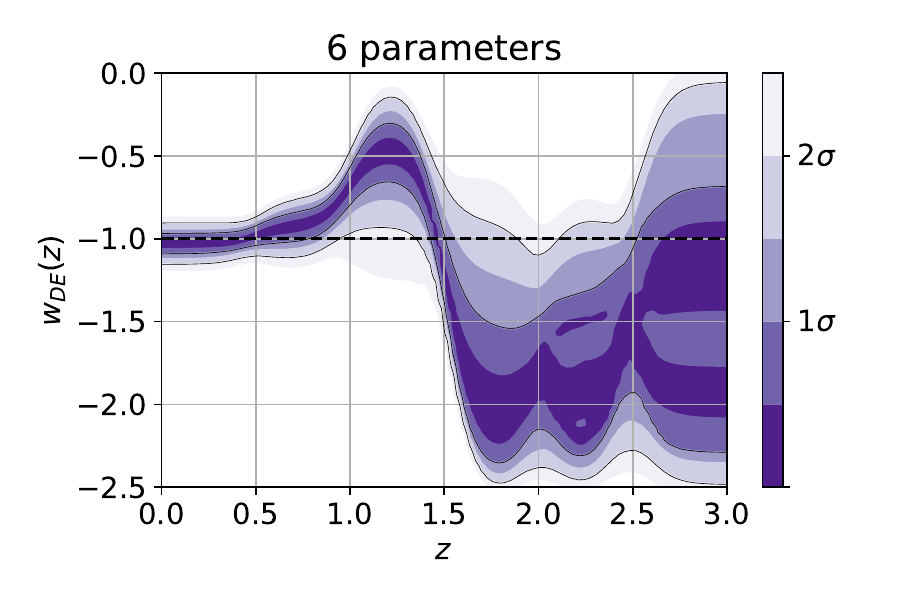}
     }
     \makebox[11cm][c]{   
     \includegraphics[trim = 5mm  0mm 25mm 0mm, clip, width=4.5cm, height=4.cm]{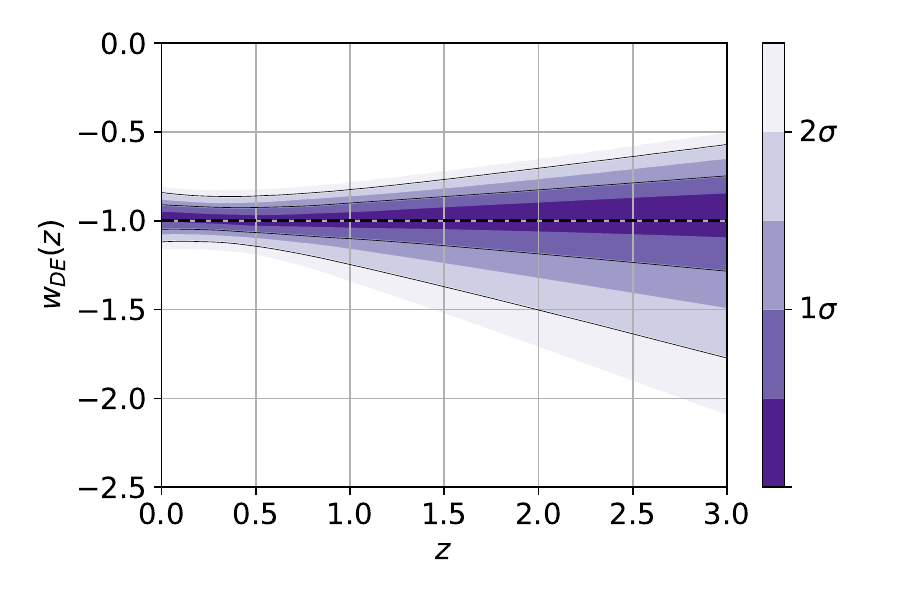}
     \includegraphics[trim = 26mm  0mm 25mm 0mm, clip, width=4.cm, height=4.cm]{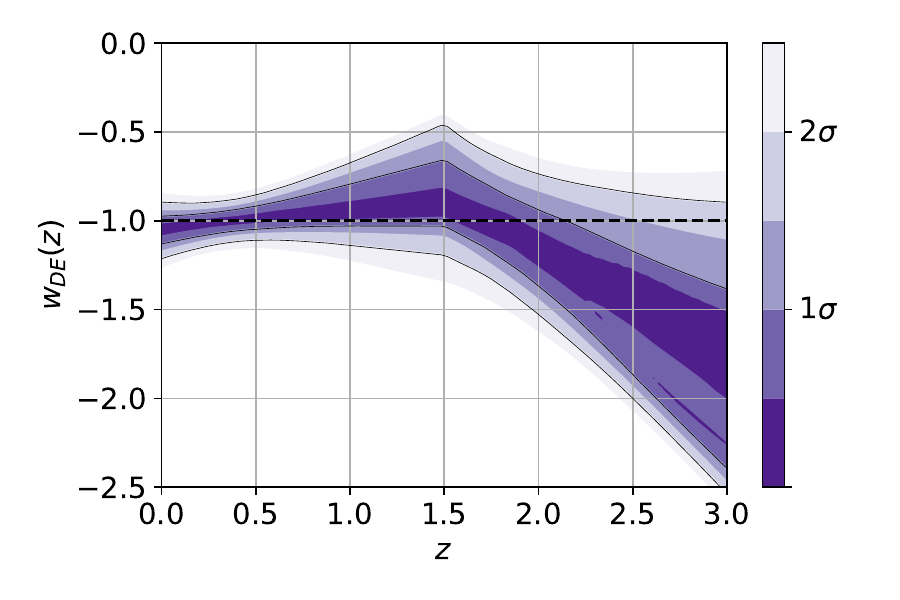}
     \includegraphics[trim = 26mm  0mm 25mm 0mm, clip, width=4.cm, height=4.cm]{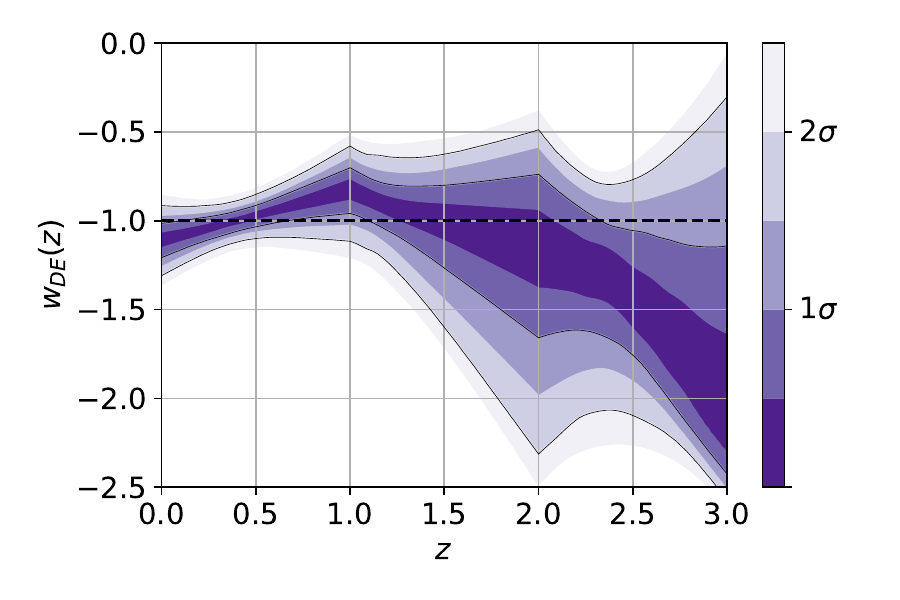}
     \includegraphics[trim = 26mm  0mm 25mm 0mm, clip, width=4.cm, height=4.cm]{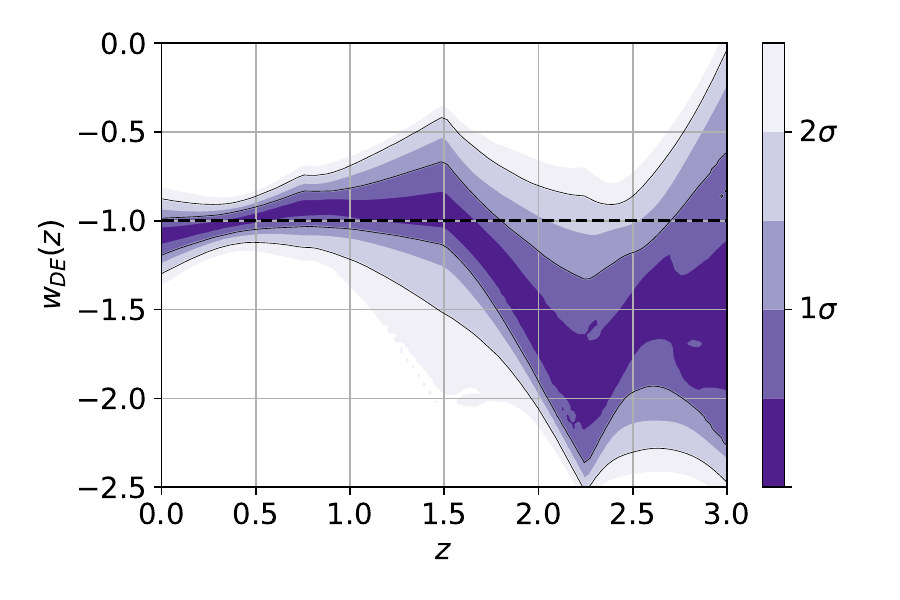}
     \includegraphics[trim = 26mm  0mm 5mm 0mm, clip, width=4.5cm, height=4.cm]{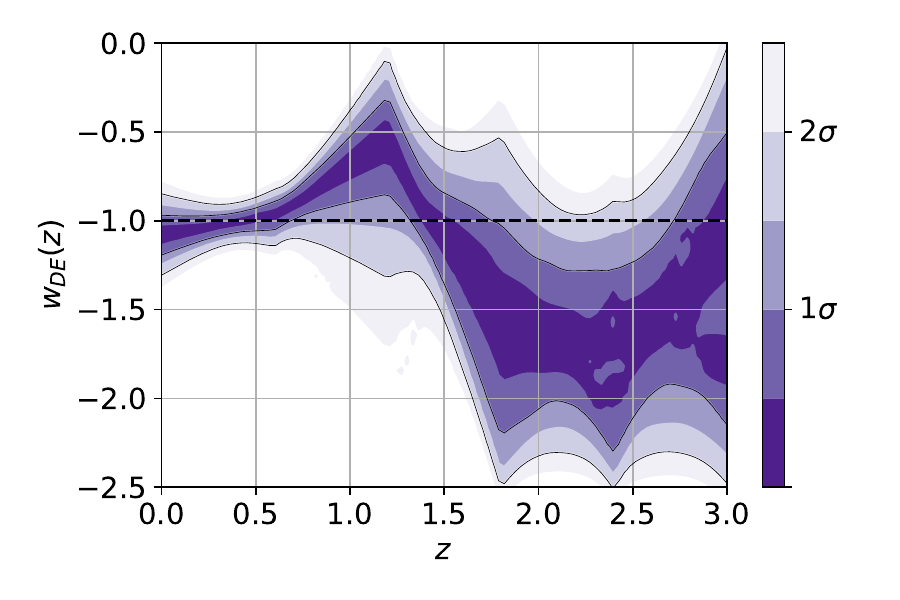}
     }
     \makebox[11cm][c]{
     \includegraphics[trim = 5mm  0mm 25mm 0mm, clip, width=4.5cm, height=4.cm]{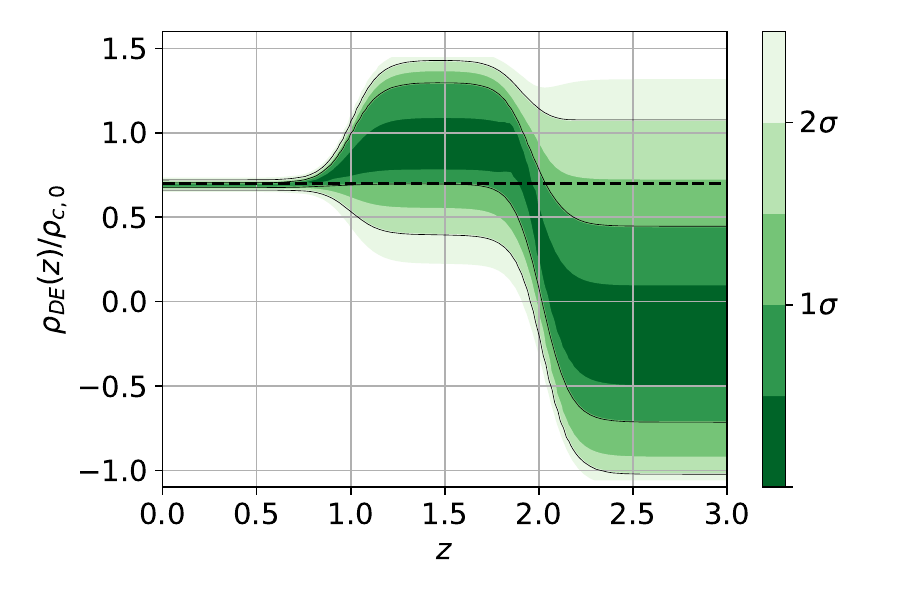}
     \includegraphics[trim = 26mm  0mm 25mm 0mm, clip, width=4.cm, height=4.cm]{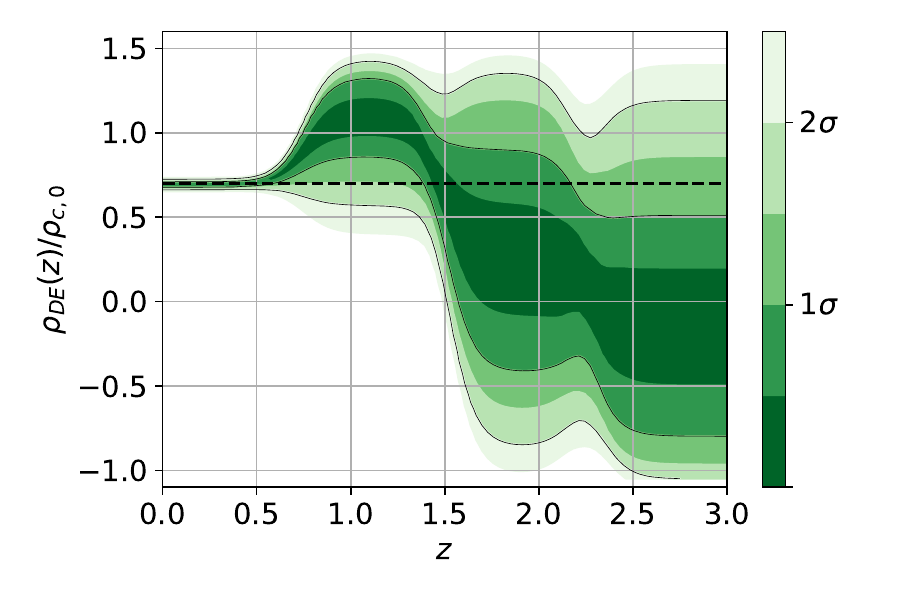}
     \includegraphics[trim = 26mm  0mm 25mm 0mm, clip, width=4.cm, height=4.cm]{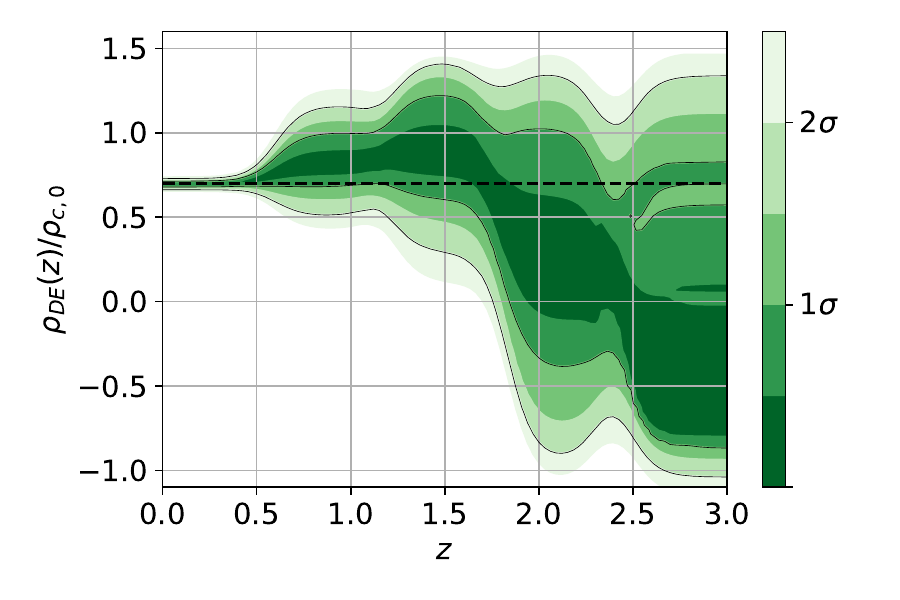}
     \includegraphics[trim = 26mm  0mm 25mm 0mm, clip, width=4.cm, height=4.cm]{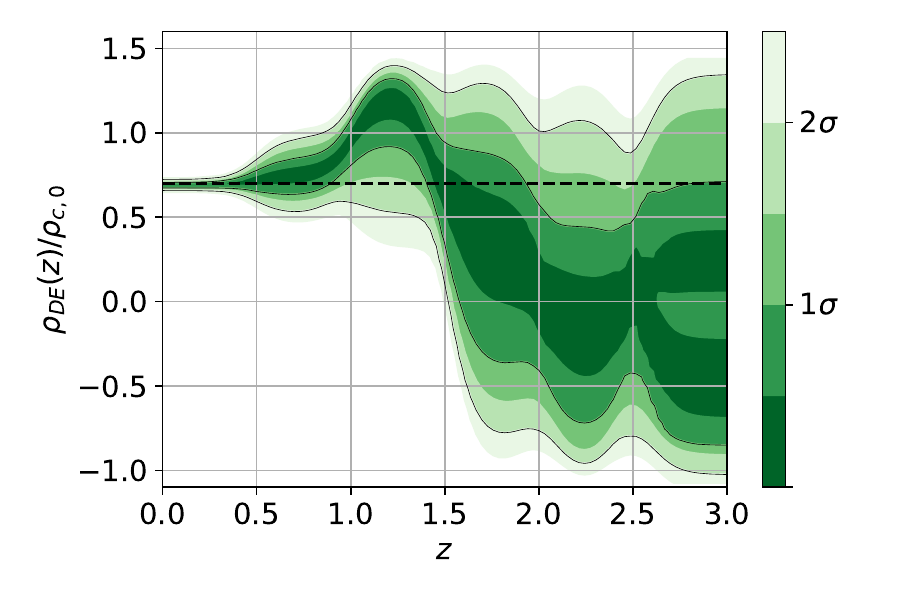}
     \includegraphics[trim = 26mm  0mm 5mm 0mm, clip, width=4.5cm, height=4.cm]{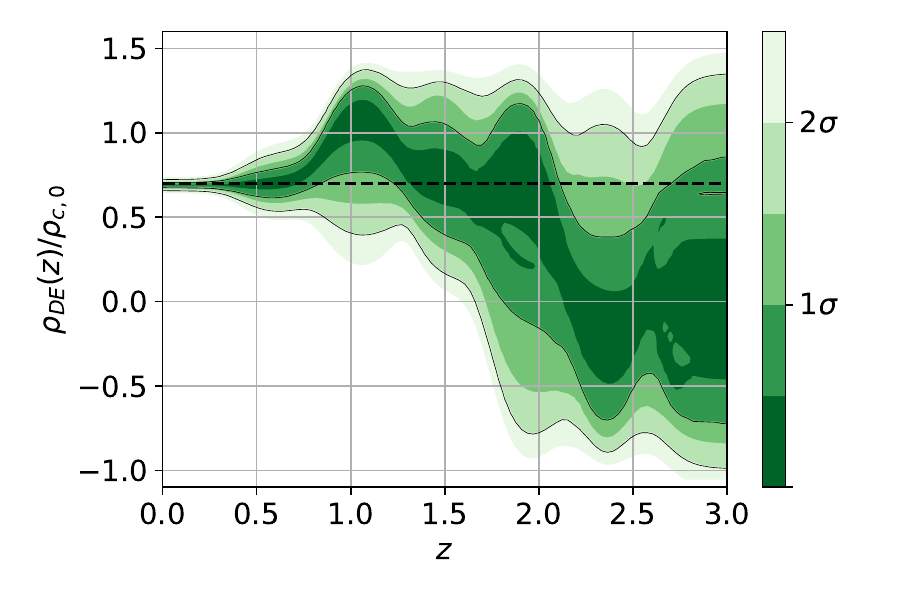}
     }
     \makebox[11cm][c]{
     \includegraphics[trim = 5mm  0mm 25mm 0mm, clip, width=4.5cm, height=4.cm]{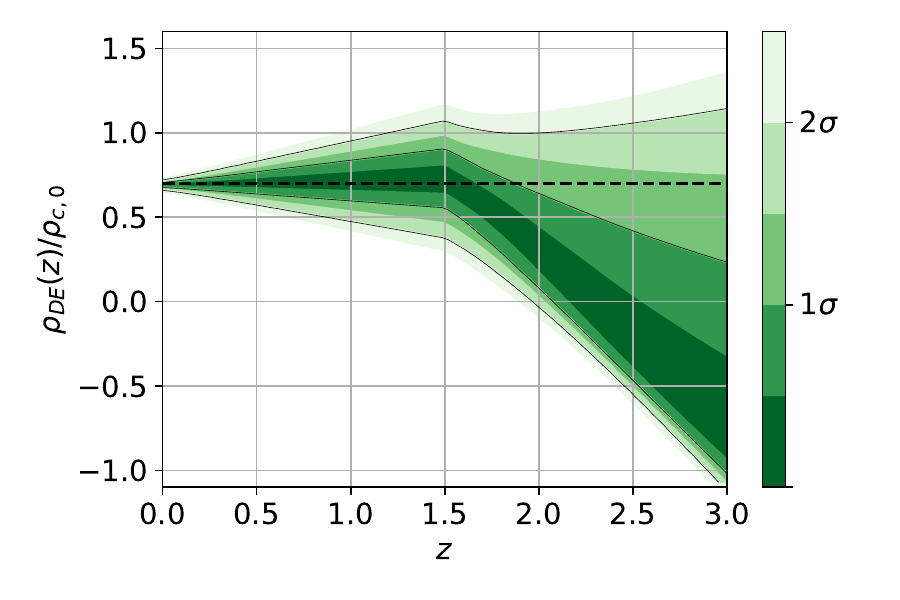}
     \includegraphics[trim = 26mm  0mm 25mm 0mm, clip, width=4.cm, height=4.cm]{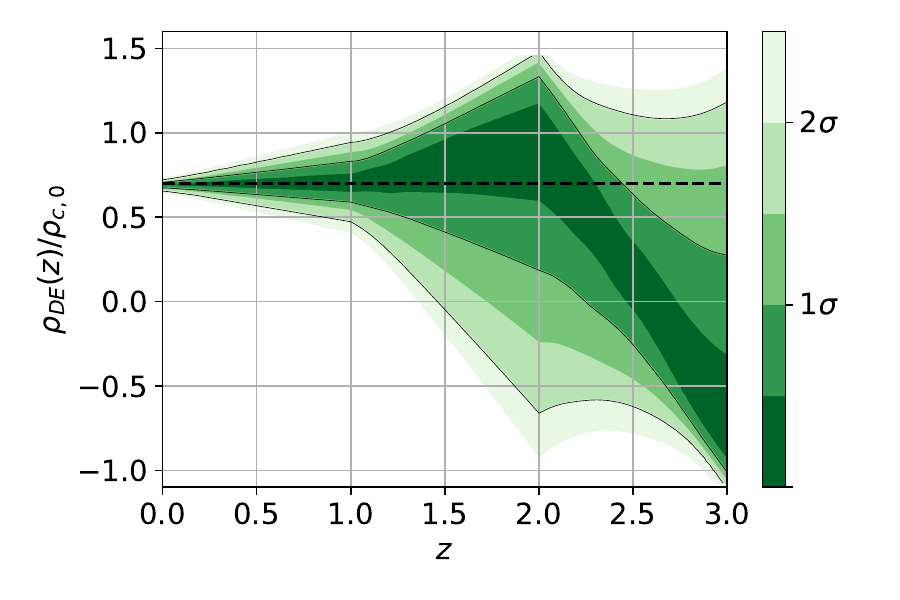}
     \includegraphics[trim = 26mm  0mm 25mm 0mm, clip, width=4.cm, height=4.cm]{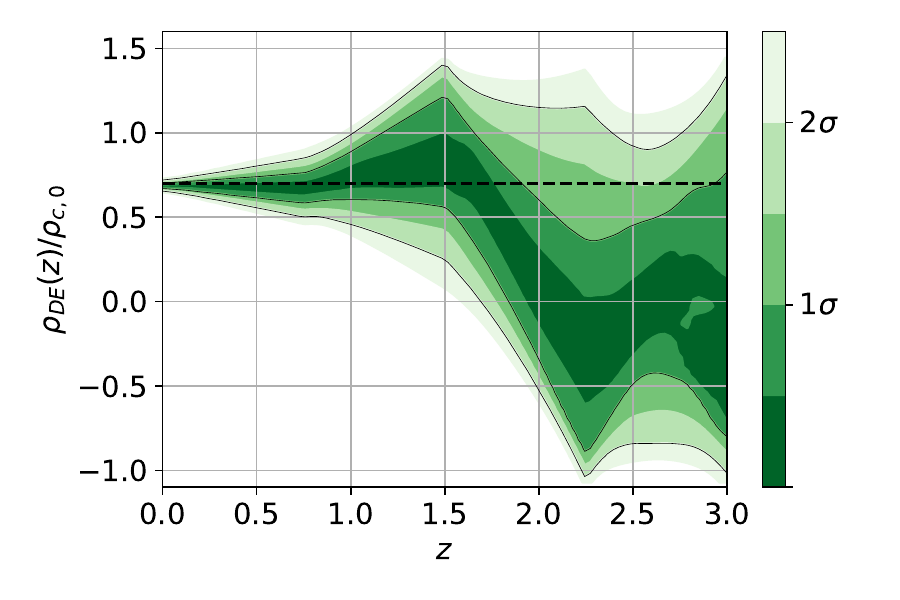}
     \includegraphics[trim = 26mm  0mm 25mm 0mm, clip, width=4.cm, height=4.cm]{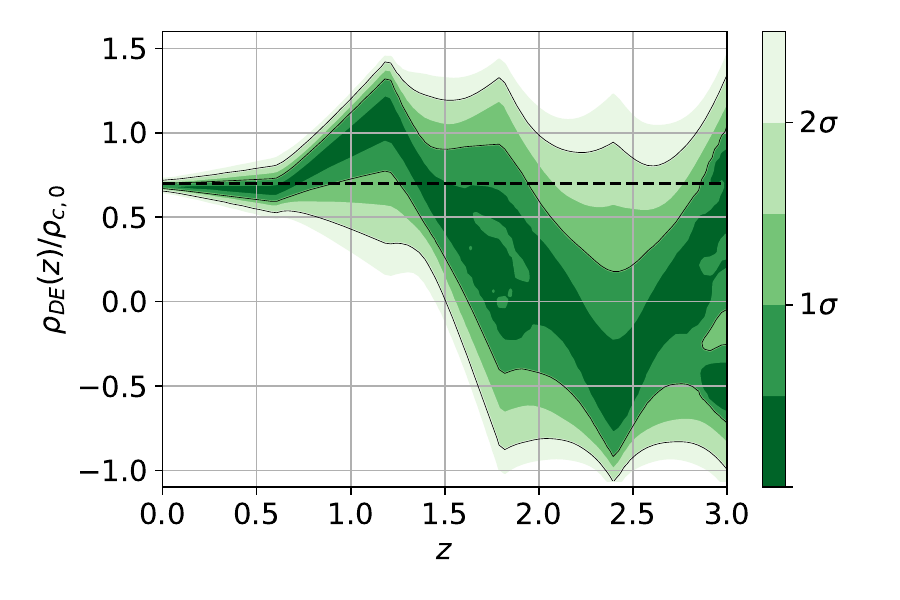}
     \includegraphics[trim = 26mm  0mm 5mm 0mm, clip, width=4.5cm, height=4.cm]{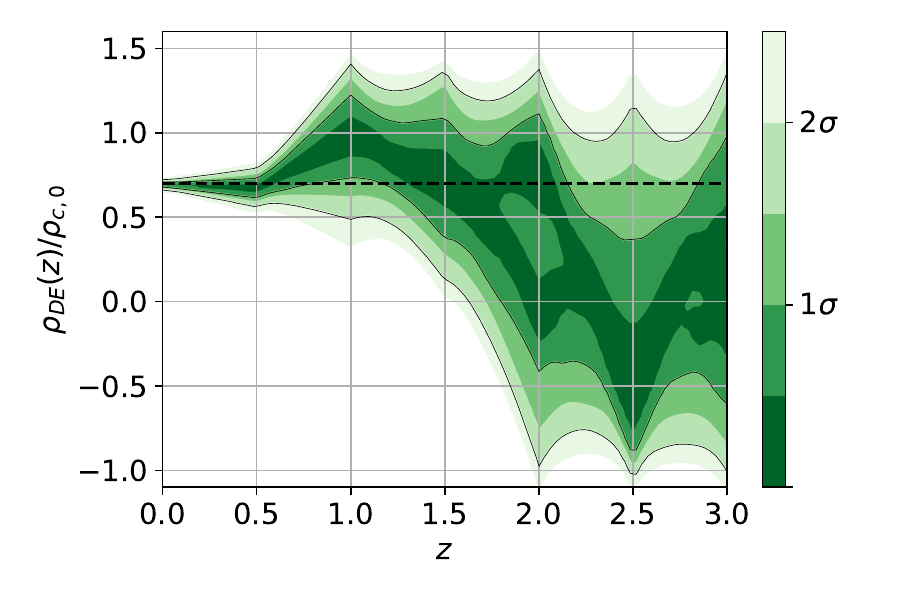}
     }
     
\caption{From top to bottom: reconstructed $w_{\tt DE}(z)$ with bins (and nodes), reconstructed $\rho_{\tt DE}(z)/\rho_{\tt c,0}$ with bins (and nodes).
 It is easy to observe there is more structure (more apparent features) in the reconstructions as the number of parameters increases (from left to right).} \label{fig:nodes_full_panth_eos}
\end{figure*}

The best-fit values (with standard deviations) for the $w$CDM and the CPL parameterization, correspond to $w_c=-0.99\pm0.06$, $w_0=-1.01\pm0.08$ and $w_a=0.12\pm0.47$ respectively. 
That is, the models with one or two parameters, $w$CDM, CPL and the 2-parameter reconstructions produce very similar results, which means they are statistically consistent with the cosmological constant, within the 68\% confidence region (see Figure \ref{fig:parameterizations} and the first column of Figure \ref{fig:nodes_full_panth_eos}). Also, these results can be validated when comparing the $\Delta \chi^2$ presented in Table \ref{tabla_evidencias}. 
\\

\begin{figure}
  \begin{center}
    \includegraphics[trim = 5mm  0mm 0mm 0mm width=9.2cm, height=6.5cm]{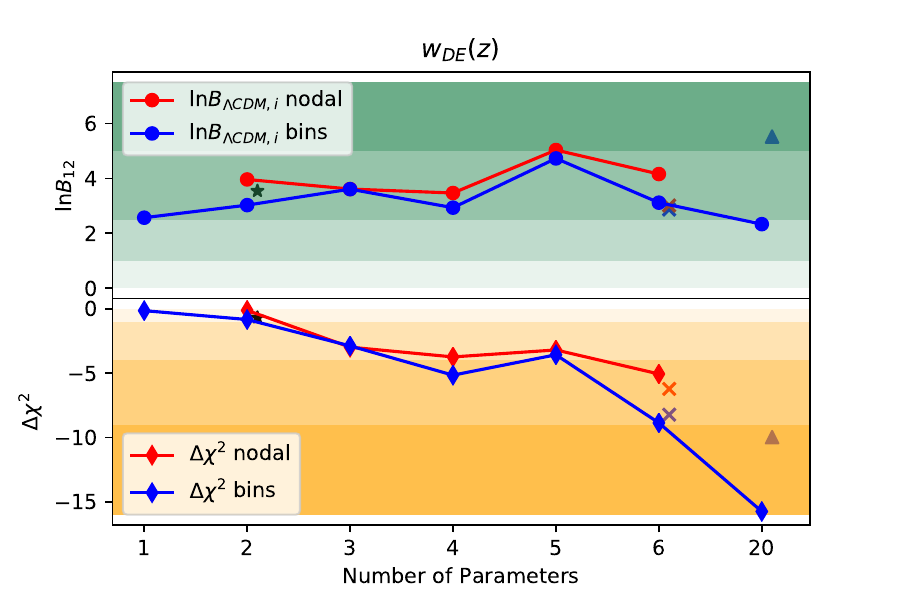}
     \includegraphics[trim = 5mm  0mm 0mm 0mm width=9.2cm, height=6.5cm]{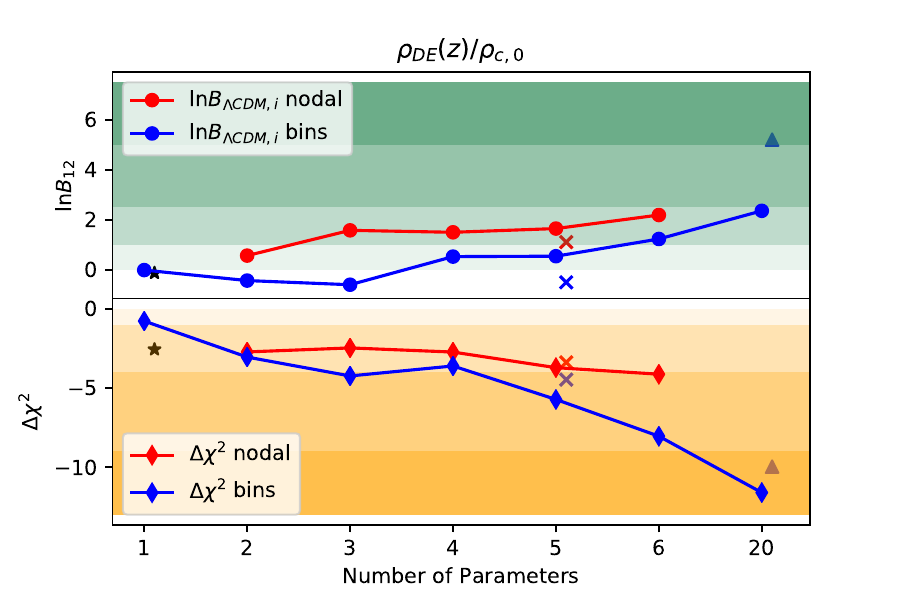}
\end{center}
    \caption{In this graph we show the differences in the $\Delta \chi^2$ and the Bayes factor between $\Lambda$CDM and our reconstructions for $w_{\tt DE}(z)$ and $\rho_{\tt DE}/\rho_{\tt c,0}$. The green shades show the strength-of-evidence according to the Jeffrey's scale and the orange shades show the 1 to 4-$\sigma$ levels of statistical significance $S/N \equiv \sqrt{|\Delta\chi^2|}$. Ideally we want the upper markers to stay as close as possible to the black line at 0 (preferably to cross it), which it is an indication of a better Bayes' factor, and the lower ones to be far away from zero, which indicates a better fit to the data. The stars indicate the fitness of the reconstruction of the CPL EoS (top) and the energy density with a sigmoid (bottom), the crosses indicate the internal $z_i$ reconstructions and the triangles the reconstructions with 20 bins plus correlation function. The binning reconstructions are plotted with blue lines, whereas the nodal with red lines.
   %
    } \label{fig:fitness_full_panth}
\end{figure}

\begin{figure*}[t!]
\captionsetup{justification=raggedright,singlelinecheck=false,font=footnotesize}
    \centering
    \makebox[11cm][c]{
     \includegraphics[trim = 0mm  0mm 25mm 0mm, clip, width=4.9cm, height=4.cm]{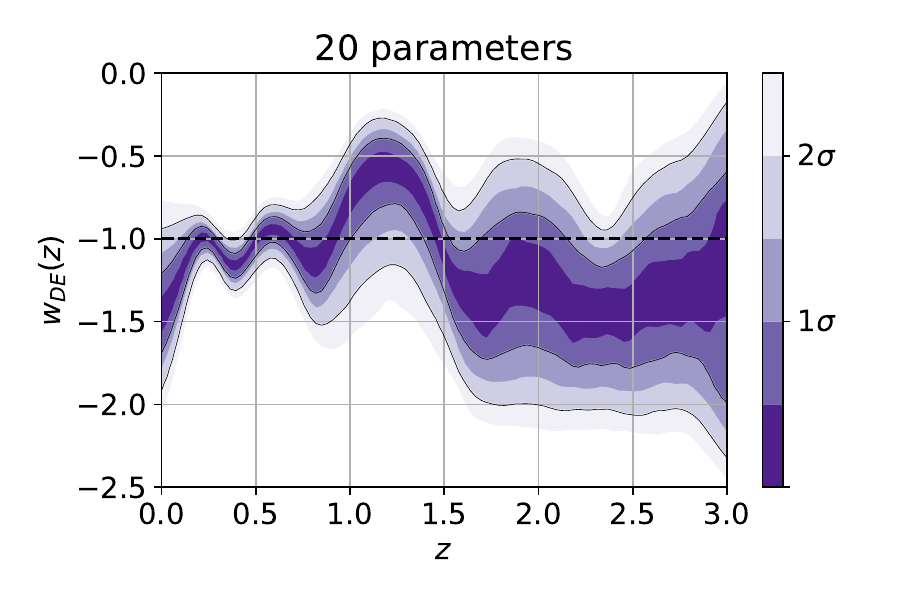}
     \includegraphics[trim = 26mm  0mm 25mm 0mm, clip, width=4.9cm, height=4.cm]{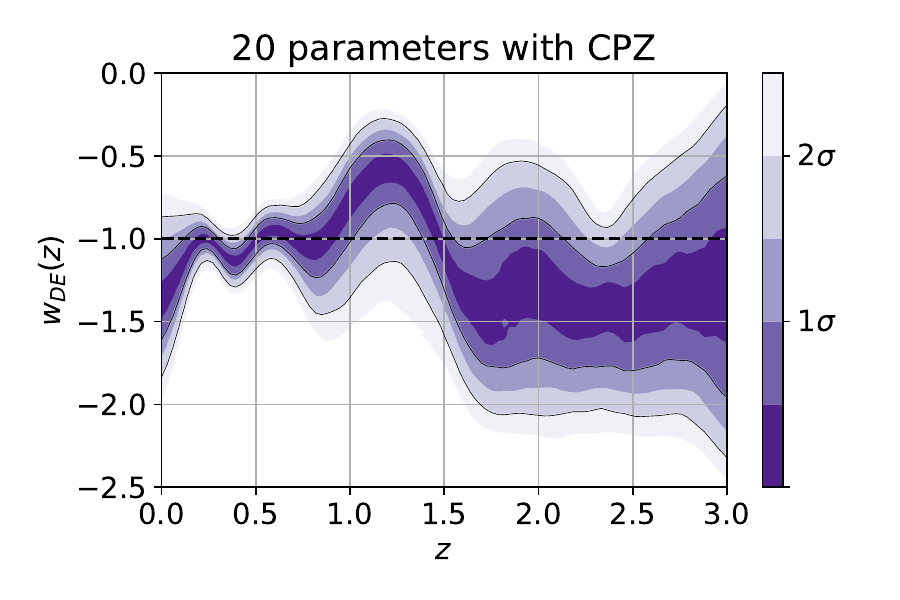}
     \includegraphics[trim = 26mm  0mm 25mm 0mm, clip, width=4.9cm, height=4.cm]{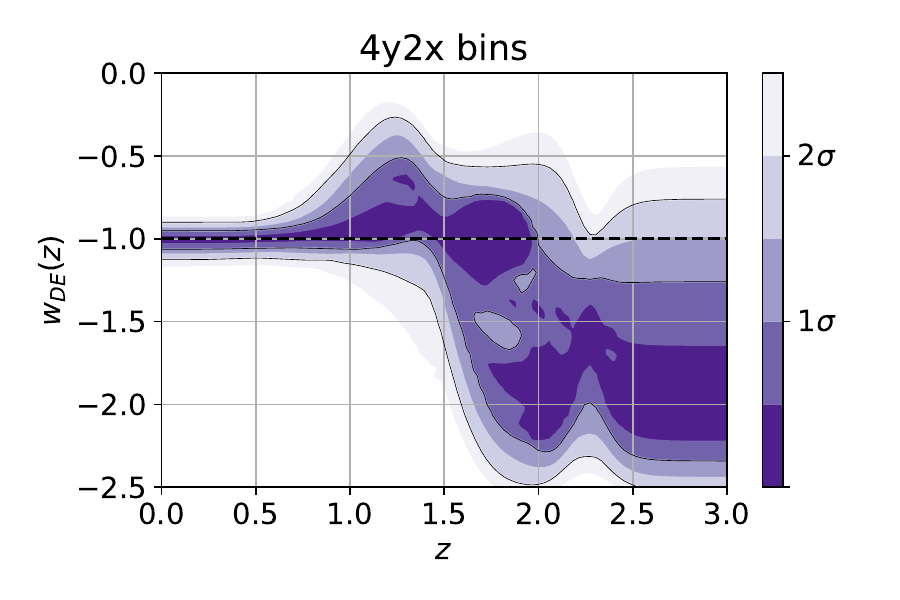}
     \includegraphics[trim = 26mm  0mm 5mm 0mm, clip, width=4.9cm, height=4.cm]{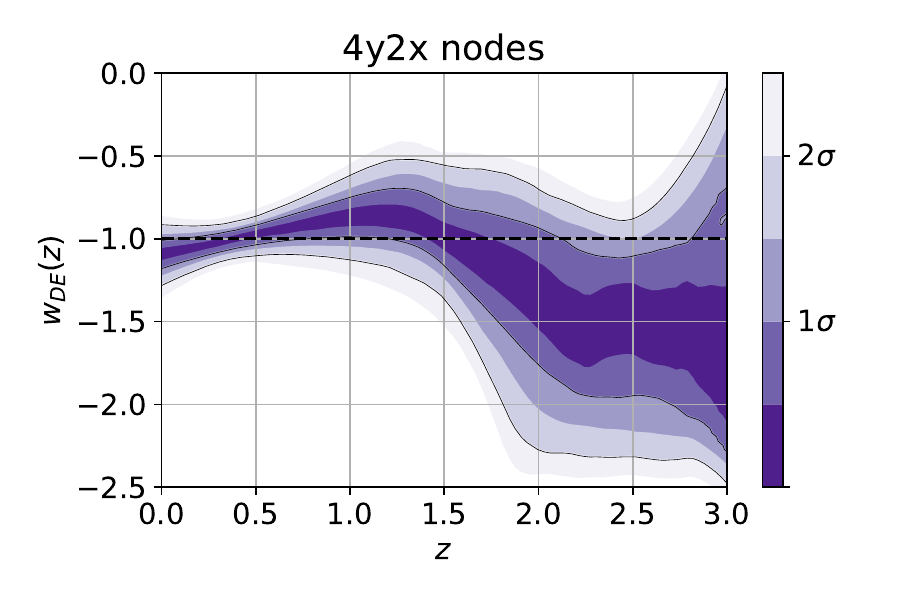}
     }
    \makebox[11cm][c]{
    \includegraphics[trim = 0mm  0mm 25mm 0mm, clip, width=4.9cm, height=4.cm]{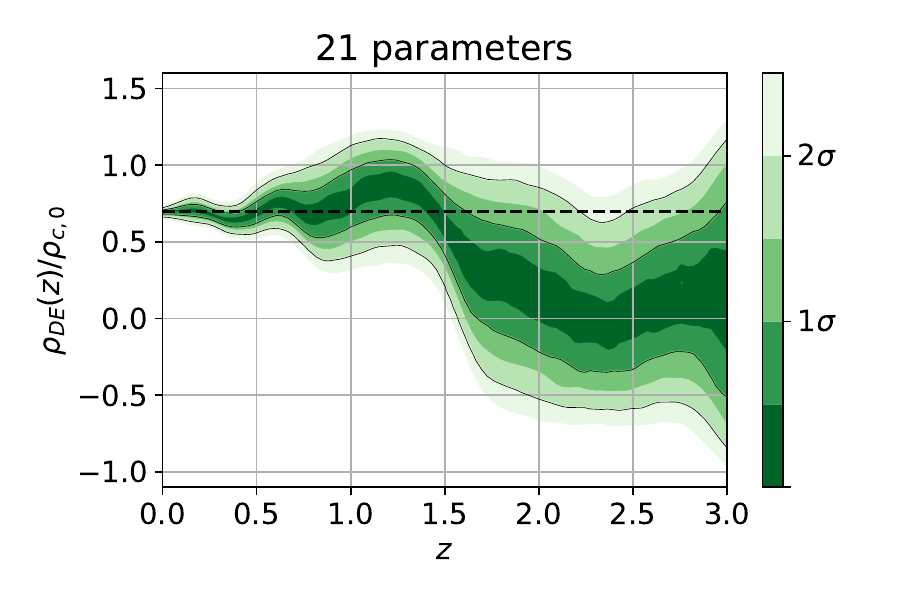}
    \includegraphics[trim = 26mm  0mm 25mm 0mm, clip, width=4.9cm, height=4.cm]{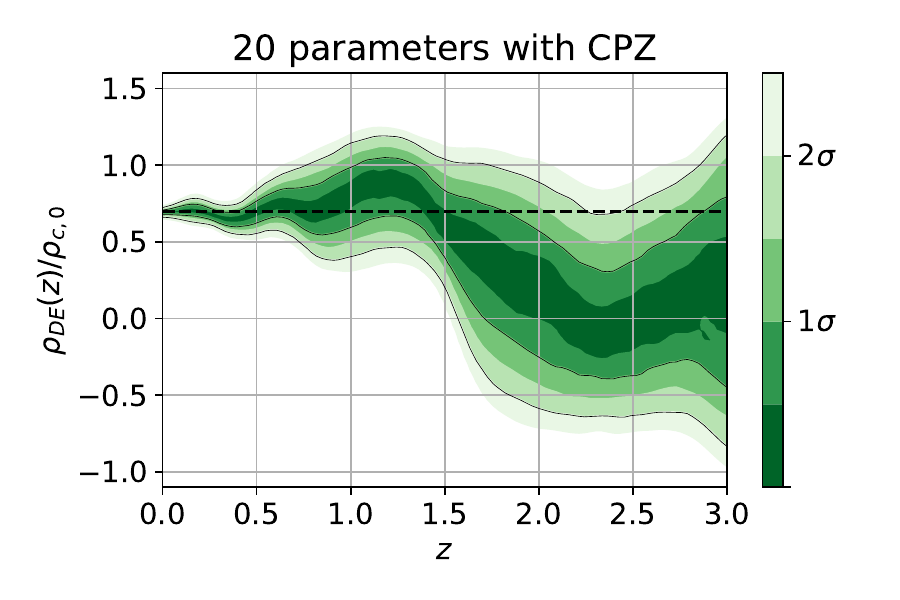}
     \includegraphics[trim = 26mm  0mm 25mm 0mm, clip, width=4.9cm, height=4.cm]{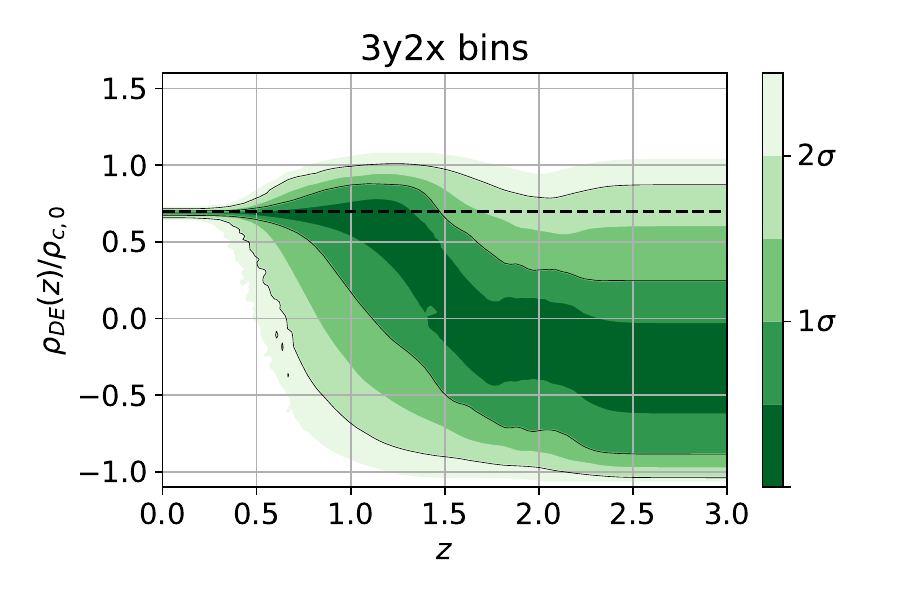} 
     \includegraphics[trim = 26mm  0mm 0mm 0mm, clip, width=4.9cm, height=4.cm]{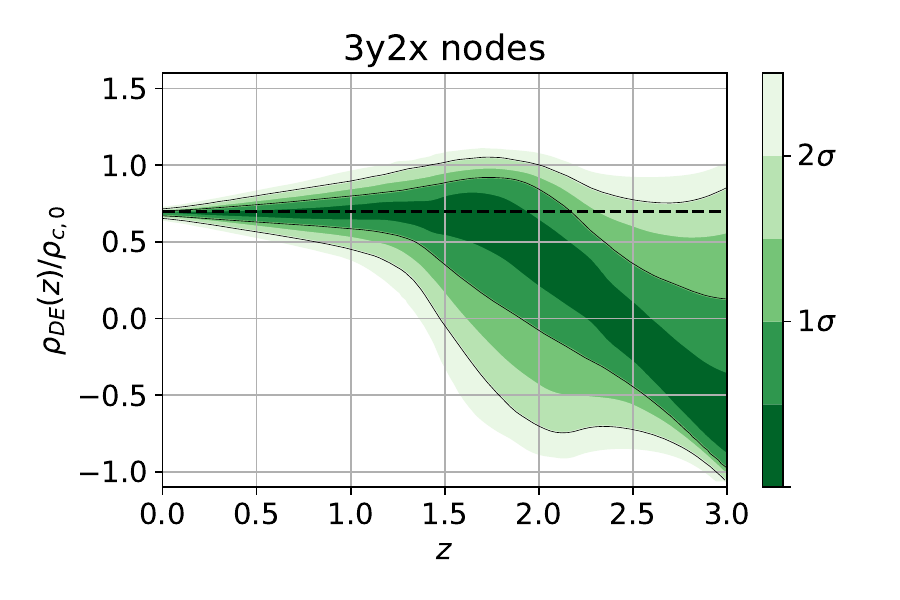} 
     }
     
\caption{From left to right: reconstructions with 20 bins, 20 bins with correlation function, internal $z_i$-bins and internal $z_i$-nodes. Purple figures correspond to the EoS whereas green to the energy density.
}\label{fig:rho_eos_corr_4y2x}
\end{figure*}

When computing the Bayes' factors of all the models (green region of the top panel in Fig. \ref{fig:fitness_full_panth}) one observes, in general,  a moderate evidence against the reconstructions regardless of the number of extra parameters, compared to the standard model. However, in the reconstructions certain features at high redshift become apparent as more parameters are added, and therefore an enhancement in the goodness of fit.  
For instance, when using three amplitudes (second column of Figure \ref{fig:nodes_full_panth_eos}), a bump-like shape appears at $z\approx 1.5$ and after that the general form prefers a crossing of the phantom-divide-line (PDL), i.e. for $z\gtrsim 2$ the amplitude values lean toward $w_{\tt DE}<-1$ outside the 68\% confidence region, which represents an improvement of $\Delta\chi^2 \sim -3$.
If we continue the process of adding amplitudes to the reconstructions, the main trend preserves a bump but now located at about $z\approx 1.2$ and a crossing of the PDL at $z\sim 1.5$. \luis{This is not true though when having 5 extra parameters (and we will have a similar problem with the density with 4 extra parameters), although the reason for this is mainly related to the reconstruction methodology, for a detailed explanation please refer to Appendix A.} We also obtained that the cosmological constant is slightly outside of the 95\% confidence contours at several places (see last column of Figure \ref{fig:nodes_full_panth_eos}), and according to the Table \ref{tabla_evidencias} by using the definition of statistical significance in terms of the standard deviation $\sigma$, that is, the signal-to-noise ratio $S/N \equiv \sqrt{|\Delta \chi^2|}$, it represents a $\sim 3\sigma$ deviation from $\Lambda$CDM based on the improvement in the fit alone.
These two features play a key role in identifying the correct dark energy model. If future surveys confirm their existence, the cosmological constant and single scalar field theories (with minimal assumptions) might be in serious problems as they cannot reproduce these essential features, and therefore alternative models should be taken into account.
Furthermore, in the internal-$z_i$ reconstructions the internal positions are able to localize the position for the bump and the PDL, and the results resemble the previous ones, see the last two columns of Figure \ref{fig:rho_eos_corr_4y2x}. Besides the presence of the bump and crossing of the PDL, we notice that at $z=0$ the 68\% confidence contour lays down right below the limits of $w_{\tt DE}<-1$. An important point to stress out is that the freedom of the internal positions led to a better fit to the data, compared to the reconstructions with the same number of parameters but fixed positions; displayed as the x-markers in the top panel of Figure \ref{fig:fitness_full_panth}. 
\\

Finally, to corroborate our findings, we include two reconstructions with 20 parameters:  binning and binning with CPZ correlation function, shown in the column 1 and 2 of Figure \ref{fig:rho_eos_corr_4y2x}. For these particular reconstructions, we have only focused on the binning method as it provides a better fit to the data.
As commented above, the correlation method is incorporated to create a function that evolves smoothly, and in general, they both share the same structure: a bump located at 
$z\approx1.2$, a crossing of the PDL at $z\approx 1.5$ and a slight preference of $w_{\tt DE}<-1$ at $z=0$. 
Besides these three features (found already in the internal models), there is also an oscillatory behaviour throughout the whole structure, which yields to a deviation of about $4\sigma$ to the $\Lambda$CDM. Even though this result may be considered as an overfitting due to the large number of parameters and small $\Delta \chi^2$, its Bayes factor is as good as the reconstruction with fewer amplitudes. Also, the authors in \cite{tamayo2019fourier} obtained a similar shape by using only three parameters in a Fourier basis and concluded a deviation of about $3\sigma$ from the cosmological constant, as we did here through a different mechanism.

\begin{figure*}[t!]
\captionsetup{justification=raggedright,singlelinecheck=false,font=footnotesize}
    \centering
    \makebox[11cm][c]{
    \includegraphics[trim = 0mm  0mm 25mm 0mm, clip, width=4.9cm, height=4.cm]{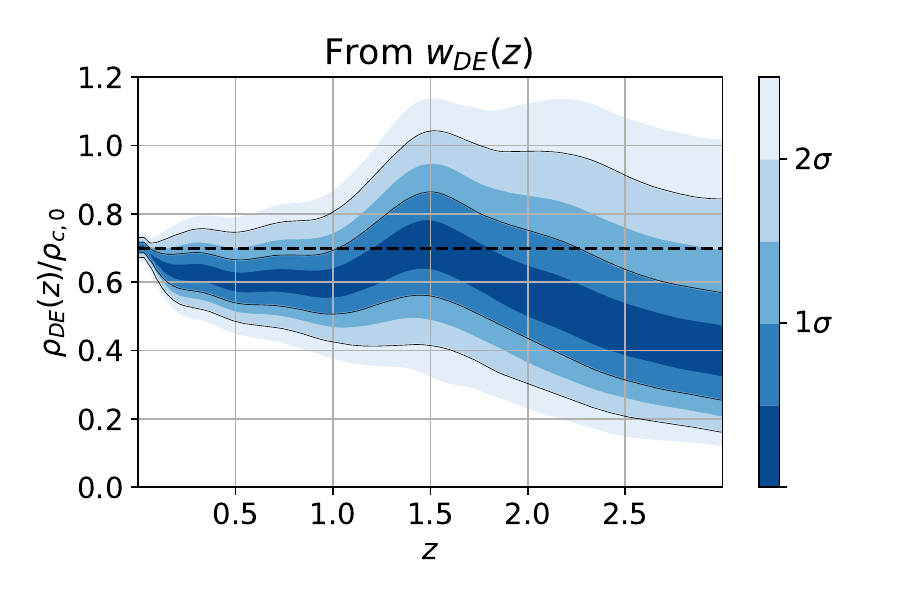}
    \includegraphics[trim = 0mm  0mm 25mm 0mm, clip, width=4.9cm, height=4.cm]{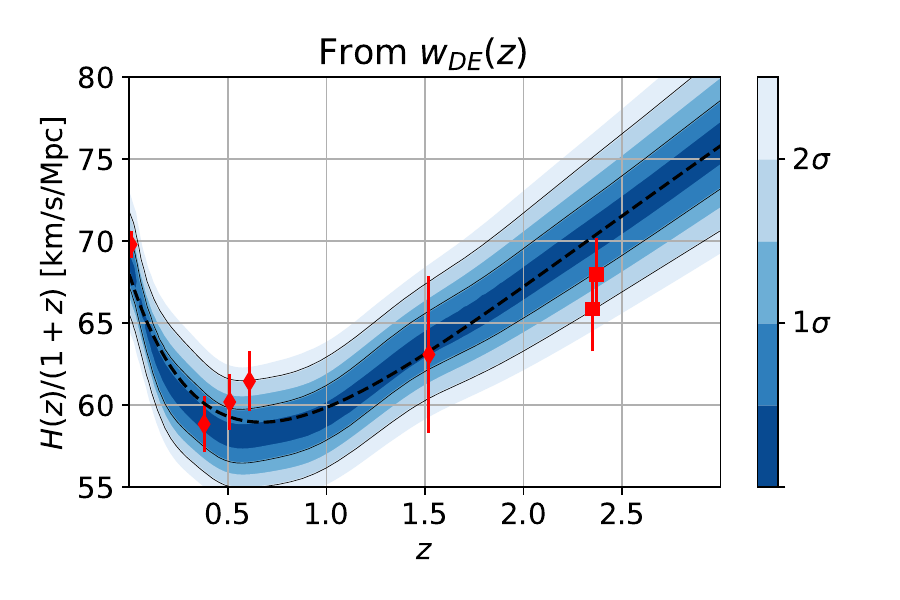}
     \includegraphics[trim = 0mm  0mm 25mm 0mm, clip, width=4.9cm, height=4.cm]{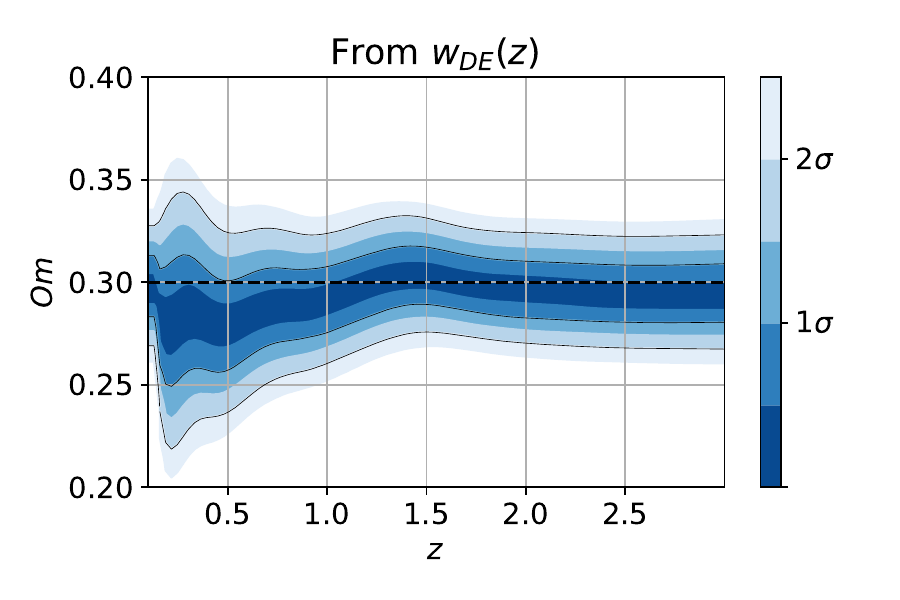}
     \includegraphics[trim = 0mm  0mm 0mm 0mm, clip, width=4.9cm, height=4.cm]{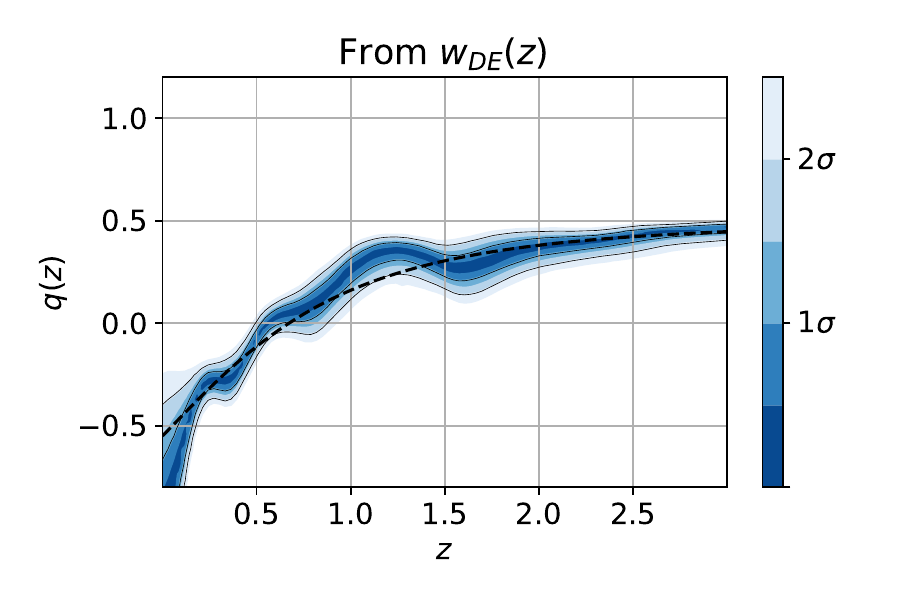}
     }
    \makebox[11cm][c]{
    \includegraphics[trim = 0mm  0mm 25mm 0mm, clip, width=4.9cm, height=4.cm]{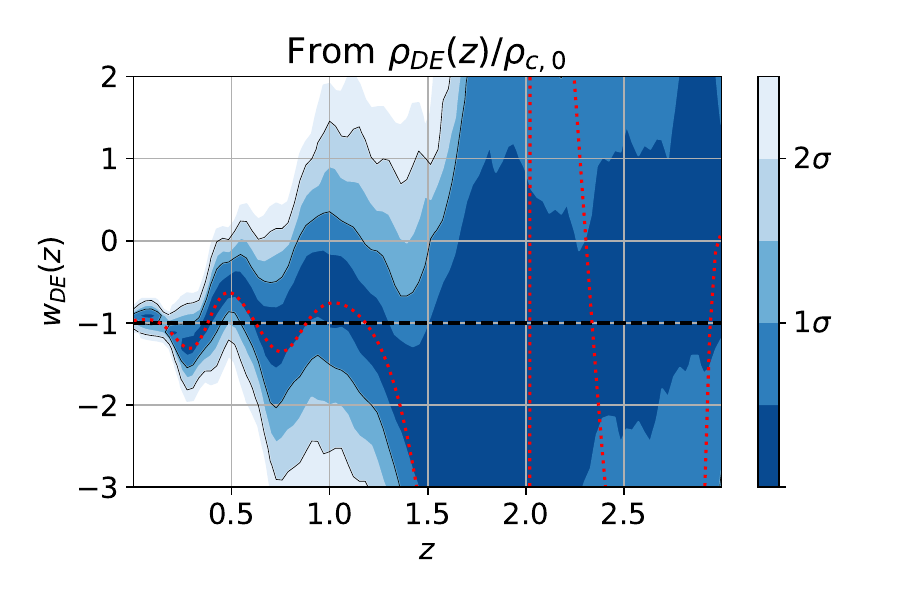}
         \includegraphics[trim = 0mm  0mm 25mm 0mm, clip, width=4.9cm, height=4.cm]{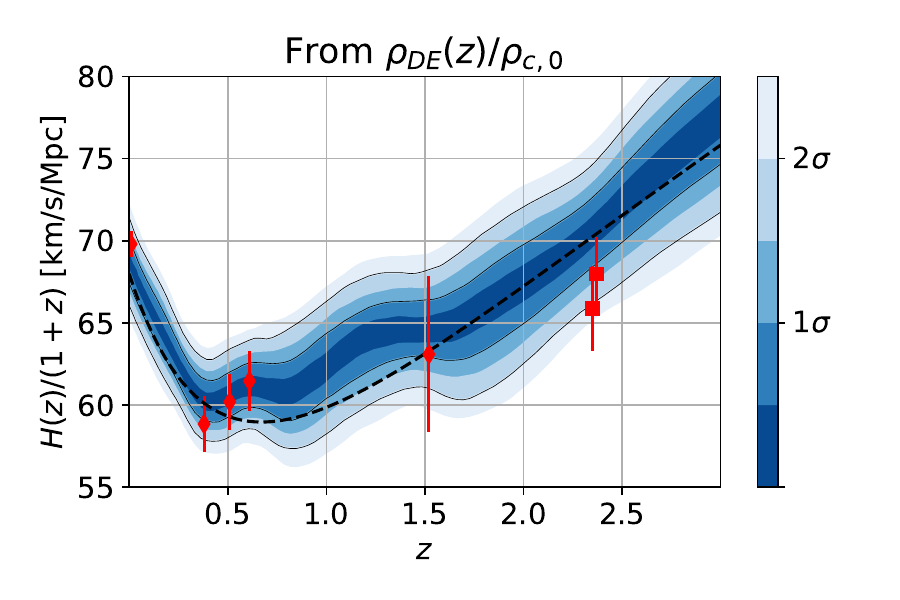}
     \includegraphics[trim = 0mm  0mm 25mm 0mm, clip, width=4.9cm, height=4.cm]{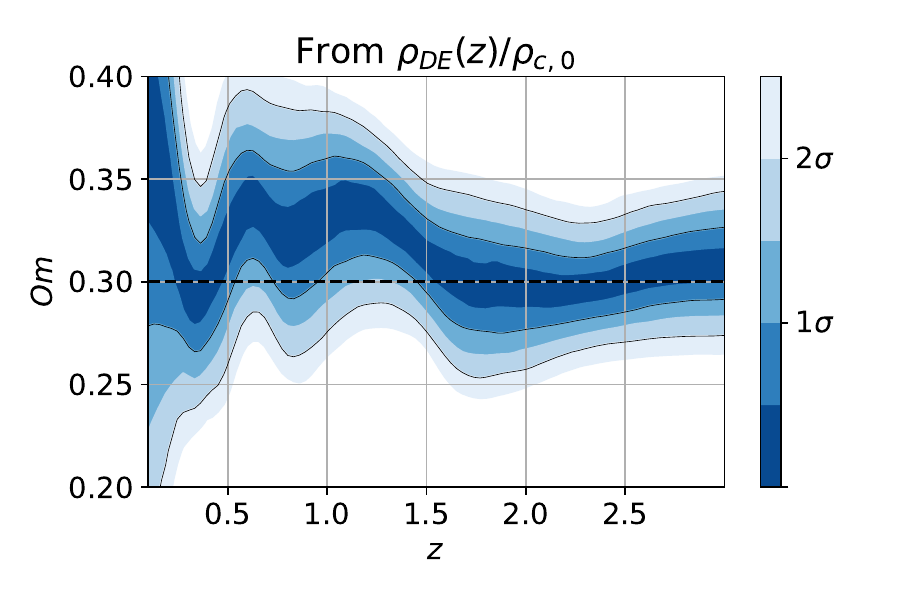}
     \includegraphics[trim = 0mm  0mm 0mm 0mm, clip, width=4.9cm, height=4.cm]{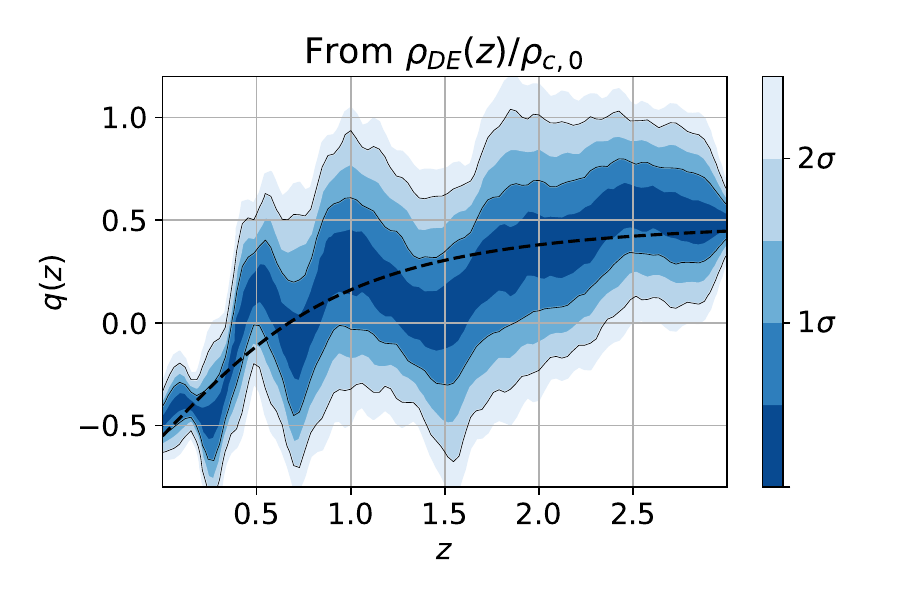}
     }
\caption{Derived functions  for the reconstructed $w_{\tt DE}(z)$ and $\rho_{\tt DE}(z)/\rho_{\tt c,0}$ with 20 bins plus correlation function. The red dotted line in the derived $w_{\tt DE}(z)$ corresponds to the best-fit values. 
}\label{fig:derived_20bins}
\end{figure*}

\subsection*{Dark Energy Density}   

Similarly to the previous section, with one extra parameter, through the $z_{\tt cut}$ in the sigmoid function for $\rho_{\tt DE}$, we have got
 a better fit to the data by more than 1$\sigma$, and its Bayes' factor results on a negative value $B_{\Lambda \text{CDM},i}= -0.124 \pm 0.131$, indicating a slight evidence in its favor, \luis{although it is still within the 1$\sigma$ for the error}. The constraints of the $z_{\tt cut}$ parameter corresponds to $2.101 \pm 0.413$, which provides an insight of a possible vanishing energy density beyond $z=2$. This feature is also noticeable  when looking at the 1-bin reconstruction where the cosmological constant is on the edge of the 68\% confidence contour for $z>1.5$   (see Fig. \ref{fig:parameterizations}).

By incorporating additional parameters to the reconstruction of $\rho_{\tt DE}(z)/\rho_{\tt c,0}$ the fit to the data becomes better than the standard model, as shown by the negative values of the $\Delta\chi^2$ on the bottom panel of Fig \ref{fig:fitness_full_panth}. 
Even though this behaviour is expected by the increased of complexity, it is accompanied by   
a penalization incorporated into the Bayes Factor, shown in the green part of Fig \ref{fig:fitness_full_panth}. 
As the number of  nodes/bins increases we expect the values of $B_{\Lambda \text{CDM},i}$ to do so too, however, an important point to note is the existence of a valley on this factor up to 4 parameters, where the first three have negative values  indicating  an evidence in favor (but still inconclusive) to our reconstructions, and also reflected on the improvement of the $\chi^2$. 
This may be happening due to the data having a preference for an energy density with a bump located at $z\approx 1.2$ and then as the redshift becomes larger $\rho_{\tt DE}(z)$ decreases until reaching a zero value, and even passing through negative values, but still statistically consistent with zero. 
That is, by having at least two amplitudes our reconstruction methodology is flexible enough to present this behavior, as seen in the first three columns of Fig \ref{fig:nodes_full_panth_eos}. 
As we continue adding more parameters, the presence of a  
possible sign change in the energy density is more pronounced. This transition occurs near $z\approx 2$, and in the region around $z\approx 2.5$ the deviation from the cosmological constant peaks. The general behaviour of our reconstructions is reflected by these two main features, which together provide a deviation up to $2.8\sigma$ from the standard model.

In the same manner as the reconstruction for the EoS, one may say that the position of these two features (the bump at $z\sim 1.2$ and the vanishing energy density behaviour for $z>1.5$) could be biased because the particular location for the amplitudes. \rev{Even so, it is stated in \cite{Akarsu:2022lhx} that these particular features and their positions are prompted by the Lyman-$\alpha$ BAO data (as we will further discuss in the PCA subsection).}
In order to find an optimal place for the internal positions, we set them free by allowing them to move around the $z$-axis. Because of this additional freedom, the internal reconstruction (or 3y2x model) is able to localize these features and provides a better fit to the data,  compared to the reconstruction with three fixed amplitudes.  
Moreover, despite having 5 extra parameters, the binning reconstruction has a negative Bayes factor which favours this model over the rest of the reconstructions (displayed as crosses in the bottom panel of Fig. \ref{fig:fitness_full_panth}).

Lastly, as was made with the EoS, we incorporated a 20 bins and 20 bins with the correlation function reconstructions.  
Both present more substructures like an oscillatory-like behavior at late times and also have a transition to a null or negative density in $z>2$, as seen in Fig \ref{fig:rho_eos_corr_4y2x}.
Analogous to the guidance offered by the oscillatory demeanor found in the EoS reconstruction, its apparent wavering nature in $z<1.5$ could encourage the study of an oscillatory basis such as a Fourier series.
Looking at the $\Delta\chi^2$ we notice a deviation from the $\Lambda$CDM of about $3.4\sigma$, and a Bayes factor comparable with the few-parameter reconstructions.

The drop-off behaviour of $\rho_{\tt DE}(z)$ and, perhaps, a transition to a negative energy density has been captured in other works \cite{mortsell2018does,marciu2016quintom, wang2018evolution, visinelli2019revisiting, akarsu2020graduated, Akarsu:2022lhx, Dutta:2018vmq}, as it seems to alleviate the tension that arises by estimating the Hubble constant $H_0$ with different datasets. \rev{It was also found in \cite{Colgain:2022rxy} that, when considering flat $\Lambda$CDM and binning the data, negative energy densities ($\Omega_{m,0} > 1$) are expected for higher redshifts.} Hence, it may be pertinent to study models with a similar demeanor. This behaviour does not necessarily imply a negative physical energy density per se, but it may be the indication of an effective energy density, i.e. similar to the one generated by the curvature component \cite{handley2021curvature, acquaviva2021simple}. 
\\

In general, and throughout the reconstruction process, we have found different features beyond the cosmological constant, which result in deviations up to $4\sigma$.
One final interesting observation is that the reconstructions when using bins are generally better than with nodes and also the Bayes factor shows an improvement. Likewise, there is a preference for the reconstruction with 20 bins over 20 bins plus the correlation function, reflected on the $\Delta\chi^2$,  in fact there is a strong evidence against using the 20 bins+prior model, according to the Jeffreys' scale.

\subsection*{Derived functions}

Once we have obtained the general form of both the EoS and the energy density we proceed to obtain their respective derived functions, these being: the Hubble parameter $H(z)/(1+z)$, the $Om$ diagnostic and the deceleration parameter $q(z)$. For the reconstructed energy density, we get as a derived the EoS and similarly for the reconstructed EoS, the energy density is an inferred one.
These derived functions correspond to the 20 bin reconstruction, which produced the best fit, and their posterior probabilities are displayed in Fig. \ref{fig:derived_20bins}. In general, the functions coming from the energy density show an enhanced oscillatory behaviour, compared to the functions derived from the EoS.  

When comparing the reconstructed EoS with the one deduced from  $\rho_{\tt DE}/\rho_{\tt c,0}$ we notice an important difference: we allowed for negative energy density values, hereby the derived EoS presents a discontinuity at about $z\approx2$, seen in the  best-fit model denoted by the red dotted line in Fig. \ref{fig:derived_20bins}. Such type of discontinuities have been found and studied in other reconstructions and different models, such as \cite{akarsu2020graduated,gomez2015background, wang2019searching}. 
Regarding the derived energy density: when reconstructing $\rho_{\tt DE}/\rho_{\tt c,0}$ directly its freedom in the parameter space allowed it to reach null values of the energy density, and even negative ones at high redshifts; but when a barotropic EoS is imposed, through the conservation equation, the derived energy density remains always positive with a bump located at $z\approx 1.5$ and a smooth drop out at high redshifts.

Considering the $H(z)/(1+z)$, the best-fit reconstructed function passes through the observational $H(z)$ values (red error bars in Fig. \ref{fig:derived_20bins}), $H_0=69.8\pm0.8\,{\rm km\,s}^{-1}{\rm Mpc}^{-1}$ from the TRGB \cite{Freedman:2019jwv}, consensus Galaxy BAO (from $z_{\rm eff}=0.38,\,0.51,\,0.61$) and DR14 Ly-$\alpha$ BAO (from $z_{\rm eff}=2.34,\,2.35$) \cite{blomqvist2019baryon,de2019baryon,anderson2014clustering}, and hence the best-fit is slightly better compared to the $\Lambda$CDM model (black dashed line). 

In general, the $Om$ diagnostic shows consistency with several parameterizations and reconstructions \cite{qi2018parameterized, mukherjee2021kinematical}. Nevertheless, in our reconstructions we have found a mixed behaviour between quintessence and phantom components, corroborated by the EoS. That is, we have certain places where $Om(z)> \Omega_{\rm m,0}$ (quintessence) and others with $Om(z)< \Omega_{\rm m,0}$ (phantom).

Finally, the deceleration parameter $q(z)$ for the reconstructed EoS gives a value for the transition to an accelerated Universe around $z\sim 0.6$, which is statistically consistent with the $\Lambda$CDM value, and with results previously obtained \cite{al2017parametric, bernal2017asymmetry, mukherjee2021kinematical}.
On the other hand, when $q(z)$ is reconstructed through the energy density, the universe goes through several short periods of acceleration-deceleration (a similar behaviour was seen in \cite{wang2018evolution}), however the main acceleration transition corresponds to $z\sim 0.45$, unlike 
in $\Lambda$CDM, where the acceleration starts at $z\sim 0.7$.

\subsection*{PCA and the Bayes' Factor}

\begin{figure}[t!]
\captionsetup{justification=raggedright,singlelinecheck=false,font=footnotesize}
    \centering
    \includegraphics[width=9.5cm, height=6.cm]{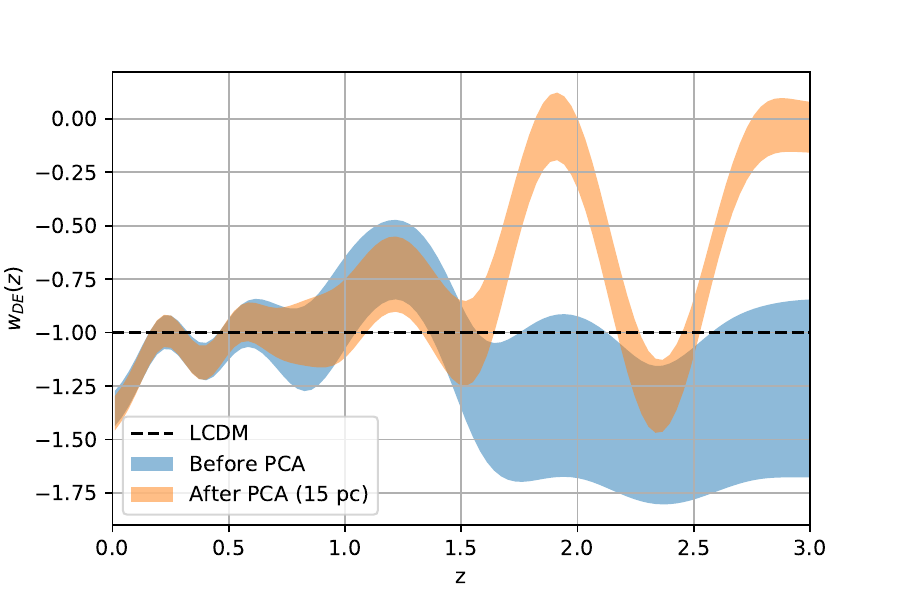}
    \includegraphics[width=9.5cm, height=6.cm]{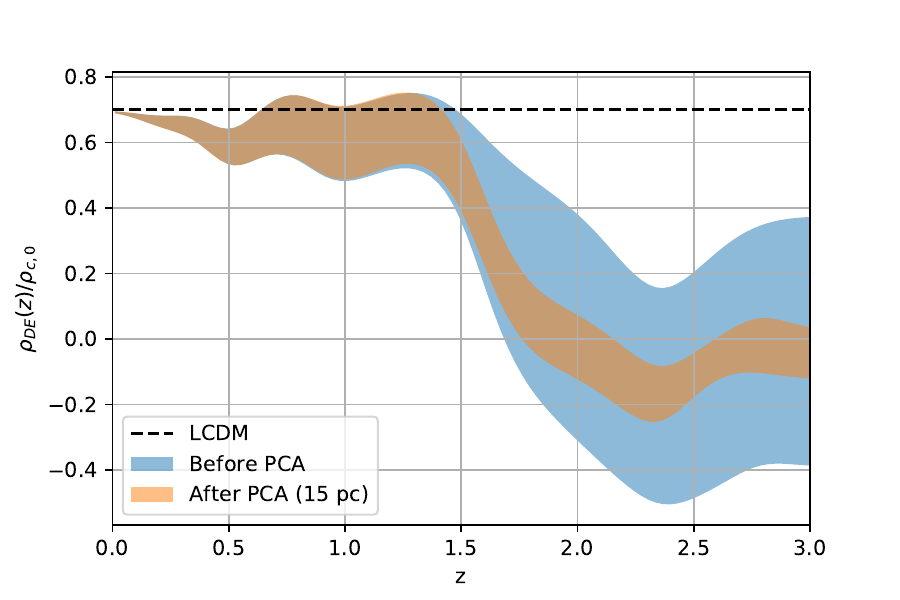}
    
\caption{Applying the principal component technique  to our reconstructions of the EoS and the energy density with 20 bins. By eliminating 5 PCs (which add up to about 5\% of the total variance for each reconstruction) we obtain the orange figures with slightly overestimated errors localized in $z>1.5$. 
    }\label{fig:pca_rho_nodal6}
\end{figure}

By applying the principal component analysis to remove the noisiest modes (enough to preserve $95\%$ of the variance) of the reconstructions, we see that the biggest changes happened mainly  at $z>2.0$. An example of this can be seen in Fig. \ref{fig:pca_rho_nodal6}, where we used the reconstruction with 20 bins for the energy density and the EoS. 
Since the only information here is coming from the Lyman-$\alpha$ BAO ($z\sim 2.3$) it may be reasonable to argue that the main tensions amongst models come from these high-redshift data (as has been also suggested in \cite{Planck:2018vyg, Evslin:2016gre, Cuceu:2019for}), or perhaps it exists the possibility that large systematic errors are present in this dataset. 
This may be an indication that the dynamical DE behavior, referring to the crossing of the PDL for the EoS and the null energy density at early times, is merely due to noisy data, although this will be confirmed until a significant amount of information is obtained in this redshift region. \luis{This also has an effect in the Bayesian Evidence. As the parameters beyond $z=1.5$ are the least constrained they contribute little to nothing to the final evidence of the reconstruction and thus means that we have to be very careful when utilizing the Bayes' Factor to directly perform model selection. Added to this is the fact that the Evidence is pretty susceptible to the prior range. Such problems are common when performing reconstructions, but a possible solution has been proposed in \cite{Keeley:2021dmx}, although it is still a work in progress.}

Nevertheless,  something that should be borne in mind is that certain characteristics pointing out to dynamics are still present even when removing the noisiest PCs. These characteristics,  in the energy density, are the oscillatory nature at low redshift and the transition to a null or negative energy density in $z\approx1.5$; with the EoS the preserved features include the oscillatory behavior, the bump in $z\approx 1.3$, a preference for values below $w_{\tt DE}=-1$ at $z=0$, and the crossing of the phantom divide line in late times. 

\section{Conclusions}\label{section:conclusions}

Throughout this work we have studied several parametric forms and model-independent reconstructions of two properties of the Dark Energy: the EoS and the energy density. 
Regarding the parametric forms, we have analyzed the $w$CDM and CPL models, whereas for the energy density we introduced a simple form given by a sigmoid function to describe a transition behaviour. Then, to introduce more flexibility for the recovered functions, we also included two types of reconstructions: based on step functions smoothly connected (bins) and linear interpolations (nodes). Each of the reconstructions may have several amplitudes as free parameters with fixed locations, as well as variable positions to localize possible features. On the top of the bin reconstruction we incorporated a floating prior that averages the set of neighboring amplitudes to behave as a mean between bins, hence  to preserve continuity among heights. 


The $w$CDM and CPL parameterizations of the EoS showed little to no improvement over $\Lambda$CDM.
However, with just a single parameter, the sigmoid function for the energy density had a better fitness to data as well as a better Bayes' factor when compared to $\Lambda$CDM. 
Even though this parameterization is phenomenological, it could be an indication that these type of models should be further studied upon.

By adding complexity in the model-independent reconstructions, through extra amplitudes with fixed positions, the  general outcome presented a dynamical behavior beyond the cosmological constant. 
In both reconstructed functions, $\rho_{\tt DE}(z)$ and $w_{\tt DE}(z)$, there is a presence of a bump located at about $z\sim 1.2$. This feature along with a crossing to the phantom-divide-line in the EoS, and a transition to a null energy density (or even negative values) yield to deviations from a constant energy density of about $3\sigma$. However, for these type of reconstructions the Bayes factor penalized the incorporation of parameters and displayed a moderate evidence against the reconstruction of $w_{\tt DE}(z)$ and an inconclusive to weak evidence for $\rho_{\tt DE}(z)$.

In order to avoid a possible bias due to the location of the amplitudes, we have let the internal points to move freely in the $z$-space. The autonomy of the internal positions led to localize the features, previously mentioned, and to an improvement on the fit of about $1\sigma$, in comparison to the same number of amplitudes with fixed positions. Another point to stress out about the internal reconstructions, is the enhancement of the Bayes' factor which could get even negative values, i.e. for $\rho_{\tt DE}(z)$ with bins and the $3y2x$ method.

Finally, when considering 20 parameters in the binned reconstruction we noticed an improvement in the fitness up to about $4\sigma$. Nevertheless, a key result to bear in mind is that for the case of 20 bins+prior the Bayes's factor showed a strong evidence against this type of reconstruction. This is also reflected on a worse $\Delta \chi^2$, when compared to a reconstruction with the same number of parameters.





In general, the derived functions inherited the behaviour from the reconstructed ones: an oscillatory shape. These functions also exhibited consistency with other reconstruction methods. For instance, a dynamical DE behaviour, a different redshift transition from a decelerating universe to an accelerating one, a better agreement to the BAO data.
From the energy density reconstruction, the derived EoS presents a discontinuity at redshift around $z\sim 2$, which is necessary if the energy density transitioned to negative values; 
found in several models and parameterizations.


By performing a principal component analysis we found that, in both types of reconstructions, the amplitudes beyond $z=1.5$ are the least constrained, where the predominant data around these redshift values is the BAO Lyman-$\alpha$. This shows that current Lyman-$\alpha$ BAO prefers a nearly null or negative energy density,  or a transition from quintessence to a phantom dark energy. However, more precise data in this region is necessary to fully discern a dynamical DE, \luis{also, considering that the parameters in this region are unconstrained, it directly affects the estimation of the Bayesian Evidence, so any remarks about the Bayes' Factor should be made carefully}. Nevertheless it was found that some features could indeed be considered as signal, like the general oscillatory behaviour, the bump located at $z\approx1.2$ and the $w_{\tt DE}<-1$ at the present redshift.
\\

Some concluding remarks can be summarized as follow: i) the binned reconstruction provided better results than the node method; ii) when incorporating the correlation function, the fitness and the Bayes' Factor worsen considerably, even when the final shape is very much alike;
iii) our model-independent reconstructions resulted in a better fit to the data (up to $4\sigma$) and, in  some cases a better Bayes' Factor; 
iv) if future surveys confirm our results, the cosmological constant and single scalar field theories (with minimal assumptions) might be in serious problems as they cannot reproduce these essential features. This could be a great incentive to study DE models with this type of dynamical behavior and encourage direct reconstruction of other functions, such as those that may lead to discontinuities in the EoS.
Finally, v) the PCA analysis showed a great promise for some of the reconstructed features, such as the oscillations and a bump at intermediate redshifts. 

As a future prospect, it remains the addition of other datasets that contain information from linear perturbations, such as the cosmic microwave background and the matter power spectrum data. Also, with the resulting shape provided by our model-independent reconstructions, we may search for a parameterization or a physical model that incorporates the important characteristics already found. It would also be interesting to study some alternative PCA methods that have a mathematical basis to truly discern important features from noise.

\appendix

\section*{Acknowledgements}

L.A.E. was supported by CONACyT M\'exico. J.A.V. acknowledges the support provided by FOSEC SEP-CONACYT Investigaci\'on B\'asica A1-S-21925, FORDECYT-PRONACES-CONACYT/304001/2020 and UNAM-DGAPA-PAPIIT IA104221.

\section*{Data availability}
The data underlying this article are available in the article and in its online supplementary material.






\bibliographystyle{apsrev4-1}
\bibliography{bibliography.bib}

\begin{thebibliography}{108}%
\makeatletter
\providecommand \@ifxundefined [1]{%
 \@ifx{#1\undefined}
}%
\providecommand \@ifnum [1]{%
 \ifnum #1\expandafter \@firstoftwo
 \else \expandafter \@secondoftwo
 \fi
}%
\providecommand \@ifx [1]{%
 \ifx #1\expandafter \@firstoftwo
 \else \expandafter \@secondoftwo
 \fi
}%
\providecommand \natexlab [1]{#1}%
\providecommand \enquote  [1]{``#1''}%
\providecommand \bibnamefont  [1]{#1}%
\providecommand \bibfnamefont [1]{#1}%
\providecommand \citenamefont [1]{#1}%
\providecommand \href@noop [0]{\@secondoftwo}%
\providecommand \href [0]{\begingroup \@sanitize@url \@href}%
\providecommand \@href[1]{\@@startlink{#1}\@@href}%
\providecommand \@@href[1]{\endgroup#1\@@endlink}%
\providecommand \@sanitize@url [0]{\catcode `\\12\catcode `\$12\catcode
  `\&12\catcode `\#12\catcode `\^12\catcode `\_12\catcode `\%12\relax}%
\providecommand \@@startlink[1]{}%
\providecommand \@@endlink[0]{}%
\providecommand \url  [0]{\begingroup\@sanitize@url \@url }%
\providecommand \@url [1]{\endgroup\@href {#1}{\urlprefix }}%
\providecommand \urlprefix  [0]{URL }%
\providecommand \Eprint [0]{\href }%
\providecommand \doibase [0]{http://dx.doi.org/}%
\providecommand \selectlanguage [0]{\@gobble}%
\providecommand \bibinfo  [0]{\@secondoftwo}%
\providecommand \bibfield  [0]{\@secondoftwo}%
\providecommand \translation [1]{[#1]}%
\providecommand \BibitemOpen [0]{}%
\providecommand \bibitemStop [0]{}%
\providecommand \bibitemNoStop [0]{.\EOS\space}%
\providecommand \EOS [0]{\spacefactor3000\relax}%
\providecommand \BibitemShut  [1]{\csname bibitem#1\endcsname}%
\let\auto@bib@innerbib\@empty
\bibitem [{\citenamefont {Sahni}(2002)}]{sahni2002cosmological}%
  \BibitemOpen
  \bibfield  {author} {\bibinfo {author} {\bibfnamefont {V.}~\bibnamefont
  {Sahni}},\ }\href {\doibase 10.1088/0264-9381/19/13/304} {\bibfield
  {journal} {\bibinfo  {journal} {Class. Quant. Grav.}\ }\textbf {\bibinfo
  {volume} {19}},\ \bibinfo {pages} {3435} (\bibinfo {year} {2002})},\ \Eprint
  {http://arxiv.org/abs/astro-ph/0202076} {arXiv:astro-ph/0202076} \BibitemShut
  {NoStop}%
\bibitem [{\citenamefont {Bull}\ \emph {et~al.}(2016)\citenamefont {Bull},
  \citenamefont {Akrami}, \citenamefont {Adamek}, \citenamefont {Baker},
  \citenamefont {Bellini}, \citenamefont {Jimenez}, \citenamefont {Bentivegna},
  \citenamefont {Camera}, \citenamefont {Clesse}, \citenamefont {Davis} \emph
  {et~al.}}]{bull2016beyond}%
  \BibitemOpen
  \bibfield  {author} {\bibinfo {author} {\bibfnamefont {P.}~\bibnamefont
  {Bull}}, \bibinfo {author} {\bibfnamefont {Y.}~\bibnamefont {Akrami}},
  \bibinfo {author} {\bibfnamefont {J.}~\bibnamefont {Adamek}}, \bibinfo
  {author} {\bibfnamefont {T.}~\bibnamefont {Baker}}, \bibinfo {author}
  {\bibfnamefont {E.}~\bibnamefont {Bellini}}, \bibinfo {author} {\bibfnamefont
  {J.~B.}\ \bibnamefont {Jimenez}}, \bibinfo {author} {\bibfnamefont
  {E.}~\bibnamefont {Bentivegna}}, \bibinfo {author} {\bibfnamefont
  {S.}~\bibnamefont {Camera}}, \bibinfo {author} {\bibfnamefont
  {S.}~\bibnamefont {Clesse}}, \bibinfo {author} {\bibfnamefont {J.~H.}\
  \bibnamefont {Davis}},  \emph {et~al.},\ }\href {\doibase
  10.1016/j.dark.2016.02.001} {\bibfield  {journal} {\bibinfo  {journal} {Phys.
  Dark Univ.}\ }\textbf {\bibinfo {volume} {12}},\ \bibinfo {pages} {56}
  (\bibinfo {year} {2016})},\ \Eprint {http://arxiv.org/abs/1512.05356}
  {arXiv:1512.05356 [astro-ph.CO]} \BibitemShut {NoStop}%
\bibitem [{\citenamefont {Zhao}\ \emph {et~al.}(2017)\citenamefont {Zhao},
  \citenamefont {Raveri}, \citenamefont {Pogosian}, \citenamefont {Wang},
  \citenamefont {Crittenden}, \citenamefont {Handley}, \citenamefont
  {Percival}, \citenamefont {Beutler}, \citenamefont {Brinkmann}, \citenamefont
  {Chuang} \emph {et~al.}}]{zhao2017dynamical}%
  \BibitemOpen
  \bibfield  {author} {\bibinfo {author} {\bibfnamefont {G.-B.}\ \bibnamefont
  {Zhao}}, \bibinfo {author} {\bibfnamefont {M.}~\bibnamefont {Raveri}},
  \bibinfo {author} {\bibfnamefont {L.}~\bibnamefont {Pogosian}}, \bibinfo
  {author} {\bibfnamefont {Y.}~\bibnamefont {Wang}}, \bibinfo {author}
  {\bibfnamefont {R.~G.}\ \bibnamefont {Crittenden}}, \bibinfo {author}
  {\bibfnamefont {W.~J.}\ \bibnamefont {Handley}}, \bibinfo {author}
  {\bibfnamefont {W.~J.}\ \bibnamefont {Percival}}, \bibinfo {author}
  {\bibfnamefont {F.}~\bibnamefont {Beutler}}, \bibinfo {author} {\bibfnamefont
  {J.}~\bibnamefont {Brinkmann}}, \bibinfo {author} {\bibfnamefont {C.-H.}\
  \bibnamefont {Chuang}},  \emph {et~al.},\ }\href {\doibase
  10.1038/s41550-017-0216-z} {\bibfield  {journal} {\bibinfo  {journal} {Nature
  Astron.}\ }\textbf {\bibinfo {volume} {1}},\ \bibinfo {pages} {627} (\bibinfo
  {year} {2017})},\ \Eprint {http://arxiv.org/abs/1701.08165} {arXiv:1701.08165
  [astro-ph.CO]} \BibitemShut {NoStop}%
\bibitem [{\citenamefont {Wong}\ \emph {et~al.}(2020)\citenamefont {Wong} \emph
  {et~al.}}]{Wong:2019kwg}%
  \BibitemOpen
  \bibfield  {author} {\bibinfo {author} {\bibfnamefont {K.~C.}\ \bibnamefont
  {Wong}} \emph {et~al.},\ }\href {\doibase 10.1093/mnras/stz3094} {\bibfield
  {journal} {\bibinfo  {journal} {Mon. Not. Roy. Astron. Soc.}\ }\textbf
  {\bibinfo {volume} {498}},\ \bibinfo {pages} {1420} (\bibinfo {year}
  {2020})},\ \Eprint {http://arxiv.org/abs/1907.04869} {arXiv:1907.04869
  [astro-ph.CO]} \BibitemShut {NoStop}%
\bibitem [{\citenamefont {Krishnan}\ \emph {et~al.}(2021)\citenamefont
  {Krishnan}, \citenamefont {Colg\'ain}, \citenamefont {Sheikh-Jabbari},\ and\
  \citenamefont {Yang}}]{Krishnan:2020vaf}%
  \BibitemOpen
  \bibfield  {author} {\bibinfo {author} {\bibfnamefont {C.}~\bibnamefont
  {Krishnan}}, \bibinfo {author} {\bibfnamefont {E.~O.}\ \bibnamefont
  {Colg\'ain}}, \bibinfo {author} {\bibfnamefont {M.~M.}\ \bibnamefont
  {Sheikh-Jabbari}}, \ and\ \bibinfo {author} {\bibfnamefont {T.}~\bibnamefont
  {Yang}},\ }\href {\doibase 10.1103/PhysRevD.103.103509} {\bibfield  {journal}
  {\bibinfo  {journal} {Phys. Rev. D}\ }\textbf {\bibinfo {volume} {103}},\
  \bibinfo {pages} {103509} (\bibinfo {year} {2021})},\ \Eprint
  {http://arxiv.org/abs/2011.02858} {arXiv:2011.02858 [astro-ph.CO]}
  \BibitemShut {NoStop}%
\bibitem [{\citenamefont {Hu}\ and\ \citenamefont {Wang}(2022)}]{Hu:2022kes}%
  \BibitemOpen
  \bibfield  {author} {\bibinfo {author} {\bibfnamefont {J.-P.}\ \bibnamefont
  {Hu}}\ and\ \bibinfo {author} {\bibfnamefont {F.~Y.}\ \bibnamefont {Wang}},\
  }\href {\doibase 10.1093/mnras/stac2728} {\bibfield  {journal} {\bibinfo
  {journal} {Mon. Not. Roy. Astron. Soc.}\ }\textbf {\bibinfo {volume} {517}},\
  \bibinfo {pages} {576} (\bibinfo {year} {2022})},\ \Eprint
  {http://arxiv.org/abs/2203.13037} {arXiv:2203.13037 [astro-ph.CO]}
  \BibitemShut {NoStop}%
\bibitem [{\citenamefont {Krishnan}\ and\ \citenamefont
  {Mondol}(2022)}]{Krishnan:2022fzz}%
  \BibitemOpen
  \bibfield  {author} {\bibinfo {author} {\bibfnamefont {C.}~\bibnamefont
  {Krishnan}}\ and\ \bibinfo {author} {\bibfnamefont {R.}~\bibnamefont
  {Mondol}},\ }\href@noop {} {\  (\bibinfo {year} {2022})},\ \Eprint
  {http://arxiv.org/abs/2201.13384} {arXiv:2201.13384 [astro-ph.CO]}
  \BibitemShut {NoStop}%
\bibitem [{\citenamefont {Colg\'ain}\ \emph {et~al.}(2022)\citenamefont
  {Colg\'ain}, \citenamefont {Sheikh-Jabbari}, \citenamefont {Solomon},
  \citenamefont {Dainotti},\ and\ \citenamefont {Stojkovic}}]{Colgain:2022rxy}%
  \BibitemOpen
  \bibfield  {author} {\bibinfo {author} {\bibfnamefont {E.~O.}\ \bibnamefont
  {Colg\'ain}}, \bibinfo {author} {\bibfnamefont {M.~M.}\ \bibnamefont
  {Sheikh-Jabbari}}, \bibinfo {author} {\bibfnamefont {R.}~\bibnamefont
  {Solomon}}, \bibinfo {author} {\bibfnamefont {M.~G.}\ \bibnamefont
  {Dainotti}}, \ and\ \bibinfo {author} {\bibfnamefont {D.}~\bibnamefont
  {Stojkovic}},\ }\href@noop {} {\  (\bibinfo {year} {2022})},\ \Eprint
  {http://arxiv.org/abs/2206.11447} {arXiv:2206.11447 [astro-ph.CO]}
  \BibitemShut {NoStop}%
\bibitem [{\citenamefont {V\'azquez}\ \emph {et~al.}(2021)\citenamefont
  {V\'azquez}, \citenamefont {Tamayo}, \citenamefont {Sen},\ and\ \citenamefont
  {Quiros}}]{Vazquez:2020ani}%
  \BibitemOpen
  \bibfield  {author} {\bibinfo {author} {\bibfnamefont {J.~A.}\ \bibnamefont
  {V\'azquez}}, \bibinfo {author} {\bibfnamefont {D.}~\bibnamefont {Tamayo}},
  \bibinfo {author} {\bibfnamefont {A.~A.}\ \bibnamefont {Sen}}, \ and\
  \bibinfo {author} {\bibfnamefont {I.}~\bibnamefont {Quiros}},\ }\href
  {\doibase 10.1103/PhysRevD.103.043506} {\bibfield  {journal} {\bibinfo
  {journal} {Phys. Rev. D}\ }\textbf {\bibinfo {volume} {103}},\ \bibinfo
  {pages} {043506} (\bibinfo {year} {2021})},\ \Eprint
  {http://arxiv.org/abs/2009.01904} {arXiv:2009.01904 [gr-qc]} \BibitemShut
  {NoStop}%
\bibitem [{\citenamefont {Caldwell}(2002)}]{caldwell2002phantom}%
  \BibitemOpen
  \bibfield  {author} {\bibinfo {author} {\bibfnamefont {R.~R.}\ \bibnamefont
  {Caldwell}},\ }\href {\doibase 10.1016/S0370-2693(02)02589-3} {\bibfield
  {journal} {\bibinfo  {journal} {Phys. Lett. B}\ }\textbf {\bibinfo {volume}
  {545}},\ \bibinfo {pages} {23} (\bibinfo {year} {2002})},\ \Eprint
  {http://arxiv.org/abs/astro-ph/9908168} {arXiv:astro-ph/9908168} \BibitemShut
  {NoStop}%
\bibitem [{\citenamefont {Caldwell}\ and\ \citenamefont
  {Linder}(2005)}]{caldwell2005limits}%
  \BibitemOpen
  \bibfield  {author} {\bibinfo {author} {\bibfnamefont {R.}~\bibnamefont
  {Caldwell}}\ and\ \bibinfo {author} {\bibfnamefont {E.~V.}\ \bibnamefont
  {Linder}},\ }\href {\doibase 10.1103/PhysRevLett.95.141301} {\bibfield
  {journal} {\bibinfo  {journal} {Phys. Rev. Lett.}\ }\textbf {\bibinfo
  {volume} {95}},\ \bibinfo {pages} {141301} (\bibinfo {year} {2005})},\
  \Eprint {http://arxiv.org/abs/astro-ph/0505494} {arXiv:astro-ph/0505494}
  \BibitemShut {NoStop}%
\bibitem [{\citenamefont {Guo}\ \emph {et~al.}(2005)\citenamefont {Guo},
  \citenamefont {Piao}, \citenamefont {Zhang},\ and\ \citenamefont
  {Zhang}}]{guo2005cosmological}%
  \BibitemOpen
  \bibfield  {author} {\bibinfo {author} {\bibfnamefont {Z.-K.}\ \bibnamefont
  {Guo}}, \bibinfo {author} {\bibfnamefont {Y.-S.}\ \bibnamefont {Piao}},
  \bibinfo {author} {\bibfnamefont {X.-M.}\ \bibnamefont {Zhang}}, \ and\
  \bibinfo {author} {\bibfnamefont {Y.-Z.}\ \bibnamefont {Zhang}},\ }\href
  {\doibase 10.1016/j.physletb.2005.01.017} {\bibfield  {journal} {\bibinfo
  {journal} {Phys. Lett. B}\ }\textbf {\bibinfo {volume} {608}},\ \bibinfo
  {pages} {177} (\bibinfo {year} {2005})},\ \Eprint
  {http://arxiv.org/abs/astro-ph/0410654} {arXiv:astro-ph/0410654} \BibitemShut
  {NoStop}%
\bibitem [{\citenamefont {Nojiri}\ and\ \citenamefont
  {Odintsov}(2011)}]{nojiri2011unified}%
  \BibitemOpen
  \bibfield  {author} {\bibinfo {author} {\bibfnamefont {S.}~\bibnamefont
  {Nojiri}}\ and\ \bibinfo {author} {\bibfnamefont {S.~D.}\ \bibnamefont
  {Odintsov}},\ }\href {\doibase 10.1016/j.physrep.2011.04.001} {\bibfield
  {journal} {\bibinfo  {journal} {Phys. Rept.}\ }\textbf {\bibinfo {volume}
  {505}},\ \bibinfo {pages} {59} (\bibinfo {year} {2011})},\ \Eprint
  {http://arxiv.org/abs/1011.0544} {arXiv:1011.0544 [gr-qc]} \BibitemShut
  {NoStop}%
\bibitem [{\citenamefont {Akarsu}\ \emph
  {et~al.}(2020{\natexlab{a}})\citenamefont {Akarsu}, \citenamefont
  {Kat\i{}rc\i{}}, \citenamefont {\"Ozdemir},\ and\ \citenamefont
  {V\'azquez}}]{Akarsu:2019pvi}%
  \BibitemOpen
  \bibfield  {author} {\bibinfo {author} {\bibfnamefont {O.}~\bibnamefont
  {Akarsu}}, \bibinfo {author} {\bibfnamefont {N.}~\bibnamefont
  {Kat\i{}rc\i{}}}, \bibinfo {author} {\bibfnamefont {N.}~\bibnamefont
  {\"Ozdemir}}, \ and\ \bibinfo {author} {\bibfnamefont {J.~A.}\ \bibnamefont
  {V\'azquez}},\ }\href {\doibase 10.1140/epjc/s10052-019-7580-z} {\bibfield
  {journal} {\bibinfo  {journal} {Eur. Phys. J. C}\ }\textbf {\bibinfo {volume}
  {80}},\ \bibinfo {pages} {32} (\bibinfo {year} {2020}{\natexlab{a}})},\
  \Eprint {http://arxiv.org/abs/1903.06679} {arXiv:1903.06679 [gr-qc]}
  \BibitemShut {NoStop}%
\bibitem [{\citenamefont {Ida}(2000)}]{ida2000brane}%
  \BibitemOpen
  \bibfield  {author} {\bibinfo {author} {\bibfnamefont {D.}~\bibnamefont
  {Ida}},\ }\href {\doibase 10.1088/1126-6708/2000/09/014} {\bibfield
  {journal} {\bibinfo  {journal} {JHEP}\ }\textbf {\bibinfo {volume} {09}},\
  \bibinfo {pages} {014} (\bibinfo {year} {2000})},\ \Eprint
  {http://arxiv.org/abs/gr-qc/9912002} {arXiv:gr-qc/9912002} \BibitemShut
  {NoStop}%
\bibitem [{\citenamefont {Brax}\ \emph {et~al.}(2004)\citenamefont {Brax},
  \citenamefont {van~de Bruck},\ and\ \citenamefont {Davis}}]{brax2004brane}%
  \BibitemOpen
  \bibfield  {author} {\bibinfo {author} {\bibfnamefont {P.}~\bibnamefont
  {Brax}}, \bibinfo {author} {\bibfnamefont {C.}~\bibnamefont {van~de Bruck}},
  \ and\ \bibinfo {author} {\bibfnamefont {A.-C.}\ \bibnamefont {Davis}},\
  }\href {\doibase 10.1088/0034-4885/67/12/R02} {\bibfield  {journal} {\bibinfo
   {journal} {Rept. Prog. Phys.}\ }\textbf {\bibinfo {volume} {67}},\ \bibinfo
  {pages} {2183} (\bibinfo {year} {2004})},\ \Eprint
  {http://arxiv.org/abs/hep-th/0404011} {arXiv:hep-th/0404011} \BibitemShut
  {NoStop}%
\bibitem [{\citenamefont {Sahni}\ and\ \citenamefont
  {Starobinsky}(2006)}]{sahni2006reconstructing}%
  \BibitemOpen
  \bibfield  {author} {\bibinfo {author} {\bibfnamefont {V.}~\bibnamefont
  {Sahni}}\ and\ \bibinfo {author} {\bibfnamefont {A.}~\bibnamefont
  {Starobinsky}},\ }\href {\doibase 10.1142/S0218271806009704} {\bibfield
  {journal} {\bibinfo  {journal} {Int. J. Mod. Phys. D}\ }\textbf {\bibinfo
  {volume} {15}},\ \bibinfo {pages} {2105} (\bibinfo {year} {2006})},\ \Eprint
  {http://arxiv.org/abs/astro-ph/0610026} {arXiv:astro-ph/0610026} \BibitemShut
  {NoStop}%
\bibitem [{\citenamefont {Gong}\ and\ \citenamefont
  {Wang}(2007)}]{gong2007reconstruction}%
  \BibitemOpen
  \bibfield  {author} {\bibinfo {author} {\bibfnamefont {Y.-G.}\ \bibnamefont
  {Gong}}\ and\ \bibinfo {author} {\bibfnamefont {A.}~\bibnamefont {Wang}},\
  }\href {\doibase 10.1103/PhysRevD.75.043520} {\bibfield  {journal} {\bibinfo
  {journal} {Phys. Rev. D}\ }\textbf {\bibinfo {volume} {75}},\ \bibinfo
  {pages} {043520} (\bibinfo {year} {2007})},\ \Eprint
  {http://arxiv.org/abs/astro-ph/0612196} {arXiv:astro-ph/0612196} \BibitemShut
  {NoStop}%
\bibitem [{\citenamefont {Mamon}\ and\ \citenamefont
  {Das}(2017)}]{al2017parametric}%
  \BibitemOpen
  \bibfield  {author} {\bibinfo {author} {\bibfnamefont {A.~A.}\ \bibnamefont
  {Mamon}}\ and\ \bibinfo {author} {\bibfnamefont {S.}~\bibnamefont {Das}},\
  }\href {\doibase 10.1140/epjc/s10052-017-5066-4} {\bibfield  {journal}
  {\bibinfo  {journal} {Eur. Phys. J. C}\ }\textbf {\bibinfo {volume} {77}},\
  \bibinfo {pages} {495} (\bibinfo {year} {2017})},\ \Eprint
  {http://arxiv.org/abs/1610.07337} {arXiv:1610.07337 [gr-qc]} \BibitemShut
  {NoStop}%
\bibitem [{\citenamefont {Yu-Ting}\ \emph {et~al.}(2010)\citenamefont
  {Yu-Ting}, \citenamefont {Li-Xin}, \citenamefont {Jian-Bo},\ and\
  \citenamefont {Yuan-Xing}}]{yu2010reconstructing}%
  \BibitemOpen
  \bibfield  {author} {\bibinfo {author} {\bibfnamefont {W.}~\bibnamefont
  {Yu-Ting}}, \bibinfo {author} {\bibfnamefont {X.}~\bibnamefont {Li-Xin}},
  \bibinfo {author} {\bibfnamefont {L.}~\bibnamefont {Jian-Bo}}, \ and\
  \bibinfo {author} {\bibfnamefont {G.}~\bibnamefont {Yuan-Xing}},\ }\href
  {\doibase 10.1088/1674-1056/19/1/019801} {\bibfield  {journal} {\bibinfo
  {journal} {Chin. Phys. B}\ }\textbf {\bibinfo {volume} {19}},\ \bibinfo
  {pages} {019801} (\bibinfo {year} {2010})},\ \Eprint
  {http://arxiv.org/abs/1004.3370} {arXiv:1004.3370 [astro-ph.CO]} \BibitemShut
  {NoStop}%
\bibitem [{\citenamefont {Hern\'andez-Almada}\ \emph
  {et~al.}(2020)\citenamefont {Hern\'andez-Almada}, \citenamefont {Leon},
  \citenamefont {Maga\~na}, \citenamefont {Garc\'\i{}a-Aspeitia},\ and\
  \citenamefont {Motta}}]{hernandez2020generalized}%
  \BibitemOpen
  \bibfield  {author} {\bibinfo {author} {\bibfnamefont {A.}~\bibnamefont
  {Hern\'andez-Almada}}, \bibinfo {author} {\bibfnamefont {G.}~\bibnamefont
  {Leon}}, \bibinfo {author} {\bibfnamefont {J.}~\bibnamefont {Maga\~na}},
  \bibinfo {author} {\bibfnamefont {M.~A.}\ \bibnamefont
  {Garc\'\i{}a-Aspeitia}}, \ and\ \bibinfo {author} {\bibfnamefont
  {V.}~\bibnamefont {Motta}},\ }\href {\doibase 10.1093/mnras/staa2052}
  {\bibfield  {journal} {\bibinfo  {journal} {Mon. Not. Roy. Astron. Soc.}\
  }\textbf {\bibinfo {volume} {497}},\ \bibinfo {pages} {1590} (\bibinfo {year}
  {2020})},\ \Eprint {http://arxiv.org/abs/2002.12881} {arXiv:2002.12881
  [astro-ph.CO]} \BibitemShut {NoStop}%
\bibitem [{\citenamefont {Mukherjee}\ and\ \citenamefont
  {Banerjee}(2016)}]{mukherjee2016parametric}%
  \BibitemOpen
  \bibfield  {author} {\bibinfo {author} {\bibfnamefont {A.}~\bibnamefont
  {Mukherjee}}\ and\ \bibinfo {author} {\bibfnamefont {N.}~\bibnamefont
  {Banerjee}},\ }\href {\doibase 10.1103/PhysRevD.93.043002} {\bibfield
  {journal} {\bibinfo  {journal} {Phys. Rev. D}\ }\textbf {\bibinfo {volume}
  {93}},\ \bibinfo {pages} {043002} (\bibinfo {year} {2016})},\ \Eprint
  {http://arxiv.org/abs/1601.05172} {arXiv:1601.05172 [gr-qc]} \BibitemShut
  {NoStop}%
\bibitem [{\citenamefont {Akarsu}\ \emph {et~al.}(2015)\citenamefont {Akarsu},
  \citenamefont {Dereli},\ and\ \citenamefont {Vazquez}}]{Akarsu:2015yea}%
  \BibitemOpen
  \bibfield  {author} {\bibinfo {author} {\bibfnamefont {O.}~\bibnamefont
  {Akarsu}}, \bibinfo {author} {\bibfnamefont {T.}~\bibnamefont {Dereli}}, \
  and\ \bibinfo {author} {\bibfnamefont {J.~A.}\ \bibnamefont {Vazquez}},\
  }\href {\doibase 10.1088/1475-7516/2015/06/049} {\bibfield  {journal}
  {\bibinfo  {journal} {JCAP}\ }\textbf {\bibinfo {volume} {06}},\ \bibinfo
  {pages} {049} (\bibinfo {year} {2015})},\ \Eprint
  {http://arxiv.org/abs/1501.07598} {arXiv:1501.07598 [astro-ph.CO]}
  \BibitemShut {NoStop}%
\bibitem [{\citenamefont {Arciniega}\ \emph {et~al.}(2021)\citenamefont
  {Arciniega}, \citenamefont {Jaber}, \citenamefont {Jaime},\ and\
  \citenamefont {Rodr\'\i{}guez-L\'opez}}]{Arciniega:2021ffa}%
  \BibitemOpen
  \bibfield  {author} {\bibinfo {author} {\bibfnamefont {G.}~\bibnamefont
  {Arciniega}}, \bibinfo {author} {\bibfnamefont {M.}~\bibnamefont {Jaber}},
  \bibinfo {author} {\bibfnamefont {L.~G.}\ \bibnamefont {Jaime}}, \ and\
  \bibinfo {author} {\bibfnamefont {O.~A.}\ \bibnamefont
  {Rodr\'\i{}guez-L\'opez}},\ }\href@noop {} {\  (\bibinfo {year} {2021})},\
  \Eprint {http://arxiv.org/abs/2102.08561} {arXiv:2102.08561 [astro-ph.CO]}
  \BibitemShut {NoStop}%
\bibitem [{\citenamefont {Yang}\ \emph {et~al.}(2019)\citenamefont {Yang},
  \citenamefont {Pan}, \citenamefont {Di~Valentino}, \citenamefont
  {Saridakis},\ and\ \citenamefont {Chakraborty}}]{yang2019observational}%
  \BibitemOpen
  \bibfield  {author} {\bibinfo {author} {\bibfnamefont {W.}~\bibnamefont
  {Yang}}, \bibinfo {author} {\bibfnamefont {S.}~\bibnamefont {Pan}}, \bibinfo
  {author} {\bibfnamefont {E.}~\bibnamefont {Di~Valentino}}, \bibinfo {author}
  {\bibfnamefont {E.~N.}\ \bibnamefont {Saridakis}}, \ and\ \bibinfo {author}
  {\bibfnamefont {S.}~\bibnamefont {Chakraborty}},\ }\href {\doibase
  10.1103/PhysRevD.99.043543} {\bibfield  {journal} {\bibinfo  {journal} {Phys.
  Rev. D}\ }\textbf {\bibinfo {volume} {99}},\ \bibinfo {pages} {043543}
  (\bibinfo {year} {2019})},\ \Eprint {http://arxiv.org/abs/1810.05141}
  {arXiv:1810.05141 [astro-ph.CO]} \BibitemShut {NoStop}%
\bibitem [{\citenamefont {Chevallier}\ and\ \citenamefont
  {Polarski}(2001)}]{chevallier2001accelerating}%
  \BibitemOpen
  \bibfield  {author} {\bibinfo {author} {\bibfnamefont {M.}~\bibnamefont
  {Chevallier}}\ and\ \bibinfo {author} {\bibfnamefont {D.}~\bibnamefont
  {Polarski}},\ }\href {\doibase 10.1142/S0218271801000822} {\bibfield
  {journal} {\bibinfo  {journal} {Int. J. Mod. Phys. D}\ }\textbf {\bibinfo
  {volume} {10}},\ \bibinfo {pages} {213} (\bibinfo {year} {2001})},\ \Eprint
  {http://arxiv.org/abs/gr-qc/0009008} {arXiv:gr-qc/0009008} \BibitemShut
  {NoStop}%
\bibitem [{\citenamefont {Li}\ and\ \citenamefont
  {Shafieloo}(2020)}]{li2020evidence}%
  \BibitemOpen
  \bibfield  {author} {\bibinfo {author} {\bibfnamefont {X.}~\bibnamefont
  {Li}}\ and\ \bibinfo {author} {\bibfnamefont {A.}~\bibnamefont {Shafieloo}},\
  }\href {\doibase 10.3847/1538-4357/abb3d0} {\bibfield  {journal} {\bibinfo
  {journal} {Astrophys. J.}\ }\textbf {\bibinfo {volume} {902}},\ \bibinfo
  {pages} {58} (\bibinfo {year} {2020})},\ \Eprint
  {http://arxiv.org/abs/2001.05103} {arXiv:2001.05103 [astro-ph.CO]}
  \BibitemShut {NoStop}%
\bibitem [{\citenamefont {Yang}\ \emph {et~al.}(2021)\citenamefont {Yang},
  \citenamefont {Di~Valentino}, \citenamefont {Pan}, \citenamefont
  {Shafieloo},\ and\ \citenamefont {Li}}]{yang2021generalized}%
  \BibitemOpen
  \bibfield  {author} {\bibinfo {author} {\bibfnamefont {W.}~\bibnamefont
  {Yang}}, \bibinfo {author} {\bibfnamefont {E.}~\bibnamefont {Di~Valentino}},
  \bibinfo {author} {\bibfnamefont {S.}~\bibnamefont {Pan}}, \bibinfo {author}
  {\bibfnamefont {A.}~\bibnamefont {Shafieloo}}, \ and\ \bibinfo {author}
  {\bibfnamefont {X.}~\bibnamefont {Li}},\ }\href@noop {} {\bibfield  {journal}
  {\bibinfo  {journal} {arXiv preprint arXiv:2103.03815}\ } (\bibinfo {year}
  {2021})}\BibitemShut {NoStop}%
\bibitem [{\citenamefont {Akarsu}\ \emph
  {et~al.}(2020{\natexlab{b}})\citenamefont {Akarsu}, \citenamefont {Barrow},
  \citenamefont {Escamilla},\ and\ \citenamefont
  {Vazquez}}]{akarsu2020graduated}%
  \BibitemOpen
  \bibfield  {author} {\bibinfo {author} {\bibfnamefont {O.}~\bibnamefont
  {Akarsu}}, \bibinfo {author} {\bibfnamefont {J.~D.}\ \bibnamefont {Barrow}},
  \bibinfo {author} {\bibfnamefont {L.~A.}\ \bibnamefont {Escamilla}}, \ and\
  \bibinfo {author} {\bibfnamefont {J.~A.}\ \bibnamefont {Vazquez}},\ }\href
  {\doibase 10.1103/PhysRevD.101.063528} {\bibfield  {journal} {\bibinfo
  {journal} {Phys. Rev. D}\ }\textbf {\bibinfo {volume} {101}},\ \bibinfo
  {pages} {063528} (\bibinfo {year} {2020}{\natexlab{b}})},\ \Eprint
  {http://arxiv.org/abs/1912.08751} {arXiv:1912.08751 [astro-ph.CO]}
  \BibitemShut {NoStop}%
\bibitem [{\citenamefont {Acquaviva}\ \emph
  {et~al.}(2021{\natexlab{a}})\citenamefont {Acquaviva}, \citenamefont
  {Akarsu}, \citenamefont {Katirci},\ and\ \citenamefont
  {Vazquez}}]{Acquaviva:2021jov}%
  \BibitemOpen
  \bibfield  {author} {\bibinfo {author} {\bibfnamefont {G.}~\bibnamefont
  {Acquaviva}}, \bibinfo {author} {\bibfnamefont {O.}~\bibnamefont {Akarsu}},
  \bibinfo {author} {\bibfnamefont {N.}~\bibnamefont {Katirci}}, \ and\
  \bibinfo {author} {\bibfnamefont {J.~A.}\ \bibnamefont {Vazquez}},\ }\href
  {\doibase 10.1103/PhysRevD.104.023505} {\bibfield  {journal} {\bibinfo
  {journal} {Phys. Rev. D}\ }\textbf {\bibinfo {volume} {104}},\ \bibinfo
  {pages} {023505} (\bibinfo {year} {2021}{\natexlab{a}})},\ \Eprint
  {http://arxiv.org/abs/2104.02623} {arXiv:2104.02623 [astro-ph.CO]}
  \BibitemShut {NoStop}%
\bibitem [{\citenamefont {Akarsu}\ \emph {et~al.}(2021)\citenamefont {Akarsu},
  \citenamefont {Kumar}, \citenamefont {\"Oz\"ulker},\ and\ \citenamefont
  {Vazquez}}]{Akarsu:2021fol}%
  \BibitemOpen
  \bibfield  {author} {\bibinfo {author} {\bibfnamefont {O.}~\bibnamefont
  {Akarsu}}, \bibinfo {author} {\bibfnamefont {S.}~\bibnamefont {Kumar}},
  \bibinfo {author} {\bibfnamefont {E.}~\bibnamefont {\"Oz\"ulker}}, \ and\
  \bibinfo {author} {\bibfnamefont {J.~A.}\ \bibnamefont {Vazquez}},\ }\href
  {\doibase 10.1103/PhysRevD.104.123512} {\bibfield  {journal} {\bibinfo
  {journal} {Phys. Rev. D}\ }\textbf {\bibinfo {volume} {104}},\ \bibinfo
  {pages} {123512} (\bibinfo {year} {2021})},\ \Eprint
  {http://arxiv.org/abs/2108.09239} {arXiv:2108.09239 [astro-ph.CO]}
  \BibitemShut {NoStop}%
\bibitem [{\citenamefont {Akarsu}\ \emph
  {et~al.}(2022{\natexlab{a}})\citenamefont {Akarsu}, \citenamefont {Kumar},
  \citenamefont {\"Oz\"ulker}, \citenamefont {Vazquez},\ and\ \citenamefont
  {Yadav}}]{Akarsu:2022typ}%
  \BibitemOpen
  \bibfield  {author} {\bibinfo {author} {\bibfnamefont {O.}~\bibnamefont
  {Akarsu}}, \bibinfo {author} {\bibfnamefont {S.}~\bibnamefont {Kumar}},
  \bibinfo {author} {\bibfnamefont {E.}~\bibnamefont {\"Oz\"ulker}}, \bibinfo
  {author} {\bibfnamefont {J.~A.}\ \bibnamefont {Vazquez}}, \ and\ \bibinfo
  {author} {\bibfnamefont {A.}~\bibnamefont {Yadav}},\ }\href@noop {} {\
  (\bibinfo {year} {2022}{\natexlab{a}})},\ \Eprint
  {http://arxiv.org/abs/2211.05742} {arXiv:2211.05742 [astro-ph.CO]}
  \BibitemShut {NoStop}%
\bibitem [{\citenamefont {Poulin}\ \emph {et~al.}(2018)\citenamefont {Poulin},
  \citenamefont {Smith}, \citenamefont {Grin}, \citenamefont {Karwal},\ and\
  \citenamefont {Kamionkowski}}]{poulin2018cosmological}%
  \BibitemOpen
  \bibfield  {author} {\bibinfo {author} {\bibfnamefont {V.}~\bibnamefont
  {Poulin}}, \bibinfo {author} {\bibfnamefont {T.~L.}\ \bibnamefont {Smith}},
  \bibinfo {author} {\bibfnamefont {D.}~\bibnamefont {Grin}}, \bibinfo {author}
  {\bibfnamefont {T.}~\bibnamefont {Karwal}}, \ and\ \bibinfo {author}
  {\bibfnamefont {M.}~\bibnamefont {Kamionkowski}},\ }\href {\doibase
  10.1103/PhysRevD.98.083525} {\bibfield  {journal} {\bibinfo  {journal} {Phys.
  Rev. D}\ }\textbf {\bibinfo {volume} {98}},\ \bibinfo {pages} {083525}
  (\bibinfo {year} {2018})},\ \Eprint {http://arxiv.org/abs/1806.10608}
  {arXiv:1806.10608 [astro-ph.CO]} \BibitemShut {NoStop}%
\bibitem [{\citenamefont {Akarsu}\ \emph {et~al.}(2019)\citenamefont {Akarsu},
  \citenamefont {Barrow}, \citenamefont {Board}, \citenamefont {Uzun},\ and\
  \citenamefont {Vazquez}}]{Akarsu:2019ygx}%
  \BibitemOpen
  \bibfield  {author} {\bibinfo {author} {\bibfnamefont {O.}~\bibnamefont
  {Akarsu}}, \bibinfo {author} {\bibfnamefont {J.~D.}\ \bibnamefont {Barrow}},
  \bibinfo {author} {\bibfnamefont {C.~V.~R.}\ \bibnamefont {Board}}, \bibinfo
  {author} {\bibfnamefont {N.~M.}\ \bibnamefont {Uzun}}, \ and\ \bibinfo
  {author} {\bibfnamefont {J.~A.}\ \bibnamefont {Vazquez}},\ }\href {\doibase
  10.1140/epjc/s10052-019-7333-z} {\bibfield  {journal} {\bibinfo  {journal}
  {Eur. Phys. J. C}\ }\textbf {\bibinfo {volume} {79}},\ \bibinfo {pages} {846}
  (\bibinfo {year} {2019})},\ \Eprint {http://arxiv.org/abs/1903.11519}
  {arXiv:1903.11519 [gr-qc]} \BibitemShut {NoStop}%
\bibitem [{\citenamefont {Keeley}\ \emph {et~al.}(2021)\citenamefont {Keeley},
  \citenamefont {Shafieloo}, \citenamefont {Zhao}, \citenamefont {Vazquez},\
  and\ \citenamefont {Koo}}]{Keeley:2020aym}%
  \BibitemOpen
  \bibfield  {author} {\bibinfo {author} {\bibfnamefont {R.~E.}\ \bibnamefont
  {Keeley}}, \bibinfo {author} {\bibfnamefont {A.}~\bibnamefont {Shafieloo}},
  \bibinfo {author} {\bibfnamefont {G.-B.}\ \bibnamefont {Zhao}}, \bibinfo
  {author} {\bibfnamefont {J.~A.}\ \bibnamefont {Vazquez}}, \ and\ \bibinfo
  {author} {\bibfnamefont {H.}~\bibnamefont {Koo}},\ }\href {\doibase
  10.3847/1538-3881/abdd2a} {\bibfield  {journal} {\bibinfo  {journal} {Astron.
  J.}\ }\textbf {\bibinfo {volume} {161}},\ \bibinfo {pages} {151} (\bibinfo
  {year} {2021})},\ \Eprint {http://arxiv.org/abs/2010.03234} {arXiv:2010.03234
  [astro-ph.CO]} \BibitemShut {NoStop}%
\bibitem [{\citenamefont {Seikel}\ \emph {et~al.}(2012)\citenamefont {Seikel},
  \citenamefont {Clarkson},\ and\ \citenamefont
  {Smith}}]{seikel2012reconstruction}%
  \BibitemOpen
  \bibfield  {author} {\bibinfo {author} {\bibfnamefont {M.}~\bibnamefont
  {Seikel}}, \bibinfo {author} {\bibfnamefont {C.}~\bibnamefont {Clarkson}}, \
  and\ \bibinfo {author} {\bibfnamefont {M.}~\bibnamefont {Smith}},\ }\href
  {\doibase 10.1088/1475-7516/2012/06/036} {\bibfield  {journal} {\bibinfo
  {journal} {JCAP}\ }\textbf {\bibinfo {volume} {06}},\ \bibinfo {pages} {036}
  (\bibinfo {year} {2012})},\ \Eprint {http://arxiv.org/abs/1204.2832}
  {arXiv:1204.2832 [astro-ph.CO]} \BibitemShut {NoStop}%
\bibitem [{\citenamefont {Holsclaw}\ \emph {et~al.}(2010)\citenamefont
  {Holsclaw}, \citenamefont {Alam}, \citenamefont {Sanso}, \citenamefont {Lee},
  \citenamefont {Heitmann}, \citenamefont {Habib},\ and\ \citenamefont
  {Higdon}}]{holsclaw2010nonparametric}%
  \BibitemOpen
  \bibfield  {author} {\bibinfo {author} {\bibfnamefont {T.}~\bibnamefont
  {Holsclaw}}, \bibinfo {author} {\bibfnamefont {U.}~\bibnamefont {Alam}},
  \bibinfo {author} {\bibfnamefont {B.}~\bibnamefont {Sanso}}, \bibinfo
  {author} {\bibfnamefont {H.}~\bibnamefont {Lee}}, \bibinfo {author}
  {\bibfnamefont {K.}~\bibnamefont {Heitmann}}, \bibinfo {author}
  {\bibfnamefont {S.}~\bibnamefont {Habib}}, \ and\ \bibinfo {author}
  {\bibfnamefont {D.}~\bibnamefont {Higdon}},\ }\href {\doibase
  10.1103/PhysRevLett.105.241302} {\bibfield  {journal} {\bibinfo  {journal}
  {Phys. Rev. Lett.}\ }\textbf {\bibinfo {volume} {105}},\ \bibinfo {pages}
  {241302} (\bibinfo {year} {2010})},\ \Eprint {http://arxiv.org/abs/1011.3079}
  {arXiv:1011.3079 [astro-ph.CO]} \BibitemShut {NoStop}%
\bibitem [{\citenamefont {Vel{\'a}squez-Toribio}\ and\ \citenamefont
  {Fabris}(2020)}]{velasquez2020growth}%
  \BibitemOpen
  \bibfield  {author} {\bibinfo {author} {\bibfnamefont {A.}~\bibnamefont
  {Vel{\'a}squez-Toribio}}\ and\ \bibinfo {author} {\bibfnamefont {J.~C.}\
  \bibnamefont {Fabris}},\ }\href@noop {} {\bibfield  {journal} {\bibinfo
  {journal} {arXiv preprint arXiv:2008.12741}\ } (\bibinfo {year}
  {2020})}\BibitemShut {NoStop}%
\bibitem [{\citenamefont {Aljaf}\ \emph {et~al.}(2020)\citenamefont {Aljaf},
  \citenamefont {Gregoris},\ and\ \citenamefont
  {Khurshudyan}}]{aljaf2020constraints}%
  \BibitemOpen
  \bibfield  {author} {\bibinfo {author} {\bibfnamefont {M.}~\bibnamefont
  {Aljaf}}, \bibinfo {author} {\bibfnamefont {D.}~\bibnamefont {Gregoris}}, \
  and\ \bibinfo {author} {\bibfnamefont {M.}~\bibnamefont {Khurshudyan}},\
  }\href@noop {} {\bibfield  {journal} {\bibinfo  {journal} {arXiv preprint
  arXiv:2005.01891}\ } (\bibinfo {year} {2020})}\BibitemShut {NoStop}%
\bibitem [{\citenamefont {Yang}\ and\ \citenamefont
  {Gong}(2020)}]{yang2020evidence}%
  \BibitemOpen
  \bibfield  {author} {\bibinfo {author} {\bibfnamefont {Y.}~\bibnamefont
  {Yang}}\ and\ \bibinfo {author} {\bibfnamefont {Y.}~\bibnamefont {Gong}},\
  }\href {\doibase 10.1088/1475-7516/2020/06/059} {\bibfield  {journal}
  {\bibinfo  {journal} {JCAP}\ }\textbf {\bibinfo {volume} {06}},\ \bibinfo
  {pages} {059} (\bibinfo {year} {2020})},\ \Eprint
  {http://arxiv.org/abs/1912.07375} {arXiv:1912.07375 [astro-ph.CO]}
  \BibitemShut {NoStop}%
\bibitem [{\citenamefont {Gerardi}\ \emph {et~al.}(2019)\citenamefont
  {Gerardi}, \citenamefont {Martinelli},\ and\ \citenamefont
  {Silvestri}}]{gerardi2019reconstruction}%
  \BibitemOpen
  \bibfield  {author} {\bibinfo {author} {\bibfnamefont {F.}~\bibnamefont
  {Gerardi}}, \bibinfo {author} {\bibfnamefont {M.}~\bibnamefont {Martinelli}},
  \ and\ \bibinfo {author} {\bibfnamefont {A.}~\bibnamefont {Silvestri}},\
  }\href {\doibase 10.1088/1475-7516/2019/07/042} {\bibfield  {journal}
  {\bibinfo  {journal} {JCAP}\ }\textbf {\bibinfo {volume} {07}},\ \bibinfo
  {pages} {042} (\bibinfo {year} {2019})},\ \Eprint
  {http://arxiv.org/abs/1902.09423} {arXiv:1902.09423 [astro-ph.CO]}
  \BibitemShut {NoStop}%
\bibitem [{\citenamefont {Yang}\ \emph {et~al.}(2015)\citenamefont {Yang},
  \citenamefont {Guo},\ and\ \citenamefont {Cai}}]{yang2015reconstructing}%
  \BibitemOpen
  \bibfield  {author} {\bibinfo {author} {\bibfnamefont {T.}~\bibnamefont
  {Yang}}, \bibinfo {author} {\bibfnamefont {Z.-K.}\ \bibnamefont {Guo}}, \
  and\ \bibinfo {author} {\bibfnamefont {R.-G.}\ \bibnamefont {Cai}},\ }\href
  {\doibase 10.1103/PhysRevD.91.123533} {\bibfield  {journal} {\bibinfo
  {journal} {Phys. Rev. D}\ }\textbf {\bibinfo {volume} {91}},\ \bibinfo
  {pages} {123533} (\bibinfo {year} {2015})},\ \Eprint
  {http://arxiv.org/abs/1505.04443} {arXiv:1505.04443 [astro-ph.CO]}
  \BibitemShut {NoStop}%
\bibitem [{\citenamefont {Holsclaw}\ \emph {et~al.}(2013)\citenamefont
  {Holsclaw}, \citenamefont {Sans{\'o}}, \citenamefont {Lee}, \citenamefont
  {Heitmann}, \citenamefont {Habib}, \citenamefont {Higdon},\ and\
  \citenamefont {Alam}}]{holsclaw2013gaussian}%
  \BibitemOpen
  \bibfield  {author} {\bibinfo {author} {\bibfnamefont {T.}~\bibnamefont
  {Holsclaw}}, \bibinfo {author} {\bibfnamefont {B.}~\bibnamefont {Sans{\'o}}},
  \bibinfo {author} {\bibfnamefont {H.~K.}\ \bibnamefont {Lee}}, \bibinfo
  {author} {\bibfnamefont {K.}~\bibnamefont {Heitmann}}, \bibinfo {author}
  {\bibfnamefont {S.}~\bibnamefont {Habib}}, \bibinfo {author} {\bibfnamefont
  {D.}~\bibnamefont {Higdon}}, \ and\ \bibinfo {author} {\bibfnamefont
  {U.}~\bibnamefont {Alam}},\ }\href@noop {} {\bibfield  {journal} {\bibinfo
  {journal} {Technometrics}\ }\textbf {\bibinfo {volume} {55}},\ \bibinfo
  {pages} {57} (\bibinfo {year} {2013})}\BibitemShut {NoStop}%
\bibitem [{\citenamefont {Wang}\ and\ \citenamefont
  {Meng}(2017)}]{wang2017improved}%
  \BibitemOpen
  \bibfield  {author} {\bibinfo {author} {\bibfnamefont {D.}~\bibnamefont
  {Wang}}\ and\ \bibinfo {author} {\bibfnamefont {X.-H.}\ \bibnamefont
  {Meng}},\ }\href {\doibase 10.1103/PhysRevD.95.023508} {\bibfield  {journal}
  {\bibinfo  {journal} {Phys. Rev. D}\ }\textbf {\bibinfo {volume} {95}},\
  \bibinfo {pages} {023508} (\bibinfo {year} {2017})},\ \Eprint
  {http://arxiv.org/abs/1708.07750} {arXiv:1708.07750 [astro-ph.CO]}
  \BibitemShut {NoStop}%
\bibitem [{\citenamefont {Wang}\ \emph {et~al.}(2020)\citenamefont {Wang},
  \citenamefont {Ma}, \citenamefont {Li},\ and\ \citenamefont
  {Xia}}]{wang2020reconstructing}%
  \BibitemOpen
  \bibfield  {author} {\bibinfo {author} {\bibfnamefont {G.-J.}\ \bibnamefont
  {Wang}}, \bibinfo {author} {\bibfnamefont {X.-J.}\ \bibnamefont {Ma}},
  \bibinfo {author} {\bibfnamefont {S.-Y.}\ \bibnamefont {Li}}, \ and\ \bibinfo
  {author} {\bibfnamefont {J.-Q.}\ \bibnamefont {Xia}},\ }\href {\doibase
  https://doi.org/10.3847/1538-4365/ab620b} {\bibfield  {journal} {\bibinfo
  {journal} {apjsupp}\ }\textbf {\bibinfo {volume} {246}},\ \bibinfo {pages}
  {13} (\bibinfo {year} {2020})}\BibitemShut {NoStop}%
\bibitem [{\citenamefont {G\'omez-Vargas}\ \emph {et~al.}(2021)\citenamefont
  {G\'omez-Vargas}, \citenamefont {V\'azquez}, \citenamefont {Esquivel},\ and\
  \citenamefont {Garc\'\i{}a-Salcedo}}]{Gomez-Vargas:2021zyl}%
  \BibitemOpen
  \bibfield  {author} {\bibinfo {author} {\bibfnamefont {I.}~\bibnamefont
  {G\'omez-Vargas}}, \bibinfo {author} {\bibfnamefont {J.~A.}\ \bibnamefont
  {V\'azquez}}, \bibinfo {author} {\bibfnamefont {R.~M.}\ \bibnamefont
  {Esquivel}}, \ and\ \bibinfo {author} {\bibfnamefont {R.}~\bibnamefont
  {Garc\'\i{}a-Salcedo}},\ }\href@noop {} {\  (\bibinfo {year} {2021})},\
  \Eprint {http://arxiv.org/abs/2104.00595} {arXiv:2104.00595 [astro-ph.CO]}
  \BibitemShut {NoStop}%
\bibitem [{\citenamefont {Tamayo}\ and\ \citenamefont
  {Vazquez}(2019)}]{tamayo2019fourier}%
  \BibitemOpen
  \bibfield  {author} {\bibinfo {author} {\bibfnamefont {D.}~\bibnamefont
  {Tamayo}}\ and\ \bibinfo {author} {\bibfnamefont {J.~A.}\ \bibnamefont
  {Vazquez}},\ }\href {\doibase 10.1093/mnras/stz1229} {\bibfield  {journal}
  {\bibinfo  {journal} {Mon. Not. Roy. Astron. Soc.}\ }\textbf {\bibinfo
  {volume} {487}},\ \bibinfo {pages} {729} (\bibinfo {year} {2019})},\ \Eprint
  {http://arxiv.org/abs/1901.08679} {arXiv:1901.08679 [astro-ph.CO]}
  \BibitemShut {NoStop}%
\bibitem [{\citenamefont {Mehrabi}\ and\ \citenamefont
  {Basilakos}(2018)}]{basilakos2018dark}%
  \BibitemOpen
  \bibfield  {author} {\bibinfo {author} {\bibfnamefont {A.}~\bibnamefont
  {Mehrabi}}\ and\ \bibinfo {author} {\bibfnamefont {S.}~\bibnamefont
  {Basilakos}},\ }\href {\doibase 10.1140/epjc/s10052-018-6368-x} {\bibfield
  {journal} {\bibinfo  {journal} {Eur. Phys. J. C}\ }\textbf {\bibinfo {volume}
  {78}},\ \bibinfo {pages} {889} (\bibinfo {year} {2018})},\ \Eprint
  {http://arxiv.org/abs/1804.10794} {arXiv:1804.10794 [astro-ph.CO]}
  \BibitemShut {NoStop}%
\bibitem [{\citenamefont {Crittenden}\ \emph {et~al.}(2012)\citenamefont
  {Crittenden}, \citenamefont {Zhao}, \citenamefont {Pogosian}, \citenamefont
  {Samushia},\ and\ \citenamefont {Zhang}}]{crittenden2012fables}%
  \BibitemOpen
  \bibfield  {author} {\bibinfo {author} {\bibfnamefont {R.~G.}\ \bibnamefont
  {Crittenden}}, \bibinfo {author} {\bibfnamefont {G.-B.}\ \bibnamefont
  {Zhao}}, \bibinfo {author} {\bibfnamefont {L.}~\bibnamefont {Pogosian}},
  \bibinfo {author} {\bibfnamefont {L.}~\bibnamefont {Samushia}}, \ and\
  \bibinfo {author} {\bibfnamefont {X.}~\bibnamefont {Zhang}},\ }\href
  {\doibase 10.1088/1475-7516/2012/02/048} {\bibfield  {journal} {\bibinfo
  {journal} {JCAP}\ }\textbf {\bibinfo {volume} {02}},\ \bibinfo {pages} {048}
  (\bibinfo {year} {2012})},\ \Eprint {http://arxiv.org/abs/1112.1693}
  {arXiv:1112.1693 [astro-ph.CO]} \BibitemShut {NoStop}%
\bibitem [{\citenamefont {Wang}\ \emph {et~al.}(2018)\citenamefont {Wang},
  \citenamefont {Pogosian}, \citenamefont {Zhao},\ and\ \citenamefont
  {Zucca}}]{wang2018evolution}%
  \BibitemOpen
  \bibfield  {author} {\bibinfo {author} {\bibfnamefont {Y.}~\bibnamefont
  {Wang}}, \bibinfo {author} {\bibfnamefont {L.}~\bibnamefont {Pogosian}},
  \bibinfo {author} {\bibfnamefont {G.-B.}\ \bibnamefont {Zhao}}, \ and\
  \bibinfo {author} {\bibfnamefont {A.}~\bibnamefont {Zucca}},\ }\href
  {\doibase 10.3847/2041-8213/aaf238} {\bibfield  {journal} {\bibinfo
  {journal} {Astrophys. J. Lett.}\ }\textbf {\bibinfo {volume} {869}},\
  \bibinfo {pages} {L8} (\bibinfo {year} {2018})},\ \Eprint
  {http://arxiv.org/abs/1807.03772} {arXiv:1807.03772 [astro-ph.CO]}
  \BibitemShut {NoStop}%
\bibitem [{\citenamefont {Yu}\ \emph {et~al.}(2013)\citenamefont {Yu},
  \citenamefont {Yuan},\ and\ \citenamefont {Zhang}}]{yu2013nonparametric}%
  \BibitemOpen
  \bibfield  {author} {\bibinfo {author} {\bibfnamefont {H.-R.}\ \bibnamefont
  {Yu}}, \bibinfo {author} {\bibfnamefont {S.}~\bibnamefont {Yuan}}, \ and\
  \bibinfo {author} {\bibfnamefont {T.-J.}\ \bibnamefont {Zhang}},\ }\href
  {\doibase 10.1103/PhysRevD.88.103528} {\bibfield  {journal} {\bibinfo
  {journal} {Phys. Rev. D}\ }\textbf {\bibinfo {volume} {88}},\ \bibinfo
  {pages} {103528} (\bibinfo {year} {2013})},\ \Eprint
  {http://arxiv.org/abs/1310.0870} {arXiv:1310.0870 [astro-ph.CO]} \BibitemShut
  {NoStop}%
\bibitem [{\citenamefont {Huterer}\ and\ \citenamefont
  {Starkman}(2003)}]{huterer2003parametrization}%
  \BibitemOpen
  \bibfield  {author} {\bibinfo {author} {\bibfnamefont {D.}~\bibnamefont
  {Huterer}}\ and\ \bibinfo {author} {\bibfnamefont {G.}~\bibnamefont
  {Starkman}},\ }\href {\doibase 10.1103/PhysRevLett.90.031301} {\bibfield
  {journal} {\bibinfo  {journal} {Phys. Rev. Lett.}\ }\textbf {\bibinfo
  {volume} {90}},\ \bibinfo {pages} {031301} (\bibinfo {year} {2003})},\
  \Eprint {http://arxiv.org/abs/astro-ph/0207517} {arXiv:astro-ph/0207517}
  \BibitemShut {NoStop}%
\bibitem [{\citenamefont {Vazquez}\ \emph {et~al.}(2012)\citenamefont
  {Vazquez}, \citenamefont {Bridges}, \citenamefont {Hobson},\ and\
  \citenamefont {Lasenby}}]{vazquez2012model}%
  \BibitemOpen
  \bibfield  {author} {\bibinfo {author} {\bibfnamefont {J.~A.}\ \bibnamefont
  {Vazquez}}, \bibinfo {author} {\bibfnamefont {M.}~\bibnamefont {Bridges}},
  \bibinfo {author} {\bibfnamefont {M.~P.}\ \bibnamefont {Hobson}}, \ and\
  \bibinfo {author} {\bibfnamefont {A.~N.}\ \bibnamefont {Lasenby}},\ }\href
  {\doibase 10.1088/1475-7516/2012/06/006} {\bibfield  {journal} {\bibinfo
  {journal} {JCAP}\ }\textbf {\bibinfo {volume} {06}},\ \bibinfo {pages} {006}
  (\bibinfo {year} {2012})},\ \Eprint {http://arxiv.org/abs/1203.1252}
  {arXiv:1203.1252 [astro-ph.CO]} \BibitemShut {NoStop}%
\bibitem [{\citenamefont {Vazquez}\ \emph {et~al.}(2013)\citenamefont
  {Vazquez}, \citenamefont {Bridges}, \citenamefont {Ma},\ and\ \citenamefont
  {Hobson}}]{Vazquez:2013dva}%
  \BibitemOpen
  \bibfield  {author} {\bibinfo {author} {\bibfnamefont {J.~A.}\ \bibnamefont
  {Vazquez}}, \bibinfo {author} {\bibfnamefont {M.}~\bibnamefont {Bridges}},
  \bibinfo {author} {\bibfnamefont {Y.-Z.}\ \bibnamefont {Ma}}, \ and\ \bibinfo
  {author} {\bibfnamefont {M.~P.}\ \bibnamefont {Hobson}},\ }\href {\doibase
  10.1088/1475-7516/2013/08/001} {\bibfield  {journal} {\bibinfo  {journal}
  {JCAP}\ }\textbf {\bibinfo {volume} {08}},\ \bibinfo {pages} {001} (\bibinfo
  {year} {2013})},\ \Eprint {http://arxiv.org/abs/1303.4014} {arXiv:1303.4014
  [astro-ph.CO]} \BibitemShut {NoStop}%
\bibitem [{\citenamefont {Handley}\ \emph {et~al.}(2019)\citenamefont
  {Handley}, \citenamefont {Lasenby}, \citenamefont {Peiris},\ and\
  \citenamefont {Hobson}}]{handley2019bayesian}%
  \BibitemOpen
  \bibfield  {author} {\bibinfo {author} {\bibfnamefont {W.~J.}\ \bibnamefont
  {Handley}}, \bibinfo {author} {\bibfnamefont {A.~N.}\ \bibnamefont
  {Lasenby}}, \bibinfo {author} {\bibfnamefont {H.~V.}\ \bibnamefont {Peiris}},
  \ and\ \bibinfo {author} {\bibfnamefont {M.~P.}\ \bibnamefont {Hobson}},\
  }\href@noop {} {\bibfield  {journal} {\bibinfo  {journal} {arXiv preprint
  arXiv:1908.00906}\ } (\bibinfo {year} {2019})}\BibitemShut {NoStop}%
\bibitem [{\citenamefont {Aslanyan}\ \emph {et~al.}(2014)\citenamefont
  {Aslanyan}, \citenamefont {Price}, \citenamefont {Abazajian},\ and\
  \citenamefont {Easther}}]{aslanyan2014knotted}%
  \BibitemOpen
  \bibfield  {author} {\bibinfo {author} {\bibfnamefont {G.}~\bibnamefont
  {Aslanyan}}, \bibinfo {author} {\bibfnamefont {L.~C.}\ \bibnamefont {Price}},
  \bibinfo {author} {\bibfnamefont {K.~N.}\ \bibnamefont {Abazajian}}, \ and\
  \bibinfo {author} {\bibfnamefont {R.}~\bibnamefont {Easther}},\ }\href
  {\doibase 10.1088/1475-7516/2014/08/052} {\bibfield  {journal} {\bibinfo
  {journal} {JCAP}\ }\textbf {\bibinfo {volume} {08}},\ \bibinfo {pages} {052}
  (\bibinfo {year} {2014})},\ \Eprint {http://arxiv.org/abs/1403.5849}
  {arXiv:1403.5849 [astro-ph.CO]} \BibitemShut {NoStop}%
\bibitem [{\citenamefont {Hee}\ \emph {et~al.}(2017)\citenamefont {Hee},
  \citenamefont {V\'azquez}, \citenamefont {Handley}, \citenamefont {Hobson},\
  and\ \citenamefont {Lasenby}}]{hee2017constraining}%
  \BibitemOpen
  \bibfield  {author} {\bibinfo {author} {\bibfnamefont {S.}~\bibnamefont
  {Hee}}, \bibinfo {author} {\bibfnamefont {J.~A.}\ \bibnamefont {V\'azquez}},
  \bibinfo {author} {\bibfnamefont {W.~J.}\ \bibnamefont {Handley}}, \bibinfo
  {author} {\bibfnamefont {M.~P.}\ \bibnamefont {Hobson}}, \ and\ \bibinfo
  {author} {\bibfnamefont {A.~N.}\ \bibnamefont {Lasenby}},\ }\href {\doibase
  10.1093/mnras/stw3102} {\bibfield  {journal} {\bibinfo  {journal} {Mon. Not.
  Roy. Astron. Soc.}\ }\textbf {\bibinfo {volume} {466}},\ \bibinfo {pages}
  {369} (\bibinfo {year} {2017})},\ \Eprint {http://arxiv.org/abs/1607.00270}
  {arXiv:1607.00270 [astro-ph.CO]} \BibitemShut {NoStop}%
\bibitem [{\citenamefont {Trotta}(2008)}]{trotta2008bayes}%
  \BibitemOpen
  \bibfield  {author} {\bibinfo {author} {\bibfnamefont {R.}~\bibnamefont
  {Trotta}},\ }\href {\doibase 10.1080/00107510802066753} {\bibfield  {journal}
  {\bibinfo  {journal} {Contemp. Phys.}\ }\textbf {\bibinfo {volume} {49}},\
  \bibinfo {pages} {71} (\bibinfo {year} {2008})},\ \Eprint
  {http://arxiv.org/abs/0803.4089} {arXiv:0803.4089 [astro-ph]} \BibitemShut
  {NoStop}%
\bibitem [{\citenamefont {Crittenden}\ \emph {et~al.}(2009)\citenamefont
  {Crittenden}, \citenamefont {Pogosian},\ and\ \citenamefont
  {Zhao}}]{crittenden2009investigating}%
  \BibitemOpen
  \bibfield  {author} {\bibinfo {author} {\bibfnamefont {R.~G.}\ \bibnamefont
  {Crittenden}}, \bibinfo {author} {\bibfnamefont {L.}~\bibnamefont
  {Pogosian}}, \ and\ \bibinfo {author} {\bibfnamefont {G.-B.}\ \bibnamefont
  {Zhao}},\ }\href {\doibase 10.1088/1475-7516/2009/12/025} {\bibfield
  {journal} {\bibinfo  {journal} {JCAP}\ }\textbf {\bibinfo {volume} {12}},\
  \bibinfo {pages} {025} (\bibinfo {year} {2009})},\ \Eprint
  {http://arxiv.org/abs/astro-ph/0510293} {arXiv:astro-ph/0510293} \BibitemShut
  {NoStop}%
\bibitem [{\citenamefont {Colg\'ain}\ \emph {et~al.}(2021)\citenamefont
  {Colg\'ain}, \citenamefont {Sheikh-Jabbari},\ and\ \citenamefont
  {Yin}}]{Colgain:2021pmf}%
  \BibitemOpen
  \bibfield  {author} {\bibinfo {author} {\bibfnamefont {E.~O.}\ \bibnamefont
  {Colg\'ain}}, \bibinfo {author} {\bibfnamefont {M.~M.}\ \bibnamefont
  {Sheikh-Jabbari}}, \ and\ \bibinfo {author} {\bibfnamefont {L.}~\bibnamefont
  {Yin}},\ }\href {\doibase 10.1103/PhysRevD.104.023510} {\bibfield  {journal}
  {\bibinfo  {journal} {Phys. Rev. D}\ }\textbf {\bibinfo {volume} {104}},\
  \bibinfo {pages} {023510} (\bibinfo {year} {2021})},\ \Eprint
  {http://arxiv.org/abs/2104.01930} {arXiv:2104.01930 [astro-ph.CO]}
  \BibitemShut {NoStop}%
\bibitem [{\citenamefont {Raveri}\ \emph {et~al.}(2017)\citenamefont {Raveri},
  \citenamefont {Bull}, \citenamefont {Silvestri},\ and\ \citenamefont
  {Pogosian}}]{Raveri:2017qvt}%
  \BibitemOpen
  \bibfield  {author} {\bibinfo {author} {\bibfnamefont {M.}~\bibnamefont
  {Raveri}}, \bibinfo {author} {\bibfnamefont {P.}~\bibnamefont {Bull}},
  \bibinfo {author} {\bibfnamefont {A.}~\bibnamefont {Silvestri}}, \ and\
  \bibinfo {author} {\bibfnamefont {L.}~\bibnamefont {Pogosian}},\ }\href
  {\doibase 10.1103/PhysRevD.96.083509} {\bibfield  {journal} {\bibinfo
  {journal} {Phys. Rev. D}\ }\textbf {\bibinfo {volume} {96}},\ \bibinfo
  {pages} {083509} (\bibinfo {year} {2017})},\ \Eprint
  {http://arxiv.org/abs/1703.05297} {arXiv:1703.05297 [astro-ph.CO]}
  \BibitemShut {NoStop}%
\bibitem [{\citenamefont {Espejo}\ \emph {et~al.}(2019)\citenamefont {Espejo},
  \citenamefont {Peirone}, \citenamefont {Raveri}, \citenamefont {Koyama},
  \citenamefont {Pogosian},\ and\ \citenamefont {Silvestri}}]{Espejo:2018hxa}%
  \BibitemOpen
  \bibfield  {author} {\bibinfo {author} {\bibfnamefont {J.}~\bibnamefont
  {Espejo}}, \bibinfo {author} {\bibfnamefont {S.}~\bibnamefont {Peirone}},
  \bibinfo {author} {\bibfnamefont {M.}~\bibnamefont {Raveri}}, \bibinfo
  {author} {\bibfnamefont {K.}~\bibnamefont {Koyama}}, \bibinfo {author}
  {\bibfnamefont {L.}~\bibnamefont {Pogosian}}, \ and\ \bibinfo {author}
  {\bibfnamefont {A.}~\bibnamefont {Silvestri}},\ }\href {\doibase
  10.1103/PhysRevD.99.023512} {\bibfield  {journal} {\bibinfo  {journal} {Phys.
  Rev. D}\ }\textbf {\bibinfo {volume} {99}},\ \bibinfo {pages} {023512}
  (\bibinfo {year} {2019})},\ \Eprint {http://arxiv.org/abs/1809.01121}
  {arXiv:1809.01121 [astro-ph.CO]} \BibitemShut {NoStop}%
\bibitem [{\citenamefont {Pogosian}\ \emph {et~al.}(2022)\citenamefont
  {Pogosian}, \citenamefont {Raveri}, \citenamefont {Koyama}, \citenamefont
  {Martinelli}, \citenamefont {Silvestri}, \citenamefont {Zhao}, \citenamefont
  {Li}, \citenamefont {Peirone},\ and\ \citenamefont
  {Zucca}}]{Pogosian:2021mcs}%
  \BibitemOpen
  \bibfield  {author} {\bibinfo {author} {\bibfnamefont {L.}~\bibnamefont
  {Pogosian}}, \bibinfo {author} {\bibfnamefont {M.}~\bibnamefont {Raveri}},
  \bibinfo {author} {\bibfnamefont {K.}~\bibnamefont {Koyama}}, \bibinfo
  {author} {\bibfnamefont {M.}~\bibnamefont {Martinelli}}, \bibinfo {author}
  {\bibfnamefont {A.}~\bibnamefont {Silvestri}}, \bibinfo {author}
  {\bibfnamefont {G.-B.}\ \bibnamefont {Zhao}}, \bibinfo {author}
  {\bibfnamefont {J.}~\bibnamefont {Li}}, \bibinfo {author} {\bibfnamefont
  {S.}~\bibnamefont {Peirone}}, \ and\ \bibinfo {author} {\bibfnamefont
  {A.}~\bibnamefont {Zucca}},\ }\href {\doibase 10.1038/s41550-022-01808-7}
  {\bibfield  {journal} {\bibinfo  {journal} {Nature Astron.}\ }\textbf
  {\bibinfo {volume} {6}},\ \bibinfo {pages} {1484} (\bibinfo {year} {2022})},\
  \Eprint {http://arxiv.org/abs/2107.12992} {arXiv:2107.12992 [astro-ph.CO]}
  \BibitemShut {NoStop}%
\bibitem [{\citenamefont {Raveri}\ \emph {et~al.}(2021)\citenamefont {Raveri},
  \citenamefont {Pogosian}, \citenamefont {Martinelli}, \citenamefont {Koyama},
  \citenamefont {Silvestri}, \citenamefont {Zhao}, \citenamefont {Li},
  \citenamefont {Peirone},\ and\ \citenamefont {Zucca}}]{Raveri:2021dbu}%
  \BibitemOpen
  \bibfield  {author} {\bibinfo {author} {\bibfnamefont {M.}~\bibnamefont
  {Raveri}}, \bibinfo {author} {\bibfnamefont {L.}~\bibnamefont {Pogosian}},
  \bibinfo {author} {\bibfnamefont {M.}~\bibnamefont {Martinelli}}, \bibinfo
  {author} {\bibfnamefont {K.}~\bibnamefont {Koyama}}, \bibinfo {author}
  {\bibfnamefont {A.}~\bibnamefont {Silvestri}}, \bibinfo {author}
  {\bibfnamefont {G.-B.}\ \bibnamefont {Zhao}}, \bibinfo {author}
  {\bibfnamefont {J.}~\bibnamefont {Li}}, \bibinfo {author} {\bibfnamefont
  {S.}~\bibnamefont {Peirone}}, \ and\ \bibinfo {author} {\bibfnamefont
  {A.}~\bibnamefont {Zucca}},\ }\href@noop {} {\  (\bibinfo {year} {2021})},\
  \Eprint {http://arxiv.org/abs/2107.12990} {arXiv:2107.12990 [astro-ph.CO]}
  \BibitemShut {NoStop}%
\bibitem [{\citenamefont {Efron}\ \emph {et~al.}(2001)\citenamefont {Efron},
  \citenamefont {Gous}, \citenamefont {Kass}, \citenamefont {Datta},\ and\
  \citenamefont {Lahiri}}]{efron2001scales}%
  \BibitemOpen
  \bibfield  {author} {\bibinfo {author} {\bibfnamefont {B.}~\bibnamefont
  {Efron}}, \bibinfo {author} {\bibfnamefont {A.}~\bibnamefont {Gous}},
  \bibinfo {author} {\bibfnamefont {R.}~\bibnamefont {Kass}}, \bibinfo {author}
  {\bibfnamefont {G.}~\bibnamefont {Datta}}, \ and\ \bibinfo {author}
  {\bibfnamefont {P.}~\bibnamefont {Lahiri}},\ }\href@noop {} {\bibfield
  {journal} {\bibinfo  {journal} {Lecture Notes-Monograph Series}\ ,\ \bibinfo
  {pages} {208}} (\bibinfo {year} {2001})}\BibitemShut {NoStop}%
\bibitem [{\citenamefont {Linares Cede\~no}\ \emph {et~al.}(2019)\citenamefont
  {Linares Cede\~no}, \citenamefont {Montiel}, \citenamefont {Hidalgo},\ and\
  \citenamefont {Germ\'an}}]{cedeno2019bayesian}%
  \BibitemOpen
  \bibfield  {author} {\bibinfo {author} {\bibfnamefont {F.~X.}\ \bibnamefont
  {Linares Cede\~no}}, \bibinfo {author} {\bibfnamefont {A.}~\bibnamefont
  {Montiel}}, \bibinfo {author} {\bibfnamefont {J.~C.}\ \bibnamefont
  {Hidalgo}}, \ and\ \bibinfo {author} {\bibfnamefont {G.}~\bibnamefont
  {Germ\'an}},\ }\href {\doibase 10.1088/1475-7516/2019/08/002} {\bibfield
  {journal} {\bibinfo  {journal} {JCAP}\ }\textbf {\bibinfo {volume} {08}},\
  \bibinfo {pages} {002} (\bibinfo {year} {2019})},\ \Eprint
  {http://arxiv.org/abs/1905.00834} {arXiv:1905.00834 [gr-qc]} \BibitemShut
  {NoStop}%
\bibitem [{\citenamefont {Nesseris}\ and\ \citenamefont
  {Garcia-Bellido}(2013)}]{Nesseris:2012cq}%
  \BibitemOpen
  \bibfield  {author} {\bibinfo {author} {\bibfnamefont {S.}~\bibnamefont
  {Nesseris}}\ and\ \bibinfo {author} {\bibfnamefont {J.}~\bibnamefont
  {Garcia-Bellido}},\ }\href {\doibase 10.1088/1475-7516/2013/08/036}
  {\bibfield  {journal} {\bibinfo  {journal} {JCAP}\ }\textbf {\bibinfo
  {volume} {08}},\ \bibinfo {pages} {036} (\bibinfo {year} {2013})},\ \Eprint
  {http://arxiv.org/abs/1210.7652} {arXiv:1210.7652 [astro-ph.CO]} \BibitemShut
  {NoStop}%
\bibitem [{\citenamefont {Keeley}\ and\ \citenamefont
  {Shafieloo}(2022)}]{Keeley:2021dmx}%
  \BibitemOpen
  \bibfield  {author} {\bibinfo {author} {\bibfnamefont {R.~E.}\ \bibnamefont
  {Keeley}}\ and\ \bibinfo {author} {\bibfnamefont {A.}~\bibnamefont
  {Shafieloo}},\ }\href {\doibase 10.1093/mnras/stac1851} {\bibfield  {journal}
  {\bibinfo  {journal} {Mon. Not. Roy. Astron. Soc.}\ }\textbf {\bibinfo
  {volume} {515}},\ \bibinfo {pages} {293} (\bibinfo {year} {2022})},\ \Eprint
  {http://arxiv.org/abs/2111.04231} {arXiv:2111.04231 [astro-ph.CO]}
  \BibitemShut {NoStop}%
\bibitem [{\citenamefont {Sahni}\ \emph {et~al.}(2008)\citenamefont {Sahni},
  \citenamefont {Shafieloo},\ and\ \citenamefont {Starobinsky}}]{sahni2008two}%
  \BibitemOpen
  \bibfield  {author} {\bibinfo {author} {\bibfnamefont {V.}~\bibnamefont
  {Sahni}}, \bibinfo {author} {\bibfnamefont {A.}~\bibnamefont {Shafieloo}}, \
  and\ \bibinfo {author} {\bibfnamefont {A.~A.}\ \bibnamefont {Starobinsky}},\
  }\href {\doibase 10.1103/PhysRevD.78.103502} {\bibfield  {journal} {\bibinfo
  {journal} {Phys. Rev. D}\ }\textbf {\bibinfo {volume} {78}},\ \bibinfo
  {pages} {103502} (\bibinfo {year} {2008})},\ \Eprint
  {http://arxiv.org/abs/0807.3548} {arXiv:0807.3548 [astro-ph]} \BibitemShut
  {NoStop}%
\bibitem [{\citenamefont {Jimenez}\ \emph {et~al.}(2003)\citenamefont
  {Jimenez}, \citenamefont {Verde}, \citenamefont {Treu},\ and\ \citenamefont
  {Stern}}]{jimenez2003constraints}%
  \BibitemOpen
  \bibfield  {author} {\bibinfo {author} {\bibfnamefont {R.}~\bibnamefont
  {Jimenez}}, \bibinfo {author} {\bibfnamefont {L.}~\bibnamefont {Verde}},
  \bibinfo {author} {\bibfnamefont {T.}~\bibnamefont {Treu}}, \ and\ \bibinfo
  {author} {\bibfnamefont {D.}~\bibnamefont {Stern}},\ }\href {\doibase
  10.1086/376595} {\bibfield  {journal} {\bibinfo  {journal} {Astrophys. J.}\
  }\textbf {\bibinfo {volume} {593}},\ \bibinfo {pages} {622} (\bibinfo {year}
  {2003})},\ \Eprint {http://arxiv.org/abs/astro-ph/0302560}
  {arXiv:astro-ph/0302560} \BibitemShut {NoStop}%
\bibitem [{\citenamefont {Simon}\ \emph {et~al.}(2005)\citenamefont {Simon},
  \citenamefont {Verde},\ and\ \citenamefont {Jimenez}}]{simon2005constraints}%
  \BibitemOpen
  \bibfield  {author} {\bibinfo {author} {\bibfnamefont {J.}~\bibnamefont
  {Simon}}, \bibinfo {author} {\bibfnamefont {L.}~\bibnamefont {Verde}}, \ and\
  \bibinfo {author} {\bibfnamefont {R.}~\bibnamefont {Jimenez}},\ }\href
  {\doibase 10.1103/PhysRevD.71.123001} {\bibfield  {journal} {\bibinfo
  {journal} {Phys. Rev. D}\ }\textbf {\bibinfo {volume} {71}},\ \bibinfo
  {pages} {123001} (\bibinfo {year} {2005})},\ \Eprint
  {http://arxiv.org/abs/astro-ph/0412269} {arXiv:astro-ph/0412269} \BibitemShut
  {NoStop}%
\bibitem [{\citenamefont {Stern}\ \emph {et~al.}(2010)\citenamefont {Stern},
  \citenamefont {Jimenez}, \citenamefont {Verde}, \citenamefont
  {Kamionkowski},\ and\ \citenamefont {Stanford}}]{stern2010cosmic}%
  \BibitemOpen
  \bibfield  {author} {\bibinfo {author} {\bibfnamefont {D.}~\bibnamefont
  {Stern}}, \bibinfo {author} {\bibfnamefont {R.}~\bibnamefont {Jimenez}},
  \bibinfo {author} {\bibfnamefont {L.}~\bibnamefont {Verde}}, \bibinfo
  {author} {\bibfnamefont {M.}~\bibnamefont {Kamionkowski}}, \ and\ \bibinfo
  {author} {\bibfnamefont {S.~A.}\ \bibnamefont {Stanford}},\ }\href {\doibase
  10.1088/1475-7516/2010/02/008} {\bibfield  {journal} {\bibinfo  {journal}
  {JCAP}\ }\textbf {\bibinfo {volume} {02}},\ \bibinfo {pages} {008} (\bibinfo
  {year} {2010})},\ \Eprint {http://arxiv.org/abs/0907.3149} {arXiv:0907.3149
  [astro-ph.CO]} \BibitemShut {NoStop}%
\bibitem [{\citenamefont {Moresco}\ \emph {et~al.}(2012)\citenamefont
  {Moresco}, \citenamefont {Verde}, \citenamefont {Pozzetti}, \citenamefont
  {Jimenez},\ and\ \citenamefont {Cimatti}}]{moresco2012new}%
  \BibitemOpen
  \bibfield  {author} {\bibinfo {author} {\bibfnamefont {M.}~\bibnamefont
  {Moresco}}, \bibinfo {author} {\bibfnamefont {L.}~\bibnamefont {Verde}},
  \bibinfo {author} {\bibfnamefont {L.}~\bibnamefont {Pozzetti}}, \bibinfo
  {author} {\bibfnamefont {R.}~\bibnamefont {Jimenez}}, \ and\ \bibinfo
  {author} {\bibfnamefont {A.}~\bibnamefont {Cimatti}},\ }\href {\doibase
  10.1088/1475-7516/2012/07/053} {\bibfield  {journal} {\bibinfo  {journal}
  {JCAP}\ }\textbf {\bibinfo {volume} {07}},\ \bibinfo {pages} {053} (\bibinfo
  {year} {2012})},\ \Eprint {http://arxiv.org/abs/1201.6658} {arXiv:1201.6658
  [astro-ph.CO]} \BibitemShut {NoStop}%
\bibitem [{\citenamefont {Zhang}\ \emph {et~al.}(2014)\citenamefont {Zhang},
  \citenamefont {Zhang}, \citenamefont {Yuan}, \citenamefont {Zhang},\ and\
  \citenamefont {Sun}}]{zhang2014four}%
  \BibitemOpen
  \bibfield  {author} {\bibinfo {author} {\bibfnamefont {C.}~\bibnamefont
  {Zhang}}, \bibinfo {author} {\bibfnamefont {H.}~\bibnamefont {Zhang}},
  \bibinfo {author} {\bibfnamefont {S.}~\bibnamefont {Yuan}}, \bibinfo {author}
  {\bibfnamefont {T.-J.}\ \bibnamefont {Zhang}}, \ and\ \bibinfo {author}
  {\bibfnamefont {Y.-C.}\ \bibnamefont {Sun}},\ }\href {\doibase
  10.1088/1674-4527/14/10/002} {\bibfield  {journal} {\bibinfo  {journal} {Res.
  Astron. Astrophys.}\ }\textbf {\bibinfo {volume} {14}},\ \bibinfo {pages}
  {1221} (\bibinfo {year} {2014})},\ \Eprint {http://arxiv.org/abs/1207.4541}
  {arXiv:1207.4541 [astro-ph.CO]} \BibitemShut {NoStop}%
\bibitem [{\citenamefont {Moresco}(2015)}]{moresco2015raising}%
  \BibitemOpen
  \bibfield  {author} {\bibinfo {author} {\bibfnamefont {M.}~\bibnamefont
  {Moresco}},\ }\href {\doibase 10.1093/mnrasl/slv037} {\bibfield  {journal}
  {\bibinfo  {journal} {Mon. Not. Roy. Astron. Soc.}\ }\textbf {\bibinfo
  {volume} {450}},\ \bibinfo {pages} {L16} (\bibinfo {year} {2015})},\ \Eprint
  {http://arxiv.org/abs/1503.01116} {arXiv:1503.01116 [astro-ph.CO]}
  \BibitemShut {NoStop}%
\bibitem [{\citenamefont {Moresco}\ \emph {et~al.}(2016)\citenamefont
  {Moresco}, \citenamefont {Pozzetti}, \citenamefont {Cimatti}, \citenamefont
  {Jimenez}, \citenamefont {Maraston}, \citenamefont {Verde}, \citenamefont
  {Thomas}, \citenamefont {Citro}, \citenamefont {Tojeiro},\ and\ \citenamefont
  {Wilkinson}}]{moresco20166}%
  \BibitemOpen
  \bibfield  {author} {\bibinfo {author} {\bibfnamefont {M.}~\bibnamefont
  {Moresco}}, \bibinfo {author} {\bibfnamefont {L.}~\bibnamefont {Pozzetti}},
  \bibinfo {author} {\bibfnamefont {A.}~\bibnamefont {Cimatti}}, \bibinfo
  {author} {\bibfnamefont {R.}~\bibnamefont {Jimenez}}, \bibinfo {author}
  {\bibfnamefont {C.}~\bibnamefont {Maraston}}, \bibinfo {author}
  {\bibfnamefont {L.}~\bibnamefont {Verde}}, \bibinfo {author} {\bibfnamefont
  {D.}~\bibnamefont {Thomas}}, \bibinfo {author} {\bibfnamefont
  {A.}~\bibnamefont {Citro}}, \bibinfo {author} {\bibfnamefont
  {R.}~\bibnamefont {Tojeiro}}, \ and\ \bibinfo {author} {\bibfnamefont
  {D.}~\bibnamefont {Wilkinson}},\ }\href {\doibase
  10.1088/1475-7516/2016/05/014} {\bibfield  {journal} {\bibinfo  {journal}
  {JCAP}\ }\textbf {\bibinfo {volume} {05}},\ \bibinfo {pages} {014} (\bibinfo
  {year} {2016})},\ \Eprint {http://arxiv.org/abs/1601.01701} {arXiv:1601.01701
  [astro-ph.CO]} \BibitemShut {NoStop}%
\bibitem [{\citenamefont {{M. Moresco}}()}]{hz}%
  \BibitemOpen
  \bibfield  {author} {\bibinfo {author} {\bibnamefont {{M. Moresco}}},\
  }\href@noop {} {}\bibinfo {howpublished}
  {\url{https://gitlab.com/mmoresco/CCcovariance}}\BibitemShut {NoStop}%
\bibitem [{\citenamefont {Scolnic}\ \emph {et~al.}(2018)\citenamefont
  {Scolnic}, \citenamefont {Jones}, \citenamefont {Rest}, \citenamefont {Pan},
  \citenamefont {Chornock}, \citenamefont {Foley}, \citenamefont {Huber},
  \citenamefont {Kessler}, \citenamefont {Narayan}, \citenamefont {Riess} \emph
  {et~al.}}]{scolnic2018complete}%
  \BibitemOpen
  \bibfield  {author} {\bibinfo {author} {\bibfnamefont {D.~M.}\ \bibnamefont
  {Scolnic}}, \bibinfo {author} {\bibfnamefont {D.}~\bibnamefont {Jones}},
  \bibinfo {author} {\bibfnamefont {A.}~\bibnamefont {Rest}}, \bibinfo {author}
  {\bibfnamefont {Y.}~\bibnamefont {Pan}}, \bibinfo {author} {\bibfnamefont
  {R.}~\bibnamefont {Chornock}}, \bibinfo {author} {\bibfnamefont
  {R.}~\bibnamefont {Foley}}, \bibinfo {author} {\bibfnamefont
  {M.}~\bibnamefont {Huber}}, \bibinfo {author} {\bibfnamefont
  {R.}~\bibnamefont {Kessler}}, \bibinfo {author} {\bibfnamefont
  {G.}~\bibnamefont {Narayan}}, \bibinfo {author} {\bibfnamefont
  {A.}~\bibnamefont {Riess}},  \emph {et~al.} (\bibinfo {collaboration}
  {Pan-STARRS1}),\ }\href {\doibase 10.3847/1538-4357/aab9bb} {\bibfield
  {journal} {\bibinfo  {journal} {Astrophys. J.}\ }\textbf {\bibinfo {volume}
  {859}},\ \bibinfo {pages} {101} (\bibinfo {year} {2018})},\ \Eprint
  {http://arxiv.org/abs/1710.00845} {arXiv:1710.00845 [astro-ph.CO]}
  \BibitemShut {NoStop}%
\bibitem [{\citenamefont {{Jones et. al. and Scolnic et.
  al.}}()}]{pantheon_data}%
  \BibitemOpen
  \bibfield  {author} {\bibinfo {author} {\bibnamefont {{Jones et. al. and
  Scolnic et. al.}}},\ }\href@noop {} {}\bibinfo {howpublished}
  {\url{https://archive.stsci.edu/prepds/ps1cosmo/}}\BibitemShut {NoStop}%
\bibitem [{\citenamefont {Dodelson}(2003)}]{dodelson2003modern}%
  \BibitemOpen
  \bibfield  {author} {\bibinfo {author} {\bibfnamefont {S.}~\bibnamefont
  {Dodelson}},\ }\href@noop {} {\emph {\bibinfo {title} {Modern cosmology}}}\
  (\bibinfo  {publisher} {Elsevier},\ \bibinfo {year} {2003})\BibitemShut
  {NoStop}%
\bibitem [{\citenamefont {Aubourg}\ \emph {et~al.}(2015)\citenamefont
  {Aubourg}, \citenamefont {Bailey}, \citenamefont {Bautista}, \citenamefont
  {Beutler}, \citenamefont {Bhardwaj}, \citenamefont {Bizyaev}, \citenamefont
  {Blanton}, \citenamefont {Blomqvist}, \citenamefont {Bolton}, \citenamefont
  {Bovy} \emph {et~al.}}]{aubourg2015cosmological}%
  \BibitemOpen
  \bibfield  {author} {\bibinfo {author} {\bibfnamefont {{\'E}.}~\bibnamefont
  {Aubourg}}, \bibinfo {author} {\bibfnamefont {S.}~\bibnamefont {Bailey}},
  \bibinfo {author} {\bibfnamefont {J.~E.}\ \bibnamefont {Bautista}}, \bibinfo
  {author} {\bibfnamefont {F.}~\bibnamefont {Beutler}}, \bibinfo {author}
  {\bibfnamefont {V.}~\bibnamefont {Bhardwaj}}, \bibinfo {author}
  {\bibfnamefont {D.}~\bibnamefont {Bizyaev}}, \bibinfo {author} {\bibfnamefont
  {M.}~\bibnamefont {Blanton}}, \bibinfo {author} {\bibfnamefont
  {M.}~\bibnamefont {Blomqvist}}, \bibinfo {author} {\bibfnamefont {A.~S.}\
  \bibnamefont {Bolton}}, \bibinfo {author} {\bibfnamefont {J.}~\bibnamefont
  {Bovy}},  \emph {et~al.},\ }\href@noop {} {\bibfield  {journal} {\bibinfo
  {journal} {Physical Review D}\ }\textbf {\bibinfo {volume} {92}},\ \bibinfo
  {pages} {123516} (\bibinfo {year} {2015})}\BibitemShut {NoStop}%
\bibitem [{\citenamefont {Alam}\ \emph {et~al.}(2017)\citenamefont {Alam},
  \citenamefont {Ata}, \citenamefont {Bailey}, \citenamefont {Beutler},
  \citenamefont {Bizyaev}, \citenamefont {Blazek}, \citenamefont {Bolton},
  \citenamefont {Brownstein}, \citenamefont {Burden}, \citenamefont {Chuang}
  \emph {et~al.}}]{alam2017clustering}%
  \BibitemOpen
  \bibfield  {author} {\bibinfo {author} {\bibfnamefont {S.}~\bibnamefont
  {Alam}}, \bibinfo {author} {\bibfnamefont {M.}~\bibnamefont {Ata}}, \bibinfo
  {author} {\bibfnamefont {S.}~\bibnamefont {Bailey}}, \bibinfo {author}
  {\bibfnamefont {F.}~\bibnamefont {Beutler}}, \bibinfo {author} {\bibfnamefont
  {D.}~\bibnamefont {Bizyaev}}, \bibinfo {author} {\bibfnamefont {J.~A.}\
  \bibnamefont {Blazek}}, \bibinfo {author} {\bibfnamefont {A.~S.}\
  \bibnamefont {Bolton}}, \bibinfo {author} {\bibfnamefont {J.~R.}\
  \bibnamefont {Brownstein}}, \bibinfo {author} {\bibfnamefont
  {A.}~\bibnamefont {Burden}}, \bibinfo {author} {\bibfnamefont {C.-H.}\
  \bibnamefont {Chuang}},  \emph {et~al.} (\bibinfo {collaboration} {BOSS}),\
  }\href {\doibase 10.1093/mnras/stx721} {\bibfield  {journal} {\bibinfo
  {journal} {Mon. Not. Roy. Astron. Soc.}\ }\textbf {\bibinfo {volume} {470}},\
  \bibinfo {pages} {2617} (\bibinfo {year} {2017})},\ \Eprint
  {http://arxiv.org/abs/1607.03155} {arXiv:1607.03155 [astro-ph.CO]}
  \BibitemShut {NoStop}%
\bibitem [{\citenamefont {de~Sainte~Agathe}\ \emph {et~al.}(2019)\citenamefont
  {de~Sainte~Agathe}, \citenamefont {Balland}, \citenamefont {du~Mas~des
  Bourboux}, \citenamefont {Blomqvist}, \citenamefont {Guy}, \citenamefont
  {Rich}, \citenamefont {Font-Ribera}, \citenamefont {Pieri}, \citenamefont
  {Bautista}, \citenamefont {Dawson} \emph {et~al.}}]{de2019baryon}%
  \BibitemOpen
  \bibfield  {author} {\bibinfo {author} {\bibfnamefont {V.}~\bibnamefont
  {de~Sainte~Agathe}}, \bibinfo {author} {\bibfnamefont {C.}~\bibnamefont
  {Balland}}, \bibinfo {author} {\bibfnamefont {H.}~\bibnamefont {du~Mas~des
  Bourboux}}, \bibinfo {author} {\bibfnamefont {M.}~\bibnamefont {Blomqvist}},
  \bibinfo {author} {\bibfnamefont {J.}~\bibnamefont {Guy}}, \bibinfo {author}
  {\bibfnamefont {J.}~\bibnamefont {Rich}}, \bibinfo {author} {\bibfnamefont
  {A.}~\bibnamefont {Font-Ribera}}, \bibinfo {author} {\bibfnamefont {M.~M.}\
  \bibnamefont {Pieri}}, \bibinfo {author} {\bibfnamefont {J.~E.}\ \bibnamefont
  {Bautista}}, \bibinfo {author} {\bibfnamefont {K.}~\bibnamefont {Dawson}},
  \emph {et~al.},\ }\href {\doibase 10.3847/1538-4357/ab1d49} {\bibfield
  {journal} {\bibinfo  {journal} {Astrophys. J.}\ }\textbf {\bibinfo {volume}
  {878}},\ \bibinfo {pages} {47} (\bibinfo {year} {2019})},\ \Eprint
  {http://arxiv.org/abs/1901.01950} {arXiv:1901.01950 [astro-ph.CO]}
  \BibitemShut {NoStop}%
\bibitem [{\citenamefont {Ata}\ \emph {et~al.}(2018)\citenamefont {Ata},
  \citenamefont {Baumgarten}, \citenamefont {Bautista}, \citenamefont
  {Beutler}, \citenamefont {Bizyaev}, \citenamefont {Blanton}, \citenamefont
  {Blazek}, \citenamefont {Bolton}, \citenamefont {Brinkmann}, \citenamefont
  {Brownstein} \emph {et~al.}}]{ata2018clustering}%
  \BibitemOpen
  \bibfield  {author} {\bibinfo {author} {\bibfnamefont {M.}~\bibnamefont
  {Ata}}, \bibinfo {author} {\bibfnamefont {F.}~\bibnamefont {Baumgarten}},
  \bibinfo {author} {\bibfnamefont {J.}~\bibnamefont {Bautista}}, \bibinfo
  {author} {\bibfnamefont {F.}~\bibnamefont {Beutler}}, \bibinfo {author}
  {\bibfnamefont {D.}~\bibnamefont {Bizyaev}}, \bibinfo {author} {\bibfnamefont
  {M.~R.}\ \bibnamefont {Blanton}}, \bibinfo {author} {\bibfnamefont {J.~A.}\
  \bibnamefont {Blazek}}, \bibinfo {author} {\bibfnamefont {A.~S.}\
  \bibnamefont {Bolton}}, \bibinfo {author} {\bibfnamefont {J.}~\bibnamefont
  {Brinkmann}}, \bibinfo {author} {\bibfnamefont {J.~R.}\ \bibnamefont
  {Brownstein}},  \emph {et~al.},\ }\href {\doibase 10.1093/mnras/stx2630}
  {\bibfield  {journal} {\bibinfo  {journal} {Mon. Not. Roy. Astron. Soc.}\
  }\textbf {\bibinfo {volume} {473}},\ \bibinfo {pages} {4773} (\bibinfo {year}
  {2018})},\ \Eprint {http://arxiv.org/abs/1705.06373} {arXiv:1705.06373
  [astro-ph.CO]} \BibitemShut {NoStop}%
\bibitem [{\citenamefont {Blomqvist}\ \emph {et~al.}(2019)\citenamefont
  {Blomqvist}, \citenamefont {Bourboux}, \citenamefont {Busca}, \citenamefont
  {Agathe}, \citenamefont {Rich}, \citenamefont {Balland}, \citenamefont
  {Bautista}, \citenamefont {Dawson}, \citenamefont {Font-Ribera},
  \citenamefont {Guy} \emph {et~al.}}]{blomqvist2019baryon}%
  \BibitemOpen
  \bibfield  {author} {\bibinfo {author} {\bibfnamefont {M.}~\bibnamefont
  {Blomqvist}}, \bibinfo {author} {\bibfnamefont {H.~d. M.~d.}\ \bibnamefont
  {Bourboux}}, \bibinfo {author} {\bibfnamefont {N.~G.}\ \bibnamefont {Busca}},
  \bibinfo {author} {\bibfnamefont {V.~d.~S.}\ \bibnamefont {Agathe}}, \bibinfo
  {author} {\bibfnamefont {J.}~\bibnamefont {Rich}}, \bibinfo {author}
  {\bibfnamefont {C.}~\bibnamefont {Balland}}, \bibinfo {author} {\bibfnamefont
  {J.~E.}\ \bibnamefont {Bautista}}, \bibinfo {author} {\bibfnamefont
  {K.}~\bibnamefont {Dawson}}, \bibinfo {author} {\bibfnamefont
  {A.}~\bibnamefont {Font-Ribera}}, \bibinfo {author} {\bibfnamefont
  {J.}~\bibnamefont {Guy}},  \emph {et~al.},\ }\href@noop {} {\bibfield
  {journal} {\bibinfo  {journal} {arXiv preprint arXiv:1904.03430}\ } (\bibinfo
  {year} {2019})}\BibitemShut {NoStop}%
\bibitem [{\citenamefont {Beutler}\ \emph {et~al.}(2011)\citenamefont
  {Beutler}, \citenamefont {Blake}, \citenamefont {Colless}, \citenamefont
  {Jones}, \citenamefont {Staveley-Smith}, \citenamefont {Campbell},
  \citenamefont {Parker}, \citenamefont {Saunders},\ and\ \citenamefont
  {Watson}}]{beutler20116df}%
  \BibitemOpen
  \bibfield  {author} {\bibinfo {author} {\bibfnamefont {F.}~\bibnamefont
  {Beutler}}, \bibinfo {author} {\bibfnamefont {C.}~\bibnamefont {Blake}},
  \bibinfo {author} {\bibfnamefont {M.}~\bibnamefont {Colless}}, \bibinfo
  {author} {\bibfnamefont {D.~H.}\ \bibnamefont {Jones}}, \bibinfo {author}
  {\bibfnamefont {L.}~\bibnamefont {Staveley-Smith}}, \bibinfo {author}
  {\bibfnamefont {L.}~\bibnamefont {Campbell}}, \bibinfo {author}
  {\bibfnamefont {Q.}~\bibnamefont {Parker}}, \bibinfo {author} {\bibfnamefont
  {W.}~\bibnamefont {Saunders}}, \ and\ \bibinfo {author} {\bibfnamefont
  {F.}~\bibnamefont {Watson}},\ }\href {\doibase
  10.1111/j.1365-2966.2011.19250.x} {\bibfield  {journal} {\bibinfo  {journal}
  {Mon. Not. Roy. Astron. Soc.}\ }\textbf {\bibinfo {volume} {416}},\ \bibinfo
  {pages} {3017} (\bibinfo {year} {2011})},\ \Eprint
  {http://arxiv.org/abs/1106.3366} {arXiv:1106.3366 [astro-ph.CO]} \BibitemShut
  {NoStop}%
\bibitem [{\citenamefont {Anderson}\ \emph {et~al.}(2014)\citenamefont
  {Anderson}, \citenamefont {Aubourg}, \citenamefont {Bailey}, \citenamefont
  {Beutler}, \citenamefont {Bhardwaj}, \citenamefont {Blanton}, \citenamefont
  {Bolton}, \citenamefont {Brinkmann}, \citenamefont {Brownstein},
  \citenamefont {Burden} \emph {et~al.}}]{anderson2014clustering}%
  \BibitemOpen
  \bibfield  {author} {\bibinfo {author} {\bibfnamefont {L.}~\bibnamefont
  {Anderson}}, \bibinfo {author} {\bibfnamefont {E.}~\bibnamefont {Aubourg}},
  \bibinfo {author} {\bibfnamefont {S.}~\bibnamefont {Bailey}}, \bibinfo
  {author} {\bibfnamefont {F.}~\bibnamefont {Beutler}}, \bibinfo {author}
  {\bibfnamefont {V.}~\bibnamefont {Bhardwaj}}, \bibinfo {author}
  {\bibfnamefont {M.}~\bibnamefont {Blanton}}, \bibinfo {author} {\bibfnamefont
  {A.~S.}\ \bibnamefont {Bolton}}, \bibinfo {author} {\bibfnamefont
  {J.}~\bibnamefont {Brinkmann}}, \bibinfo {author} {\bibfnamefont {J.~R.}\
  \bibnamefont {Brownstein}}, \bibinfo {author} {\bibfnamefont
  {A.}~\bibnamefont {Burden}},  \emph {et~al.} (\bibinfo {collaboration}
  {BOSS}),\ }\href {\doibase 10.1093/mnras/stu523} {\bibfield  {journal}
  {\bibinfo  {journal} {Mon. Not. Roy. Astron. Soc.}\ }\textbf {\bibinfo
  {volume} {441}},\ \bibinfo {pages} {24} (\bibinfo {year} {2014})},\ \Eprint
  {http://arxiv.org/abs/1312.4877} {arXiv:1312.4877 [astro-ph.CO]} \BibitemShut
  {NoStop}%
\bibitem [{\citenamefont {{A. Slosar and J. A. Vazquez}}()}]{simplemc}%
  \BibitemOpen
  \bibfield  {author} {\bibinfo {author} {\bibnamefont {{A. Slosar and J. A.
  Vazquez}}},\ }\href@noop {} {}\bibinfo {howpublished}
  {\url{https://github.com/ja-vazquez/SimpleMC}}\BibitemShut {NoStop}%
\bibitem [{\citenamefont {Speagle}(2020)}]{speagle2020dynesty}%
  \BibitemOpen
  \bibfield  {author} {\bibinfo {author} {\bibfnamefont {J.~S.}\ \bibnamefont
  {Speagle}},\ }\href {\doibase 10.1093/mnras/staa278} {\bibfield  {journal}
  {\bibinfo  {journal} {Mon. Not. Roy. Astron. Soc.}\ }\textbf {\bibinfo
  {volume} {493}},\ \bibinfo {pages} {3132} (\bibinfo {year} {2020})},\ \Eprint
  {http://arxiv.org/abs/1904.02180} {arXiv:1904.02180 [astro-ph.IM]}
  \BibitemShut {NoStop}%
\bibitem [{\citenamefont {{J. Speagle}}()}]{dynesty}%
  \BibitemOpen
  \bibfield  {author} {\bibinfo {author} {\bibnamefont {{J. Speagle}}},\
  }\href@noop {} {}\bibinfo {howpublished}
  {\url{https://dynesty.readthedocs.io/en/stable/index.html}}\BibitemShut
  {NoStop}%
\bibitem [{\citenamefont {Handley}(2019)}]{handley2019fgivenx}%
  \BibitemOpen
  \bibfield  {author} {\bibinfo {author} {\bibfnamefont {W.}~\bibnamefont
  {Handley}},\ }\href@noop {} {\bibfield  {journal} {\bibinfo  {journal} {arXiv
  preprint arXiv:1908.01711}\ } (\bibinfo {year} {2019})}\BibitemShut {NoStop}%
\bibitem [{\citenamefont {Padilla}\ \emph {et~al.}(2021)\citenamefont
  {Padilla}, \citenamefont {Tellez}, \citenamefont {Escamilla},\ and\
  \citenamefont {Vazquez}}]{Padilla:2019mgi}%
  \BibitemOpen
  \bibfield  {author} {\bibinfo {author} {\bibfnamefont {L.~E.}\ \bibnamefont
  {Padilla}}, \bibinfo {author} {\bibfnamefont {L.~O.}\ \bibnamefont {Tellez}},
  \bibinfo {author} {\bibfnamefont {L.~A.}\ \bibnamefont {Escamilla}}, \ and\
  \bibinfo {author} {\bibfnamefont {J.~A.}\ \bibnamefont {Vazquez}},\ }\href
  {\doibase 10.3390/universe7070213} {\bibfield  {journal} {\bibinfo  {journal}
  {Universe}\ }\textbf {\bibinfo {volume} {7}},\ \bibinfo {pages} {213}
  (\bibinfo {year} {2021})},\ \Eprint {http://arxiv.org/abs/1903.11127}
  {arXiv:1903.11127 [astro-ph.CO]} \BibitemShut {NoStop}%
\bibitem [{\citenamefont {Akarsu}\ \emph
  {et~al.}(2022{\natexlab{b}})\citenamefont {Akarsu}, \citenamefont {Colgain},
  \citenamefont {\"Ozulker}, \citenamefont {Thakur},\ and\ \citenamefont
  {Yin}}]{Akarsu:2022lhx}%
  \BibitemOpen
  \bibfield  {author} {\bibinfo {author} {\bibfnamefont {O.}~\bibnamefont
  {Akarsu}}, \bibinfo {author} {\bibfnamefont {E.~O.}\ \bibnamefont {Colgain}},
  \bibinfo {author} {\bibfnamefont {E.}~\bibnamefont {\"Ozulker}}, \bibinfo
  {author} {\bibfnamefont {S.}~\bibnamefont {Thakur}}, \ and\ \bibinfo {author}
  {\bibfnamefont {L.}~\bibnamefont {Yin}},\ }\href@noop {} {\  (\bibinfo {year}
  {2022}{\natexlab{b}})},\ \Eprint {http://arxiv.org/abs/2207.10609}
  {arXiv:2207.10609 [astro-ph.CO]} \BibitemShut {NoStop}%
\bibitem [{\citenamefont {M\"ortsell}\ and\ \citenamefont
  {Dhawan}(2018)}]{mortsell2018does}%
  \BibitemOpen
  \bibfield  {author} {\bibinfo {author} {\bibfnamefont {E.}~\bibnamefont
  {M\"ortsell}}\ and\ \bibinfo {author} {\bibfnamefont {S.}~\bibnamefont
  {Dhawan}},\ }\href {\doibase 10.1088/1475-7516/2018/09/025} {\bibfield
  {journal} {\bibinfo  {journal} {JCAP}\ }\textbf {\bibinfo {volume} {09}},\
  \bibinfo {pages} {025} (\bibinfo {year} {2018})},\ \Eprint
  {http://arxiv.org/abs/1801.07260} {arXiv:1801.07260 [astro-ph.CO]}
  \BibitemShut {NoStop}%
\bibitem [{\citenamefont {Marciu}(2016)}]{marciu2016quintom}%
  \BibitemOpen
  \bibfield  {author} {\bibinfo {author} {\bibfnamefont {M.}~\bibnamefont
  {Marciu}},\ }\href {\doibase 10.1103/PhysRevD.93.123006} {\bibfield
  {journal} {\bibinfo  {journal} {Phys. Rev. D}\ }\textbf {\bibinfo {volume}
  {93}},\ \bibinfo {pages} {123006} (\bibinfo {year} {2016})}\BibitemShut
  {NoStop}%
\bibitem [{\citenamefont {Visinelli}\ \emph {et~al.}(2019)\citenamefont
  {Visinelli}, \citenamefont {Vagnozzi},\ and\ \citenamefont
  {Danielsson}}]{visinelli2019revisiting}%
  \BibitemOpen
  \bibfield  {author} {\bibinfo {author} {\bibfnamefont {L.}~\bibnamefont
  {Visinelli}}, \bibinfo {author} {\bibfnamefont {S.}~\bibnamefont {Vagnozzi}},
  \ and\ \bibinfo {author} {\bibfnamefont {U.}~\bibnamefont {Danielsson}},\
  }\href {\doibase 10.3390/sym11081035} {\bibfield  {journal} {\bibinfo
  {journal} {Symmetry}\ }\textbf {\bibinfo {volume} {11}},\ \bibinfo {pages}
  {1035} (\bibinfo {year} {2019})},\ \Eprint {http://arxiv.org/abs/1907.07953}
  {arXiv:1907.07953 [astro-ph.CO]} \BibitemShut {NoStop}%
\bibitem [{\citenamefont {Dutta}\ \emph {et~al.}(2020)\citenamefont {Dutta},
  \citenamefont {Ruchika}, \citenamefont {Roy}, \citenamefont {Sen},\ and\
  \citenamefont {Sheikh-Jabbari}}]{Dutta:2018vmq}%
  \BibitemOpen
  \bibfield  {author} {\bibinfo {author} {\bibfnamefont {K.}~\bibnamefont
  {Dutta}}, \bibinfo {author} {\bibnamefont {Ruchika}}, \bibinfo {author}
  {\bibfnamefont {A.}~\bibnamefont {Roy}}, \bibinfo {author} {\bibfnamefont
  {A.~A.}\ \bibnamefont {Sen}}, \ and\ \bibinfo {author} {\bibfnamefont
  {M.~M.}\ \bibnamefont {Sheikh-Jabbari}},\ }\href {\doibase
  10.1007/s10714-020-2665-4} {\bibfield  {journal} {\bibinfo  {journal} {Gen.
  Rel. Grav.}\ }\textbf {\bibinfo {volume} {52}},\ \bibinfo {pages} {15}
  (\bibinfo {year} {2020})},\ \Eprint {http://arxiv.org/abs/1808.06623}
  {arXiv:1808.06623 [astro-ph.CO]} \BibitemShut {NoStop}%
\bibitem [{\citenamefont {Handley}(2021)}]{handley2021curvature}%
  \BibitemOpen
  \bibfield  {author} {\bibinfo {author} {\bibfnamefont {W.}~\bibnamefont
  {Handley}},\ }\href {\doibase 10.1103/PhysRevD.103.L041301} {\bibfield
  {journal} {\bibinfo  {journal} {Phys. Rev. D}\ }\textbf {\bibinfo {volume}
  {103}},\ \bibinfo {pages} {L041301} (\bibinfo {year} {2021})},\ \Eprint
  {http://arxiv.org/abs/1908.09139} {arXiv:1908.09139 [astro-ph.CO]}
  \BibitemShut {NoStop}%
\bibitem [{\citenamefont {Acquaviva}\ \emph
  {et~al.}(2021{\natexlab{b}})\citenamefont {Acquaviva}, \citenamefont
  {Akarsu}, \citenamefont {Katirci},\ and\ \citenamefont
  {Vazquez}}]{acquaviva2021simple}%
  \BibitemOpen
  \bibfield  {author} {\bibinfo {author} {\bibfnamefont {G.}~\bibnamefont
  {Acquaviva}}, \bibinfo {author} {\bibfnamefont {O.}~\bibnamefont {Akarsu}},
  \bibinfo {author} {\bibfnamefont {N.}~\bibnamefont {Katirci}}, \ and\
  \bibinfo {author} {\bibfnamefont {J.~A.}\ \bibnamefont {Vazquez}},\ }\href
  {\doibase 10.1103/PhysRevD.104.023505} {\bibfield  {journal} {\bibinfo
  {journal} {Phys. Rev. D}\ }\textbf {\bibinfo {volume} {104}},\ \bibinfo
  {pages} {023505} (\bibinfo {year} {2021}{\natexlab{b}})},\ \Eprint
  {http://arxiv.org/abs/2104.02623} {arXiv:2104.02623 [astro-ph.CO]}
  \BibitemShut {NoStop}%
\bibitem [{\citenamefont {Gomez-Valent}\ \emph {et~al.}(2015)\citenamefont
  {Gomez-Valent}, \citenamefont {Karimkhani},\ and\ \citenamefont
  {Sola}}]{gomez2015background}%
  \BibitemOpen
  \bibfield  {author} {\bibinfo {author} {\bibfnamefont {A.}~\bibnamefont
  {Gomez-Valent}}, \bibinfo {author} {\bibfnamefont {E.}~\bibnamefont
  {Karimkhani}}, \ and\ \bibinfo {author} {\bibfnamefont {J.}~\bibnamefont
  {Sola}},\ }\href {\doibase 10.1088/1475-7516/2015/12/048} {\bibfield
  {journal} {\bibinfo  {journal} {JCAP}\ }\textbf {\bibinfo {volume} {12}},\
  \bibinfo {pages} {048} (\bibinfo {year} {2015})},\ \Eprint
  {http://arxiv.org/abs/1509.03298} {arXiv:1509.03298 [gr-qc]} \BibitemShut
  {NoStop}%
\bibitem [{\citenamefont {Wang}\ \emph {et~al.}(2019)\citenamefont {Wang},
  \citenamefont {Zhang},\ and\ \citenamefont {Meng}}]{wang2019searching}%
  \BibitemOpen
  \bibfield  {author} {\bibinfo {author} {\bibfnamefont {D.}~\bibnamefont
  {Wang}}, \bibinfo {author} {\bibfnamefont {W.}~\bibnamefont {Zhang}}, \ and\
  \bibinfo {author} {\bibfnamefont {X.-H.}\ \bibnamefont {Meng}},\ }\href
  {\doibase 10.1140/epjc/s10052-019-6726-3} {\bibfield  {journal} {\bibinfo
  {journal} {Eur. Phys. J. C}\ }\textbf {\bibinfo {volume} {79}},\ \bibinfo
  {pages} {211} (\bibinfo {year} {2019})},\ \Eprint
  {http://arxiv.org/abs/1903.08913} {arXiv:1903.08913 [astro-ph.CO]}
  \BibitemShut {NoStop}%
\bibitem [{\citenamefont {Freedman}\ \emph {et~al.}(2019)\citenamefont
  {Freedman} \emph {et~al.}}]{Freedman:2019jwv}%
  \BibitemOpen
  \bibfield  {author} {\bibinfo {author} {\bibfnamefont {W.~L.}\ \bibnamefont
  {Freedman}} \emph {et~al.},\ }\href {\doibase 10.3847/1538-4357/ab2f73} {\
  (\bibinfo {year} {2019}),\ 10.3847/1538-4357/ab2f73},\ \Eprint
  {http://arxiv.org/abs/1907.05922} {arXiv:1907.05922 [astro-ph.CO]}
  \BibitemShut {NoStop}%
\bibitem [{\citenamefont {Qi}\ \emph {et~al.}(2018)\citenamefont {Qi},
  \citenamefont {Cao}, \citenamefont {Biesiada}, \citenamefont {Xu},
  \citenamefont {Wu}, \citenamefont {Zhang},\ and\ \citenamefont
  {Zhu}}]{qi2018parameterized}%
  \BibitemOpen
  \bibfield  {author} {\bibinfo {author} {\bibfnamefont {J.-Z.}\ \bibnamefont
  {Qi}}, \bibinfo {author} {\bibfnamefont {S.}~\bibnamefont {Cao}}, \bibinfo
  {author} {\bibfnamefont {M.}~\bibnamefont {Biesiada}}, \bibinfo {author}
  {\bibfnamefont {T.}~\bibnamefont {Xu}}, \bibinfo {author} {\bibfnamefont
  {Y.}~\bibnamefont {Wu}}, \bibinfo {author} {\bibfnamefont {S.}~\bibnamefont
  {Zhang}}, \ and\ \bibinfo {author} {\bibfnamefont {Z.-H.}\ \bibnamefont
  {Zhu}},\ }\href@noop {} {\bibfield  {journal} {\bibinfo  {journal} {arXiv
  preprint arXiv:1803.04109}\ } (\bibinfo {year} {2018})}\BibitemShut {NoStop}%
\bibitem [{\citenamefont {Mukherjee}(2021)}]{mukherjee2021kinematical}%
  \BibitemOpen
  \bibfield  {author} {\bibinfo {author} {\bibfnamefont {A.}~\bibnamefont
  {Mukherjee}},\ }\href {\doibase 10.1140/epjp/s13360-021-01269-3} {\bibfield
  {journal} {\bibinfo  {journal} {Eur. Phys. J. Plus}\ }\textbf {\bibinfo
  {volume} {136}},\ \bibinfo {pages} {300} (\bibinfo {year} {2021})},\ \Eprint
  {http://arxiv.org/abs/2002.12063} {arXiv:2002.12063 [astro-ph.CO]}
  \BibitemShut {NoStop}%
\bibitem [{\citenamefont {Bernal}\ \emph {et~al.}(2017)\citenamefont {Bernal},
  \citenamefont {Cardenas},\ and\ \citenamefont {Motta}}]{bernal2017asymmetry}%
  \BibitemOpen
  \bibfield  {author} {\bibinfo {author} {\bibfnamefont {C.}~\bibnamefont
  {Bernal}}, \bibinfo {author} {\bibfnamefont {V.~H.}\ \bibnamefont
  {Cardenas}}, \ and\ \bibinfo {author} {\bibfnamefont {V.}~\bibnamefont
  {Motta}},\ }\href {\doibase 10.1016/j.physletb.2016.12.008} {\bibfield
  {journal} {\bibinfo  {journal} {Phys. Lett. B}\ }\textbf {\bibinfo {volume}
  {765}},\ \bibinfo {pages} {163} (\bibinfo {year} {2017})},\ \Eprint
  {http://arxiv.org/abs/1606.07333} {arXiv:1606.07333 [astro-ph.CO]}
  \BibitemShut {NoStop}%
\bibitem [{\citenamefont {Aghanim}\ \emph {et~al.}(2020)\citenamefont {Aghanim}
  \emph {et~al.}}]{Planck:2018vyg}%
  \BibitemOpen
  \bibfield  {author} {\bibinfo {author} {\bibfnamefont {N.}~\bibnamefont
  {Aghanim}} \emph {et~al.} (\bibinfo {collaboration} {Planck}),\ }\href
  {\doibase 10.1051/0004-6361/201833910} {\bibfield  {journal} {\bibinfo
  {journal} {Astron. Astrophys.}\ }\textbf {\bibinfo {volume} {641}},\ \bibinfo
  {pages} {A6} (\bibinfo {year} {2020})},\ \bibinfo {note} {[Erratum:
  Astron.Astrophys. 652, C4 (2021)]},\ \Eprint
  {http://arxiv.org/abs/1807.06209} {arXiv:1807.06209 [astro-ph.CO]}
  \BibitemShut {NoStop}%
\bibitem [{\citenamefont {Evslin}(2017)}]{Evslin:2016gre}%
  \BibitemOpen
  \bibfield  {author} {\bibinfo {author} {\bibfnamefont {J.}~\bibnamefont
  {Evslin}},\ }\href {\doibase 10.1088/1475-7516/2017/04/024} {\bibfield
  {journal} {\bibinfo  {journal} {JCAP}\ }\textbf {\bibinfo {volume} {04}},\
  \bibinfo {pages} {024} (\bibinfo {year} {2017})},\ \Eprint
  {http://arxiv.org/abs/1604.02809} {arXiv:1604.02809 [astro-ph.CO]}
  \BibitemShut {NoStop}%
\bibitem [{\citenamefont {Cuceu}\ \emph {et~al.}(2019)\citenamefont {Cuceu},
  \citenamefont {Farr}, \citenamefont {Lemos},\ and\ \citenamefont
  {Font-Ribera}}]{Cuceu:2019for}%
  \BibitemOpen
  \bibfield  {author} {\bibinfo {author} {\bibfnamefont {A.}~\bibnamefont
  {Cuceu}}, \bibinfo {author} {\bibfnamefont {J.}~\bibnamefont {Farr}},
  \bibinfo {author} {\bibfnamefont {P.}~\bibnamefont {Lemos}}, \ and\ \bibinfo
  {author} {\bibfnamefont {A.}~\bibnamefont {Font-Ribera}},\ }\href {\doibase
  10.1088/1475-7516/2019/10/044} {\bibfield  {journal} {\bibinfo  {journal}
  {JCAP}\ }\textbf {\bibinfo {volume} {10}},\ \bibinfo {pages} {044} (\bibinfo
  {year} {2019})},\ \Eprint {http://arxiv.org/abs/1906.11628} {arXiv:1906.11628
  [astro-ph.CO]} \BibitemShut {NoStop}%
\end{thebibliography}%


%

\appendix

\section{Capturing features with the right amount of parameters}

\luis{Given the mathematical structure of the binning scheme and linear interpolation of our reconstructions there are some subtleties that are not obvious at first glance. In particular one that has an effect to this work: a bigger number of parameters does not necessarily means a better fitness to the data. We can only guarantee it to be true when the bigger number of parameters is a multiple of the one being compared. For example if we compare a reconstruction with $2$ parameters we can only guarantee that the ones with $2n, n \in \mathbb{N}$ parameters will perform better at fitting the data (and this is true for any number of parameters, not only 2). This is because the bins/nodes are all equally spaced in the redshift range $[0,3]$. When we use, lets say 2 and 4 parameters, the reconstruction with 2 having parameter positions of $(1.5,3.0)$ is a special case of the one with 4 with parameters positions of $(0.75, 1.5, 2.25, 3.0)$, and the same is true for any multiple of 2.}

\begin{figure}[h!]
    \begin{center}
     \includegraphics[ width=9.cm, height=5.cm]{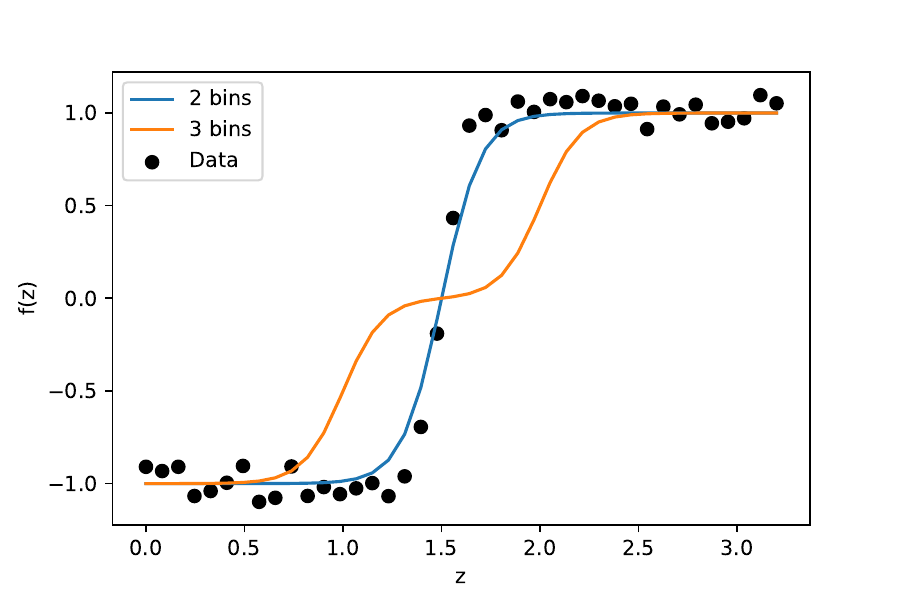}
    \end{center}
    \caption{Visual representation of the possibility of underfitting data with more parameters.
    }\label{fig:visual_example}
\end{figure}

\luis{This also means that there may be some features that a certain number of bins/nodes will not be able to capture since they are located in a region not accessible by the larger number of bins/nodes. As seen in Fig \ref{fig:visual_example} where the data (black dots) are being modeled with 2 and 3 bins. The data clearly show a transition in $z=1.5$ and the 3 bins cannot, by design, correctly model this transition since the positions of the bins interfere. It is also important to note that these problems appear with an interpolation reconstruction (linear, cubic and so forth). This happens almost exclusively to reconstructions with a low number of nodes/bins because as more parameters are utilized the resolution of the reconstruction becomes a lot better.}

\luis{Everything here discussed also applies for the nodes in the interpolation and for the reconstructed density.}

\begin{figure*}[h!]
\captionsetup{justification=raggedright,singlelinecheck=false,font=footnotesize}
    \centering
    \makebox[11cm][c]{
    \includegraphics[trim = 5mm  0mm 25mm 0mm, clip, width=6.cm, height=5cm]{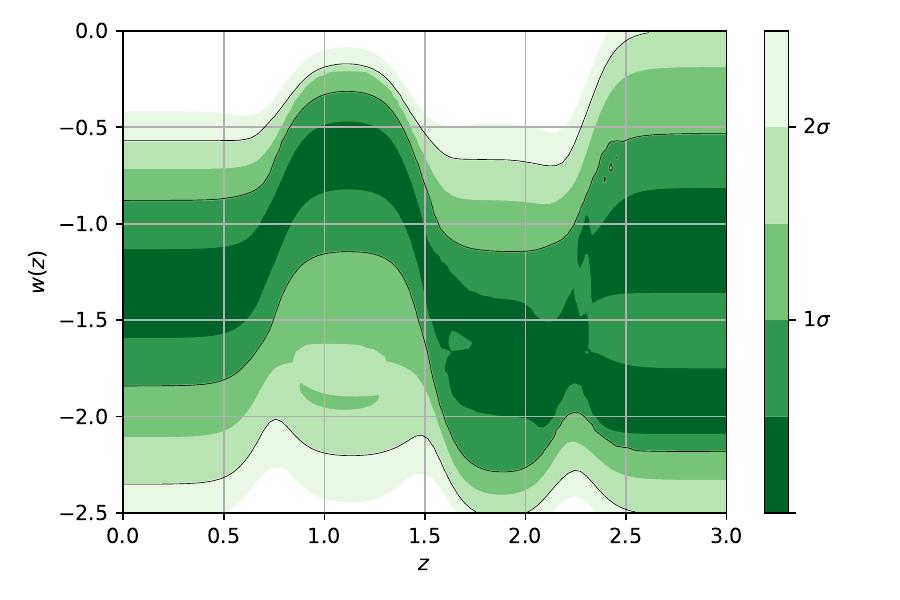}
    \includegraphics[trim = 5mm  0mm 25mm 0mm, clip, width=6.cm, height=5cm]{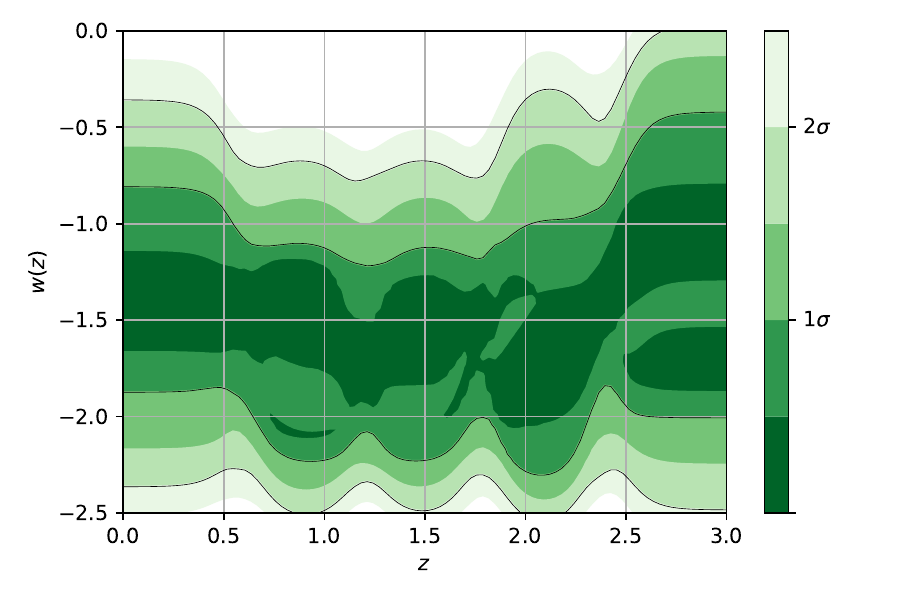}
    \includegraphics[trim = 5mm  0mm 5mm 0mm, clip, width=6.cm, height=5cm]{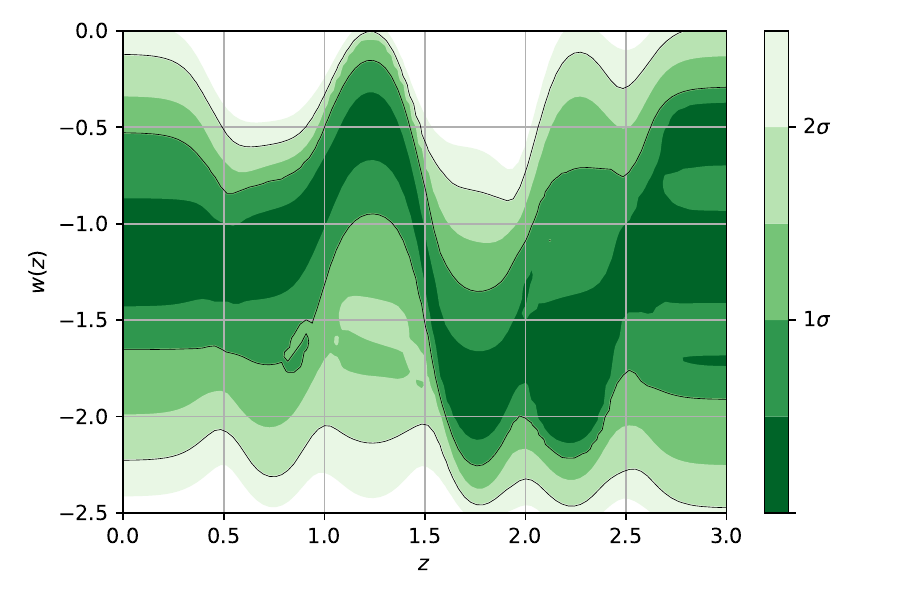}
    }
\caption{ Functional posteriors of the EoS reconstruction, from up to down, with 4, 5 and 6 bins using only Hubble parameter data. 
    }\label{fig:comparison_456bins}
\end{figure*}

\luis{Even if these effects could seem generally unimportant they are quite relevant to this work since they are both present. When reconstructing either the EoS or the density with 4 and 5 parameters we should expect a better fit to the data with 5 parameters because it has more degrees of freedom, but the $\Delta \chi^2$ says otherwise. By separating and analyzing the $\Delta \chi^2$ in its components via equation \ref{eq:chi_sq_components} we see some differences as expected, but the component responsible for the bad fitness when utilizing 5 bins is the $\Delta \chi^2_{H}$ with a difference of $3.36$ when compared to the 4 bin reconstruction (for reference we also have $\Delta \chi^2_{SN}=-1.38$ and $\Delta \chi^2_{BAO}=-0.4$ in favor of the 5 bin reconstruction). This indicates that there might be some feature present in the 4 bin reconstruction which favours it specifically with the Hubble parameter data, and it is not present in the 5 bin one. This feature is also present in the 6 bin reconstruction, as represented by its $\Delta \chi^2_{H}=5.26$ (when compared with the 5 bin one). }

\luis{The absent feature becomes obvious when reconstructing the EoS with 4, 5 and 6 bins with only Hubble parameter data, which is the data where 5 bins has trouble with. The functional posterior of these reconstructions can be seen in Fig \ref{fig:comparison_456bins}. Paying attention to the 1$\sigma$ region of 4 and 6 bins a bump can be seen followed by a slump in the interval $0.7<z<2.0$. The 5 bin reconstruction is completely missing this bump and subsequent slump. As explained at the start of the appendix, the reason for the inability with 5 bins to reproduce this behaviour comes from its bins' positions. The important bin is the one that starts in 1.2 and ends in 1.8, since the transition from bump to slump happens in $z=1.5$ it is impossible for this bin to correctly capture such trait.}


\end{document}